\documentclass[a4paper]{article}

\usepackage[authoryear,round]{natbib}
\usepackage{lscape}
\usepackage{array}
\usepackage{bbold}
\usepackage{enumitem}
\usepackage{bbm}
\usepackage{dsfont}
\usepackage{tikz}
\usepackage{url}
\usepackage{pgfplots}
\usepackage{amsmath}
\usepackage{layout}
\usepackage{geometry}
\usepackage{xcolor}
\usepackage{authblk}
\usepackage{hyperref}
\usepackage[export]{adjustbox}

\definecolor{paraviewwhite}{RGB}{207,204,197}
\definecolor{bluecolorrgb}{RGB}{74,134,173}

\DeclareMathAlphabet\mathbfcal{OMS}{cmsy}{b}{n}

\hypersetup{
    colorlinks=true,
    linkcolor=bluecolorrgb,
    citecolor=bluecolorrgb,
}

\begin{document}



\title{A unified two-scale gas-liquid multi-fluid model with capillarity and interface regularization through a mass transfer between scales}  



%

\author[1]{Arthur Loison}
\author[2]{Samuel Kokh}
\author[1]{Teddy Pichard}
\author[1]{Marc Massot}
\affil[1]{
    CMAP, CNRS, Ecole polytechnique, Institut Polytechnique de Paris\\
    91120 Palaiseau, France
}
\affil[2]{
    Université Paris-Saclay, CEA, Service de Génie Logiciel pour la Simulation\\
    91191 Gif-sur-Yvette, France
}
\date{}                     
\setcounter{Maxaffil}{0}
\renewcommand\Affilfont{\itshape\small}

\maketitle
\begin{abstract}
    In this contribution, we derive a gas-liquid two-scale multi-fluid model with capillarity effects to enable a novel interface regularization approach for multi-fluid models.
    As this unified modelling is capable of switching from the interface representation of a separated to a  disperse regime it lays a new way of modelling regime transitions as it occurs in atomization processes.
    %
    Above a preset length threshold at \textit{large scale}, a multi-fluid diffuse interface model resolves the dynamics of the interface while, at \textit{small-scale}, a set of geometric variables is used to characterize the interface geometry. 
    These variables result from a reduced-order modelling of the small-scale kinetic equation that describes a collection of liquid inclusions.
    The flow model can be viewed as a two-phase two-scale mixture, and the equations of motion are obtained thanks to the Hamilton's Stationary Action Principle, which requires to specify the kinetic and potential energies at play.
    We particularly focus on modelling the effects of capillarity on the mixture's energy by including dependencies on additional variables accounting for the interface's geometry at both scales.
    The regularization of the large-scale interface is then introduced as a local and dissipative process.
    The local curvature is limited \textit{via} a relaxation toward a modified Laplace equilibrium such that an inter-scale mass transfer is triggered when the mean curvature is too high.
    We propose an original numerical method and assess the properties and potential of the modelling strategy on the relevant test-case of a two-dimensional liquid column in a compressible gas flow.
\end{abstract}

\newcommand{\vPot}{\phi} 
\newcommand{\surfPot}{q} 
\newcommand{\adimSurfEl}{d\Omega(s)} 
\newcommand{\eulerChar}{\varsigma}

\newcommand{\SpaceDomain}{\Omega}
\newcommand{\SDF}{F}
\newcommand{\DSDF}{F^d}
\newcommand{\NDF}{f}
\newcommand{\NDFwoV}{n}

\newcommand{\harmonicOrder}{l}

\newcommand{\vel}{\boldsymbol{u}}
\newcommand{\lag}{\mathcal{L}}
\newcommand{\tEntropy}{\mathcal{H}}

\newcommand{\coorPhi}{\varphi}

\newcommand{\We}{\mbox{\textit{We}}}
\newcommand{\Ma}{\mbox{\textit{Ma}}}
\newcommand{\St}{\mbox{\textit{St}}}

\newcommand{\ie}{\textit{i.e. } }
\newcommand{\eg}{\textit{e.g. } }
\newcommand{\via}{\textit{via }}
\newcommand{\apriori}{\textit{a priori }}

\newcommand{\SigG}{\Sigma\left<G\right>}
\newcommand{\SigH}{\Sigma\left<H\right>}
\newcommand{\SigPerpEq}{\Sigma_{\perp,0}}
\newcommand{\DSigPerp}{(\Delta\Sigma_{\perp})}
\newcommand{\DtSigPerp}{(\Delta_t\Sigma_{\perp})}
\newcommand{\SigHPerpEq}{(\Sigma\left<H\right>)_{\perp,0}}
\newcommand{\DSigHPerp}{(\Delta\Sigma\left<H\right>)_{\perp}}
\newcommand{\DtSigHPerp}{(\Delta_t\Sigma\left<H\right>)_{\perp}}
\newcommand{\VFracd}{\alpha_1^d}

\newcommand{\VarTraj}{\boldsymbol{\eta}}
\newcommand{\Action}{\mathcal{A}}
\newcommand{\lagDynVar}{\chi}

\newcommand{\bnabla}{\boldsymbol{\nabla}}
\newcommand{\bcdot}{\boldsymbol{\cdot}}

\newcommand{\bK}{\boldsymbol{K}}
\newcommand{\Id}{\boldsymbol{I}}

\newcommand{\br}{\boldsymbol{r}}
\newcommand{\bq}{\boldsymbol{q}}
\newcommand{\bF}{\boldsymbol{F}}

\section{Introduction}\label{sec:intro}
%
%
%
%
%
Two-phase flows of liquid and gaseous phases appear in a variety of industrial applications such as the injection of liquid in combustion chambers \citep{tomar_multiscale_2010,providakis_characterization_2012,fiorina_modeling_2016,shinjo_recent_2018,sakano_evaluation_2022,hoarau_direct_2023}.
These flows are characterized by a complex multiscale interface dynamics which either holds too much information if solved entirely or is efficiently modelled thanks to the assumption of a specific regime \citep{ishii_thermo-fluid_1975}.
In the \textit{separated regime}, the length-scale of the interface dynamics is comparable or larger than the length-scales of the bulk phase in the sense that no arbitrary small length-scales arise from its dynamics.
Such arbitrarily small length-scales occur in the \textit{mixed regime} when the interface surface undergoes topological changes through pinching, filament break-up or apparition of holes.
Finally, the \textit{disperse regime} correspond to small inclusions of one phase carried by the other.
The complexity of two-phase flow dynamics, that we are here interested in, stands in the transition between all these different regimes which prevents the use of an efficient modelling strategy dedicated to a specific regime.

The goal of this contribution lies in a new modelling approach which allows to control the level of details available in the separated regime with a length-scale threshold and lay the basis for a proper and efficient description of transitions between the separated and disperse flow regimes.
The accurate modelling of all these flow regimes is critical in order to obtain reliable simulations that are valuable for the design of industrial components such as injection nozzles \citep{reitz_dependence_1979,bode_influence_2014,janodet_massively_2022}.
%
%

%
%
The choice of the two-phase flow model is usually adapted to the regime of interest and proposes different levels of details for the description of the interface and its associated capillarity effects.
Let us briefly describe each class of models from the smallest length-scale level of description of the interface to the largest.
When including in the modelling the full spectrum of length-scales, a family of models that can be referred to as 
\textit{Phase Field Diffuse Interface Model (DIM)} proposes to describe continuously the transition from one phase to the other and involves a potentially very small length scale, which will have to be resolved, which is the thickness of the interface. 
Many models fall within this category, such as Cahn-Hilliard-type models~\citep{cahn_free_1958}, Korteweg materials~\citep{korteweg_sur_1901,dunn_interstitial_1986,dunn_thermomechanics_1986} or second-gradient models~\citep{gouin_second_1996, seppecher_second-gradient_2002}.
Despite relying on a solid thermodynamical model (see the recent derivation from the kinetic level of description 
by  \cite{giovangigli_kinetic_2021}) and thus a proper mathematical structure \citep{giovangigli_symmetrization_2022}, they are of limited use in ambient conditions, where the physical thickness of the interface only reaches a few nanometers.
At a larger scale, that is if the thickness of the interface is not described in the model, \textit{sharp interface models} \citep{sussman_level_1994,vaudor_consistent_2017} enable
a non-ambiguous location of the interface seen as a discontinuity. 
These strategies can be viewed as a single-fluid system coupling across a sharp boundary.
This approach is sometimes used in what can be referred to as Direct Numerical Simulations (DNS - even if resolving all the scales is far from granted) regarding the capture of the interface.
However, for cases involving multiple interface topology regimes, implementing these approaches requires to reconstruct the interface at all relevant scales.
This can lead to an unreasonably high computational cost for challenging setups such as atomizations where mesh convergence can rarely be reached \citep{herrmann_detailed_2009,shinjo_simulation_2010,ling_spray_2017}.
Finally, \textit{multi-fluid models} stands at the largest scale and are typically derived from ensemble-averaging processes of local equations \citep{drew_mathematical_1983} or Hamilton's Stationary Action Principle (SAP)
\citep{gavrilyuk_hyperbolic_1998,gouin_hamiltons_1999,gavrilyuk_mathematical_2002,gouin_hamilton_2009, burtea_hamiltons_2021}.
For such models, the description of the interface usually assumes a unique flow regime set of assumptions.
In the separated regime, one can adopt a \textit{multi-fluid DIM}, where both immiscible phases coexist within an artificial mixture, and one usually considers the interface to be approximately captured in the computational domain by the transition zone from $0$ to $1$ of a colour function, which also provides an estimate of the interface area density (IAD)\footnote{Let us note that such a model, except if some specific interface compression techniques are added to the model \citep{shukla_interface_2010}, does not involve any interface thickness length scale as opposed to Phase Field DIM.}.
Following the Continuum Surface Force (CSF) model of Brackbill \citep{brackbill_continuum_1992}, one can model capillarity as a source term based on a colour function.
An alternate approach involves an equivalent flux form also referred to as the Continuum Surface Stress (CSS) model \cite{lafaurie_modelling_1994, gueyffier_volume--fluid_1999, perigaud_compressible_2005, grenier_accurate_2013,schmidmayer_model_2017}.
Other methods based on \textit{second-gradient DIM} \citep{jamet_second_2001,bueno_liquid-vapor_2016} are reminiscent of the Phase Field \citep{cahn_free_1958,jacqmin_calculation_1999} approach that relies on an adapted thermodynamic model in order to control the thickness of the interface.
However, both \textit{multi-fluid DIM} or \textit{second-gradient DIM} methods cannot be used to capture fine geometrical details that are smaller than the resolution of the bulk scale, potentially related to the  interface width in the second-gradient approach.
As a result, small features or small fluid inclusions are naturally out of reach for amenable mesh resolution.

At the other end of the spectrum of scales, in the disperse regime, the exact locations of the droplets or bubbles are unknown, and a mixture statistical description with only one volume fraction for the disperse phase can be retained in a \textit{multi-fluid disperse model} \citep{baer_two-phase_1986,raviart_non-conservative_1995,saurel_modelling_2017,drui_small-scale_2019}.
If more information about the distribution of the inclusions (\textit{e.g.} in sizes, shapes, temperatures) is desired, the inclusions can be modelled with a Number Density Function (NDF) which accounts for these additional characteristics in a multidimensional phase-space \citep{williams_spray_1958} governed by a generalized population balance equation.
A method of moments can then be used to reduce this high dimensional problem into an Eulerian reduced-order model with the transport of a finite set of moments \citep{massot_counterflow_1998,laurent_multi-fluid_2001,fox_multiphase_2007,massot_eulerian_2007}.

In order to adapt the modelling choice to the various flow regimes in a single physics, coupling strategies have been developed \citep{lebas_numerical_2009,herrmann_parallel_2010,le_touze_compressible_2020}, but the transfers between models are difficult to manage and parameter-dependent, and their mathematical properties are usually hard to study.
Another strategy consists in a two-scale modelling approach where a reduced-order model of the small-scale dynamics is used.
First attempts of such models have been proposed by \cite{gavrilyuk_mathematical_2002,drui_small-scale_2019} using the Hamilton's SAP, but only account for a disperse flow regime of bubbles in a carrier liquid phase.
Then, attempts of unified models describing both the separated and disperse regime have been proposed in \cite{devassy_atomization_2015,cordesse_derivation_2019,cordesse_diffuse_2020,di_battista_towards_2021}, with the introduction of some small-scale geometrical quantities.
In these last three works, Hamilton's SAP has been used to combine a \textit{large-scale} multi-fluid DIM model, adapted to the separated phase regime above a preset length threshold, with a \textit{small-scale} model adapted to the disperse phase regime below that threshold, which aims at enriching the geometric description of the interface below the scales resolved by the large-scale model and relies on geometric variables interpretable for any regime \citep{essadki_adaptive_2016}.
However, the proper combination of the two levels of modelling at small and large scales in order to build numerical schemes and conduct significant numerical simulations still requires key modelling features, namely the definition of a scale threshold and a mass transfer from one model to another.
%
%

%
%
In this contribution, we propose to alleviate this stumbling block of the combination of scales thanks to a unified two-scale model as well as a consistent mass transfer which introduces a length-scale threshold by regularizing the large-scale interface. The main contributions are:
1- The derivation of a unified two-scale model with Hamilton's SAP and geometric variables;
2- The definition of a dissipative mass transfer process that allows to regularize the large-scale interface consistently with capillarity models at both scales;
3- The proposition of an adequate numerical strategy along with a demonstrative test-case illustrating the properties and potential of the model.
First, we propose a novel unified two-scale approach combining two multi-fluid models.
At large-scale, we choose a multi-fluid DIM with a capillarity model based on an estimate of the IAD using a colour function along the lines of \cite{perigaud_compressible_2005,schmidmayer_model_2017}.
For the small-scale model, we assume the interface to be described as inclusions and small-scale IAD to account for capillarity is obtained as a moment of a population balance equation.
Although the IAD is the only information directly involved in the modelling of capillarity, the description of this small-scale collection of inclusions can be enriched with more geometrical information about its distribution in sizes or shapes \citep{essadki_high_2018,loison_two-scale_2023}, and thus aims at describing both the mixed and the disperse regimes.
The unified two-scale model with both multi-fluid DIM and disperse models is then derived thanks to the Hamilton's SAP by combining the energies of each model in a two-scale mixture along the lines of \cite{cordesse_contribution_2020}.
Second, the mass transfer between scales can be viewed as a regularization of the interface at the bulk scale that acts as a local dissipative process in the system.
Such a method offers a totally new point of view for dealing with interface smoothing that is usually performed as a non-local process during simulation \citep{bonometti_interface-capturing_2007,le_martelot_towards_2014} or tuned using discretization parameters like the local mesh grid size \citep{desjardins_accurate_2008,shukla_interface_2010}.
Even though our new approach still involves case-dependent parameters such as a length-scale cut-off, the mathematical properties of the overall model can be studied more thoroughly.
Third, numerical methods are gathered through a time splitting strategy to solve the convective and capillarity fluxes with adequate methods while an original implicit-explicit method is proposed for the instantaneous relaxation of pressures.
A demonstrative test-case is proposed to assess the modelling abilities of the two-scale approach such as the regularization property of the mass transfer or the IAD models at both scales.
%

Section~\ref{sec:twoscale} is dedicated to the derivation of the unified two-scale model with Hamilton's SAP including capillarity modelling at both scales.
Then, we introduce in Section~\ref{sec:masstransfer} the length-scale threshold and the regularizing mass transfer source terms, and we assess the dissipative nature of the process.
We follow with the description of the numerical strategy in Section~\ref{sec:numerics}.
The properties of the model are then observed and discussed thanks to a demonstrative test-case in Section~\ref{sec:simulations}.
Finally, we provide conclusions and perspective of this work in Section~\ref{sec:conclusion}.
\section{Two-scale model with capillarity}
\label{sec:twoscale}
We consider a two-phase multi-fluid DIM where both liquid and gaseous phases locally coexist.
Similarly to the two-scale approaches proposed in \cite{devassy_atomization_2015,cordesse_diffuse_2020}, we model the interface both in the separated and disperse regimes with a two-scale approach.
However, we endow each scale with its own capillarity model.
The \textit{large-scale} model sees only the compressible liquid and gaseous phases separated by an interface which is regular enough to be located through the field of volume fraction and its local geometry, \textit{e.g}. the mean curvature, can also be estimated with that field.
The \textit{small-scale} model accounts for smaller details of the interface geometry thanks to a set of scalar quantities gathered in a reduced-order model (see for instance \cite{essadki_new_2016,loison_two-scale_2023}).
Typically, we assume numerous liquid inclusions carried by the gaseous phase that we describe with a kinetic model, but such assumption is not restrictive and the approach only relies on the availability of a small-scale IAD equation of evolution.
In order to exhibit the key elements of our two-scale model, we purposely discuss the modelling of capillarity at both large and small scales under a set of simplifying assumptions to provide a building block model upon which the regularizing transfer is built in Section \ref{sec:masstransfer}.
The generalization of the model is discussed in the concluding remarks in Section \ref{sec:conclusion}.
Once the energies of the two-scale mixture are identified, the two-scale model is derived with Hamilton's SAP.
\subsection{Modelling assumptions for the two-scale mixture}
In this first section, we make the following assumptions about the two-scale mixture :
\begin{align}
    &\text{\textbullet\: all the phases, liquid or gas, large or small scale have the same velocity $\vel$;}  \label{hyp:same_vel} \tag{H1a}\\
    &\text{\textbullet\:  all the phases are equipped with a barotropic equation of state (EOS);}  \label{hyp:large_scale_eos} \tag{H1b}\\
    &\text{\textbullet\:  there is no mass exchanges between the phases.}  \label{hyp:no_mass_transfer} \tag{H1c}
\end{align}
Remark that \eqref{hyp:no_mass_transfer} is assumed in this first section where we only focus on the local coexistence of the two models with different modelling of capillarity.
It is further lifted and discussed in Section \ref{sec:masstransfer} where mass transfer from the large-scale liquid phase to the small-scale one is considered.

Let us denote the quantities related to the large-scale liquid and gaseous phases respectively by the indices $1$ and $2$, while the small-scale has both an index $1$ for its liquid nature, and an exponent $d$ as it accounts for a disperse regime.
For a phase $k$, we write its volume fraction $\alpha_k$ and its density $\rho_k$. 
Then, the barotropic EOS is modelled by the specific barotropic potential, the specific free energy $e_k(\rho_k)$.
The pressure is defined by $p_k:=\rho_k^2 e_k'(\rho_k)$ and the sound velocity by $c_k:=(p_k')^{1/2}$.
Denote also $m_k:=\alpha_k \rho_k$ the effective density for each phase $k=1,2,1^d$, the mass of each phase is then conserved following \eqref{hyp:same_vel} and \eqref{hyp:no_mass_transfer},
\begin{equation}
    \label{eq:mass_conservation}
    \partial_t m_k + \bnabla \bcdot (m_k \vel) = 0.
\end{equation}
If we define $\rho:=m_1+m_2+m_1^d$ the density of the medium, summing the equation above for $k=1,2,1^d$ enables to retrieve the total mass conservation equation $\partial_t \rho + \bnabla \bcdot (\rho \vel) = 0$.
The total volume occupancy of the phases in the mixture also enforces
\begin{equation}
    \alpha_1 + \alpha_2 + \alpha_1^d = 1.
\end{equation}
\subsection{Two-scale modelling of capillarity}
\label{sec:model_capillarity}
Following the assumption of the flow regime at each scale, we model now the geometry of the mixture's interface along with its associated capillarity energy.
\subsubsection{Large-scale capillarity model}
At the large scale, we construct a new colour function, the large-scale volume fraction defined by
\begin{equation}
    \overline{\alpha}_k = \frac{\alpha_k}{1-\alpha_1^d},
    \qquad
    k=1,2,
\end{equation}
such that $\overline{\alpha}_1 + \overline{\alpha}_2 = 1$.
It describes the large-scale interface geometry by taking out the influence of the small-scale volume fraction.
This new variable is used here to estimate the large-scale IAD with $\Vert\bnabla\overline{\alpha}_1\Vert$.
Such a choice extends the choice of the volume fraction proposed in \cite{perigaud_compressible_2005} which is recovered in the limit where there is no small-scale, \ie $\alpha_1^d=0$.
Furthermore, one can use this quantity that implicitly describes the large-scale interface to estimate geometric quantities such as the mean curvature \cite{goldman_curvature_2005}
\begin{equation}
    \label{eq:implicit-mean-curvature}
    \overline{H}(\overline{\alpha}_1):=-\bnabla\bcdot\left(\frac{\bnabla\overline{\alpha}_1}{\Vert\bnabla\overline{\alpha}_1\Vert}\right).
\end{equation}
Then, similarly as \cite{schmidmayer_model_2017,cordesse_contribution_2020}, we add a capillarity energy term based on the large-scale IAD estimator in the large-scale mixture free energy $\rho e$ such that
\begin{equation}
    \rho e = m_1 e_1(\rho_1) + m_2 e_2(\rho_2) + \sigma \Vert\bnabla\overline{\alpha}_1\Vert,
\end{equation}
where $\sigma$ is the capillarity coefficient. 
\subsubsection{Small-scale capillarity model}
We now focus on the modelling of the small scale that we assume here to be in the disperse regime made of small liquid inclusions carried in the gaseous phase.
Therefore, the small-scale inclusions can be modelled with the NDF $(\boldsymbol{x},t,\widehat{m})\mapsto n(\boldsymbol{x},t,\widehat{m})$ that counts the number of inclusions within the mixture in a small volume around $\boldsymbol{x}$ at time $t$, the mass of which is in a neighbourhood of $\widehat{m}$.
As we aim at proposing a small-scale model describing the inclusions created in the mixed regime, the shape of the inclusions is not prescribed yet, and we introduce the isoperimetric ratio $q:=S^3/V^2$ to characterize their shapes.
Again, we consider a minimal framework for the modelling of the small-scale inclusions by assuming that :
\label{sec:small-scale-interface}
\begin{align}
    &\text{\textbullet\:  there is no break-up or coalescence of the small-scale inclusions;}  \label{hyp:breakup} \tag{H2a}\\
    &\text{\textbullet\:  the small scale is made of inclusions characterized by an isoperimetric ratio $q$;}  \label{hyp:isop_inclusions} \tag{H2b}\\
    &\text{\textbullet\:  the small-scale liquid phase is incompressible.}  \label{hyp:small_scale_incomp} \tag{H2c}
\end{align}
We draw up outlooks in Section \ref{sec:conclusion} to lift the first two hypotheses, while the latter is only assumed at the very last of this discussion to show where the model with compressible inclusions would differ.
First, with \eqref{hyp:no_mass_transfer} and \eqref{hyp:breakup}, the dynamics of the NDF follows the following population balance equation
\begin{equation}\label{eq:WBE}
    \partial_t n + \nabla \cdot (n \boldsymbol{u}) = 0,
\end{equation}
Despite the simplicity of the formulation, the NDF remains a multidimensional function describing the local polydispersity in mass of the small-scale inclusions.
Consequently, we choose to reduce the kinetic description of the spray to a set of geometric variables following GeoMOM \citep{essadki_high_2018,loison_two-scale_2023}.
The polydispersity in mass or size of the spray can typically be accounted for with the volume fraction $\alpha_1^d$, the surface-weighted mean and Gauss curvature densities $\Sigma\left<H\right>$, $\Sigma\left<G\right>$ and the small-scale IAD denoted $\Sigma$.
However, we only keep here the small-scale IAD and volume fraction as it is the minimal description required for modelling capillarity.
Thanks to \eqref{hyp:isop_inclusions}, the small-scale IAD is obtained by adding up the surface areas of all the inclusions by integrating \eqref{eq:WBE} against the surface area $S(m, \rho_1^d)=q^{1/3} (m/\rho_1^d)^{2/3}$ of a droplet of mass $m$.
It results in
\begin{equation}
        \Sigma := \int_{\widehat{m}} S(\widehat{m}, \rho_1^d) \: n \: d\widehat{m}
        =\int_{\widehat{m}} q^{1/3} \left(\rho_1^d\right)^{-2/3} \widehat{m}^{2/3} n \:d\widehat{m}.
\end{equation}
As $\Sigma$ represents the total area of inclusions within the two-scale mixture and the capillarity energy simply reads $\sigma\Sigma$.
Before deriving the two-scale model, let us focus on the dynamics of $\Sigma$ which is obtained by integrating \eqref{eq:WBE} against $S(\widehat{m}, \rho_1^d)$.
It yields
\begin{equation}
    \label{eq:Sigma}
    \partial_t ((\rho_1^d)^{2/3}\Sigma) + \bnabla \bcdot ((\rho_1^d)^{2/3}\Sigma \vel) = 0,
\end{equation}
which can be recast into either
\begin{equation}
    \partial_t \Sigma + \bnabla \bcdot (\Sigma \vel) = \frac{2}{3} \Sigma \bnabla \bcdot u + \frac{2}{3}\frac{\Sigma}{\alpha_1^d}D_t \alpha_1^d,
    \quad
    \text{or}
    \quad
    D_t z = 0,
\end{equation}
with $z := (\rho_1^d)^{2/3}\Sigma/m_1^d$ as identified by \cite{di_battista_towards_2021}, and where the closure of the dynamics of $D_t \alpha_1^d$ would propose an equation of evolution for $\Sigma$ reminiscent of the one obtained with an averaging approach by \cite{lhuillier_evolution_2004}.
Now assume the incompressibility of the small-scale \eqref{hyp:small_scale_incomp}.
Denote the material derivative by $D_t~(\cdot)~:=~\partial_t~(\cdot)~+~\vel~\bcdot~\bnabla~(\cdot)$, the incompressibility of the small-scale liquid phase \eqref{hyp:small_scale_incomp} gives
\begin{equation}
    \label{eq:incomp}
    D_t \rho_1^d = 0,
    \quad
    \text{and}
    \quad
    \partial_t \alpha_1^d + \bnabla\bcdot(\alpha_1^d \vel)=0,
\end{equation}
and \eqref{eq:Sigma} boils down to
\begin{equation}
    \label{eq:Sigma_cons}
    \partial_t \Sigma + \bnabla \bcdot (\Sigma \vel) = 0.
\end{equation}
Remark that integrating the population balance equation \eqref{eq:WBE} against $\widehat{m}$ also recovers the mass conservation of the small scale \eqref{eq:mass_conservation}.
\subsection{Derivation of the two-scale model}
We derive now the dynamics of the two-scale mixture thanks to Hamilton's SAP \citep{herivel_derivation_1955,serrin_mathematical_1959,salmon_practical_1983,bedford_hamiltons_1985,fosdick_kinks_1991,gavrilyuk_hyperbolic_1998,gouin_hamiltons_1999, gavrilyuk_mathematical_2002,berdichevsky_variational_2009,gouin_introduction_2020,burtea_hamiltons_2021} which requires to define the mixture Lagrangian.
The action associated to this Lagrangian is then minimized to obtain the equations of motion.
Then, dissipative processes are added by studying the mathematical entropy production rate of the derived system.
\subsubsection{Model at pressure equilibrium derived with Hamilton's SAP}
The Lagrangian, denoted $\lag$, is a scalar function dimensioned as an energy which contains the model characteristics and is defined as the difference between the kinetic and potential energies.
Here, our two-scale approach is notably distinguished by its capillarity energies provided at each scale as modelled in Section \ref{sec:model_capillarity}.
We set
\begin{equation}
    \label{eq:lagrangian}
        \lag := \underbrace{\frac{1}{2}m_1 \vel^2 - m_1 e_1\left(\frac{m_1}{\alpha_1}\right)}_{\lag_1}+\underbrace{\frac{1}{2}m_2 \vel^2 - m_2 e_2\left(\frac{m_2}{\alpha_2}\right)}_{\lag_2}
        \underbrace{-\sigma\Vert\bnabla\overline{\alpha}_1\Vert\vphantom{\left(\frac{m_2}{\alpha_2}\right)}}_{\lag_{cap}}+\underbrace{\frac{1}{2}m_1^d \vel^2 - m_1^d e_1\left(\rho_1^d\right)-\sigma\Sigma}_{\lag_1^d}.
\end{equation}
Here, $\mathcal{L}_1$, $\mathcal{L}_2$ and $\mathcal{L}_1^d$ correspond respectively to the contributions to the Lagrangian of the large-scale liquid and gas and the small-scale liquid.
The large scale capillary effects are modelled through $\mathcal{L}_{cap}$, while the small-scale ones $\sigma\Sigma$ are included within $\mathcal{L}_1^d$.
Remark that we used the same barotropic EOS for the liquid phases, but the free energies are evaluated for independent densities.
Following  the methodology of Hamilton's SAP detailed in Appendix \ref{app:SAP}, the above Lagrangian defines a minimization problem under the conservation constraints \eqref{eq:mass_conservation}-\eqref{eq:incomp}-\eqref{eq:Sigma_cons}, and its solution reads
\begin{equation}
    \label{eq:non-diss-system}
    \begin{cases}
        \begin{aligned}
            &\partial_t m_k &+& \bnabla \bcdot (m_k \vel) &=& 0, \qquad k=1,2,1^d,\\
            &\partial_t \alpha_1^d &+& \bnabla \bcdot (\alpha_1^d\vel) &=& 0,\\
            &\partial_t \Sigma &+& \bnabla \bcdot (\Sigma\vel) &=& 0,\\
        \end{aligned}\\
        \partial_t (\rho\vel) + \bnabla \bcdot \Big(\rho\vel\otimes \vel
        +(\overline{p}-\sigma\Vert\bnabla\overline{\alpha}_1\Vert)\mathbf{I}+\sigma \frac{\bnabla\overline{\alpha}_1\otimes\bnabla\overline{\alpha}_1}{\Vert\bnabla\overline{\alpha}_1\Vert}\Big)= \boldsymbol{0},
    \end{cases}
\end{equation}
with
\begin{equation}
    \overline{p}
    :=
    \overline{\alpha}_1p_1\left(\frac{m_1}{\overline{\alpha}_1(1-\alpha_1^d)}\right)
    +
    \overline{\alpha}_2p_2
    \left(\frac{m_2}{(1-\overline{\alpha}_1)(1-\alpha_1^d)}\right),
\end{equation}
and $\overline{\alpha}_1$ defined by the implicit Laplace equilibrium
\begin{equation}
    \label{eq:laplace_eq}
    p_1\left(\frac{m_1}{\overline{\alpha}_1(1-\alpha_1^d)}\right)
    -p_2\left(\frac{m_2}{(1-\overline{\alpha}_1)(1-\alpha_1^d)}\right)
    =\frac{\sigma}{1-\alpha_1^d}\overline{H}(\overline{\alpha}_1),
\end{equation}
where $\overline{H}(\overline{\alpha}_1)$ is defined by \eqref{eq:implicit-mean-curvature}.
This system admits a supplementary equation of conservation for $\mathcal{H}:=\boldsymbol{K}\bcdot\vel - \lag$ (see details in appendix \ref{app:math_entropy_prod}) that reads
\begin{equation}
    \label{eq:supp_eq_math_entropy}
    \partial_t \mathcal{H}+ \bnabla \bcdot \left[(\mathcal{H}+\overline{p}-\sigma\Vert\bnabla\overline{\alpha}_1\Vert)\vel +\sigma\frac{\bnabla\overline{\alpha}_1\otimes\bnabla\overline{\alpha}_1}{\Vert\bnabla\overline{\alpha}_1\Vert}\bcdot\vel
    -\sigma \frac{\bnabla\overline{\alpha}_1}{\Vert\bnabla\overline{\alpha}_1\Vert}D_t\overline{\alpha}_1\right]=0,
\end{equation}
with
\begin{equation}
    \mathcal{H} = \frac{1}{2}\rho \vel^2 + \sum_{k=1,2,1^d}m_ke_k(\rho_k)+\sigma \Vert\bnabla\overline{\alpha}_1\Vert + \sigma \Sigma.
\end{equation}
The material time derivative $D_t\overline{\alpha}_1$ in the flux is implicitly obtained by taking the time material derivative of the Laplace equilibrium \eqref{eq:laplace_eq}.
Remark then that the system \eqref{eq:non-diss-system} and the equation \eqref{eq:supp_eq_math_entropy} are conservation equations with fluxes depending on the gradient of $\overline{\alpha}_1$.
Nevertheless, we still refer to $\mathcal{H}$ as a ``mathematical entropy'' as it naturally extends its usual definition. 
The relation between the mathematical entropy and the physical one is obtained as the isothermal limit of the Euler-Fourier model in \cite{serre_structure_2010}.
It is showed to be convex and linked to the physical entropy of the mixture $s$ with $\mathcal{H}=\rho(\varepsilon-Ts) + \tfrac{1}{2}\rho\Vert\vel\Vert^2$ where $\varepsilon = e + Ts$, and $T$ are respectively the internal energy and the temperature of the mixture.
\subsubsection{Model at pressure disequilibrium and dissipative relaxation}
Let us consider now the case where the Laplace pressure equilibrium is not fulfilled, and the dynamics of $\overline{\alpha}_1$ is not prescribed.
Then, we introduce instead the following unclosed equation
\begin{equation}
    \partial_t \overline{\alpha}_1 + \vel \bcdot \bnabla \overline{\alpha}_1 
    = R_{\overline{\alpha}_1},
\end{equation}
where $R_{\overline{\alpha}_1}$ is a source term yet to be determined.
Considering this dynamics for $\overline{\alpha}_1$ along with the system \eqref{eq:non-diss-system}, we have that 
\begin{equation}
    \label{eq:entropy_prod_alpha_1}
    \partial_t \mathcal{H} + \bnabla \bcdot \mathbfcal{G}
    =\varsigma,
\end{equation}
with $\varsigma=((1-\alpha_1^d)(p_1-p_2)-\sigma\overline{H})R_{\overline{\alpha}_1}$ (see details in appendix \ref{app:math_entropy_prod}) and the flux
\begin{equation}
    \mathbfcal{G}=\mathcal{H}\vel+\mathbf{P}\vel-\sigma \frac{\bnabla\overline{\alpha}_1}{\Vert\bnabla\overline{\alpha}_1\Vert}R_{\overline{\alpha}_1},
    \qquad
    \mathbf{P}=(\overline{\alpha}_1p_1+\overline{\alpha}_2p_2-\sigma\Vert\bnabla\overline{\alpha}_1\Vert)\mathbf{I}+\sigma\frac{\bnabla\overline{\alpha}_1\otimes\bnabla\overline{\alpha}_1}{\Vert\bnabla\overline{\alpha}_1\Vert}.
\end{equation}
The dissipation of the system is then ensured if $\varsigma~\le~0$.
Remark that assuming the Laplace equilibrium satisfied gives $\varsigma=0$ such that system \eqref{eq:non-diss-system} together with Laplace equilibrium \eqref{eq:laplace_eq} is non-dissipative.
We propose now to define $R_{\overline{\alpha}_1}$ as a pressure relaxation source term that drives the system towards the Laplace equilibrium~\eqref{eq:laplace_eq}
\begin{equation}\label{eq:alpha_dissipation}
    R_{\overline{\alpha}_1} = \epsilon^{-1}\left(p_1-p_2-\frac{\sigma}{1-\alpha_1^d}\overline{H}\right),
\end{equation}
where $\epsilon$ has the dimension of a viscosity.
Note that the equilibrium \eqref{eq:laplace_eq} is recovered for the instantaneous limit case when $\epsilon~\rightarrow~0$.
With such dynamics for $\overline{\alpha}_1$, the system now reads
\begin{equation}
    \setlength\arraycolsep{2pt}
    \label{eq:relaxed-system}
    \begin{cases}
        \begin{array}{lclcl}
            \partial_t m_k &+& \bnabla \bcdot (m_k \vel) &=& 0, \qquad k=1,2,1^d,\\
            \partial_t \alpha_1^d &+& \bnabla \bcdot (\alpha_1^d\vel) &=& 0,\\
            \partial_t \Sigma &+& \bnabla \bcdot (\Sigma\vel) &=& 0,\\
            \partial_t \overline{\alpha}_1 &+& \vel \bcdot \bnabla \overline{\alpha}_1 &=& \epsilon^{-1}\left(p_1-p_2-\frac{\sigma}{1-\alpha_1^d}\overline{H}\right),\\
        \end{array}\\
        \partial_t (\rho\vel) + \bnabla \bcdot \Big(\rho\vel\otimes \vel+(\overline{p}-\sigma\Vert\bnabla\overline{\alpha}_1\Vert)\mathbf{I}+\sigma \frac{\bnabla\overline{\alpha}_1\otimes\bnabla\overline{\alpha}_1}{\Vert\bnabla\overline{\alpha}_1\Vert}\Big)= \boldsymbol{0},
    \end{cases}
\end{equation}
and is dissipative in the sense that, following \eqref{eq:entropy_prod_alpha_1}, we have a negative mathematical entropy production rate 
\begin{equation}
    \label{eq:math_ent_prod_relaxed}
    \varsigma = -\epsilon^{-1}(1-\alpha_1^d)\left(p_1-p_2-\frac{\sigma}{1-\alpha_1^d}\overline{H}\right)^2 \le 0.
\end{equation}
Remark that, with the relaxation \eqref{eq:alpha_dissipation}, the entropy flux $\mathbfcal{G}$ in \eqref{eq:entropy_prod_alpha_1} is now explicit.
\subsection{Discussion of the two-scale models}
The conservative system~\eqref{eq:non-diss-system} and the dissipative system~\eqref{eq:relaxed-system} both extend the models of \cite{chanteperdrix_modelisation_2004,caro_dinmod_2005}.
They are recovered in the limit where $\alpha_1^d\rightarrow 0$.
Furthermore, when the capillarity effects are neglected, the systems \eqref{eq:non-diss-system} and  \eqref{eq:relaxed-system} are hyperbolic with respective sound velocities $c_W^d$ and $c_F^d$, which are the usual Wood and frozen sound velocities  \citep{caro_dinmod_2005} increased by a factor $(1-\alpha_1^d)^{-1}$ such that
\begin{equation}
    c_W^d = \frac{1}{1-\alpha_1^d}\left(\rho\left(\frac{\alpha_1}{\rho_1c_1^2}+\frac{\alpha_1}{\rho_2c_2^2}\right)\right)^{-1/2} = \frac{c_W}{1-\alpha_1^d},
    \qquad
    c_F^d = \frac{1}{1-\alpha_1^d}\sqrt{Y_1c_1^2+Y_2c_2^2} = \frac{c_F}{1-\alpha_1^d},
\end{equation}
where $Y_k=\alpha_k\rho_k/\rho$ are the mass fractions.
All the other eigenvalues evaluates to the material velocity $\vel$ with linearly degenerate eigenvectors.
Note that assuming the incompressibility of small-scale inclusion discards here any non-physical sound propagation in the disperse liquid phase as remarked in \cite{saurel_modelling_2017}.

Let us focus now on the impact of capillarity on the properties of the two-scale models.
Because of the additional tensor in the momentum flux, the system has not a usual conservative form with fluxes depending on the local state only.
But, if we authorize flux dependencies on $\bnabla\overline{\alpha}_1$, the model \eqref{eq:non-diss-system} with Laplace equilibrium \eqref{eq:laplace_eq} is shown to be conservative thanks to energy balance~\eqref{eq:supp_eq_math_entropy}, while the model \eqref{eq:relaxed-system} with the relaxation source term involves a dissipative process.
Furthermore, the hyperbolicity study of the model \eqref{eq:relaxed-system} is not possible as it involves second-order space derivatives.
Nevertheless, we propose here some elements of such a study for a comparable model relying on the same physical assumptions, but a different mathematical structure detailed in Appendix \ref{app:hyperbolicity}.
This model, detailed in \eqref{eq:relaxed-system-homo-app}, is an augmented model of \eqref{eq:relaxed-system} where an equation on variable $\boldsymbol{w} = \bnabla\overline{\alpha}_1$ is added to the system to recover first-order space derivatives only.
Let us fist underline that the augmented model~\eqref{eq:relaxed-system-homo-app} is not rotational invariant. For a particular normalized direction $\boldsymbol{\omega}$, we can study the eigenstructure of the augmented model.
Let us first note
\begin{equation}
    u_{\boldsymbol{\omega}} := \vel\bcdot\boldsymbol{\omega},
    \quad
    \boldsymbol{n}:=
    \frac{\bnabla\overline{\alpha}_1}{\Vert\bnabla\overline{\alpha}_1\Vert},
    \quad
    \psi = \sigma\Vert\bnabla\overline{\alpha}_1\Vert/(\rho (c_F^d)^2),
\end{equation}
respectively the velocity, the large-scale normal and a geometrical-physical parameter.
In the diffuse interface with moderate capillarity effects in comparison with acoustics, \ie $\psi\ll1$, we only keep the first-order terms in $\psi$.
In this case, the characteristic velocities of the augmented model are
\begin{equation}
    \label{eq:eigenvalues-2}
    u_{\boldsymbol{\omega}},
    \qquad
    u_{\boldsymbol{\omega}} \pm c_F^d (1-(\boldsymbol{\omega}\bcdot \boldsymbol{n})^2)\sqrt{
        \psi
        },
        \qquad
    u_{\boldsymbol{\omega}} \pm c_F^d
    \left(
    1+\frac{1}{2}\psi(\boldsymbol{\omega}\bcdot \boldsymbol{n})^2(1  - 
        (\boldsymbol{\omega}\bcdot \boldsymbol{n})^2)
    \right).
\end{equation}
Remark then that when the capillarity effects are negligible with respect to the acoustics ones \ie $\psi\gg 1$ or when we are oriented towards the surface normal $(\boldsymbol{\omega}\bcdot \boldsymbol{n})^2=1$, we recover at the zeroth order the two-scale frozen speed of sound $c_F^d$.
Otherwise, these velocities are \textit{a priori} distinct but, as showed in Appendix \ref{app:hyperbolicity}, the augmented system is weakly hyperbolic.
Besides, for any direction, $0\le(\boldsymbol{\omega}\bcdot \boldsymbol{n})^2\le1$, note that the absolute value of the augmented model's eigenvalues \eqref{eq:eigenvalues-2} can be upper bounded by
\begin{equation}
    \label{eq:eigenvalues-majorant}
    \lambda_{max}:=\Vert\vel\Vert+c_F^d\left(1+\frac{1}{8}\psi\right).
\end{equation}
Finally, without the small-scale modelling, this augmented system shares similarities with the one proposed in \cite{schmidmayer_model_2017}.
One difference lies in the modelling of a pressure relaxation instead of assuming a dynamics on $\overline{\alpha}_1$ that preserves the pressure equilibrium.
Consequently, we obtained eigenvalues related to the frozen sound velocity rather than the Wood sound velocity.

\section{Introducing a length-scale threshold via a regularizing inter-scale mass transfer}
\label{sec:masstransfer}
\begin{figure}
    \centering
    \includegraphics[width=.35\textwidth]{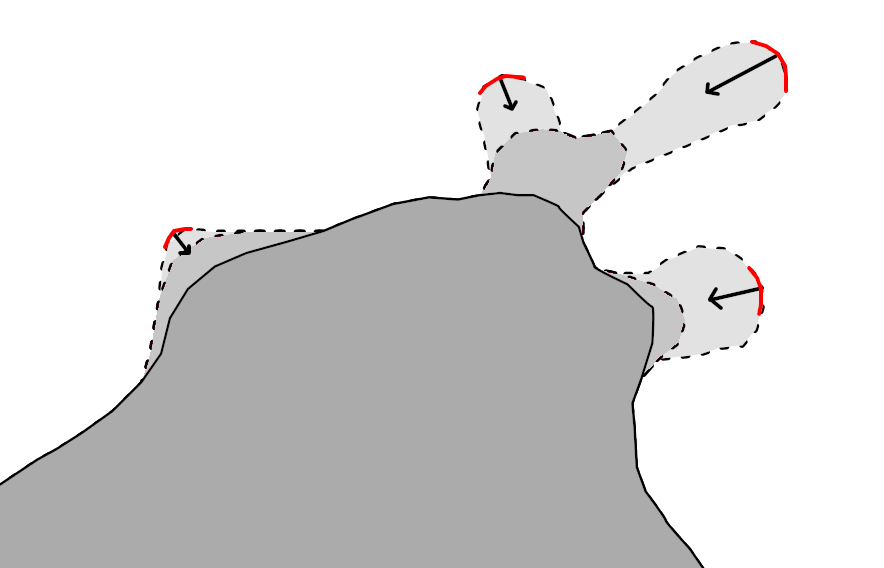}
    \caption{Schematic representation of the regularization of the large-scale interface. 
    The grey region corresponds to the liquid $\overline{\alpha}_1=1$ and the white region is the gas $\overline{\alpha}_1=0$. The red zones represent the locations where the mean curvature $\overline{H}$ is higher than a prescribed maximal mean curvature $H_{max}$.
    The creation of small-scale inclusions that occurs during the process is not represented here.}
    \label{fig:regularization}
\end{figure}

Thanks to the two-scale unified model, we can now focus on the main contribution of this work by lifting the assumption \eqref{hyp:no_mass_transfer} by now allowing that
\begin{align}
    &\text{\textbullet\: there is a mass exchange only from the large-scale to the small-scale liquid phase.} \label{hyp:mass_transfer} \tag{H2a}
\end{align}
Particularly, we are interested in modelling the liquid transfer from the large scale to the small scale following three simultaneous goals:
1- introducing a length-scale threshold separating the two scales,
2- limiting locally the large-scale interface curvature through a dissipation process,
3- modelling the transition from the separated regime to the disperse regime.
This last goal is mentioned here as a perspective and only a partial study is here proposed by an elementary parametrization of the process.
Such a regularizing process is represented in Fig. \ref{fig:regularization}: mass transfer from the large scale to the small scale initiates at points where the mean curvature is the most pronounced, advancing until the mean curvature criterion is satisfied everywhere on the large-scale interface.
Under the chosen convention, the local normal $\bnabla \overline{\alpha}_1/\Vert\bnabla \overline{\alpha}_1\Vert$ is oriented inward the liquid phase and the curvature $\overline{H}$ has a positive value in the red areas.
\subsection{Mathematical entropy production of the inter-scale mass transfer}
Let us consider a modified version of the two-scale model with capillarity~\eqref{eq:relaxed-system} by adding source terms for each quantity involved in the mass transfer.
Particularly, the evolution equation of $\overline{\alpha}_1$ now features a source term that will be specified in the sequel.
The model writes
\begin{equation}
    \label{eq:model_unclosed_mass_transfer}
    \begin{cases}
        \begin{aligned}
            &\partial_t m_1 &+& \bnabla \bcdot (m_1 \vel) &=& R_{m_1},\\
            &\partial_t m_1^d &+& \bnabla \bcdot (m_1^d \vel) &=& R_{m_1^d},\\
            &\partial_t m_2 &+& \bnabla \bcdot (m_2 \vel) &=& 0,\\
            &\partial_t \alpha_1^d &+& \bnabla \bcdot (\alpha_1^d\vel) &=& R_{\alpha_1^d},\\
            &\partial_t \Sigma &+& \bnabla \bcdot (\Sigma\vel) &=& R_{\Sigma},\\
            &\partial_t \overline{\alpha}_1 &+& \vel \bcdot \bnabla \overline{\alpha}_1 &=& R_{\overline{\alpha}_1},\\
        \end{aligned}\\
        \partial_t (\rho\vel) + \bnabla \bcdot \Big(\rho\vel\otimes \vel+(\overline{p}-\sigma\Vert\bnabla\overline{\alpha}_1\Vert)\mathbf{I}+\sigma \frac{\bnabla\overline{\alpha}_1\otimes\bnabla\overline{\alpha}_1}{\Vert\bnabla\overline{\alpha}_1\Vert}\Big)= \boldsymbol{R}_{\vel}.
    \end{cases}
\end{equation}
The equation on total momentum also features a source term $\boldsymbol{R}_{\vel}$ to balance the expected gain of capillarity energy at small scale with a loss of kinetic energy at large scale.
Remark that the source term $R_{\overline{\alpha}_1}$ depends on both the mass transfer and the process that balances the Laplace equilibrium, but they are here considered together.

Let us first relate $R_{m_1^d}$, $R_{\alpha_1^d}$, $R_{\Sigma}$ to $R_{m_1}$.
Given the total liquid mass conservation, we immediately have that
\begin{equation}
    R_{m_1^d} = -R_{m_1}.
\end{equation}
Then considering the incompressibility of the small-scale $D_t~\rho_1^d~=~0$ and dividing the equation on $m_1^d$ by $\rho_1^d$ leads to
\begin{equation}
    R_{\alpha_1^d}=-\frac{R_{m_1}}{\rho_1^d}.
\end{equation}
For the source term $R_{\Sigma}$, we consider the underlying kinetic equation \eqref{eq:WBE} with an additional source term $R_n$ accounting for the creation of droplets, \ie
\begin{equation}
    \label{eq:WBE_and_source_terms}
    \partial_t n + \bnabla\bcdot(n\vel) = R_n.
\end{equation}
As $R_n$ depends on $\widehat{m}$, it produces droplets of average size and mass
\begin{equation}
    S_{avg}:=\frac{\int_{\widehat{m}}S(\widehat{m},\rho_1^d)R_n \: d\widehat{m}}{\int_{\widehat{m}}R_n \: d\widehat{m}},
    \qquad
    m_{avg}:=\frac{\int_{\widehat{m}}\widehat{m}R_n \: d\widehat{m}}{\int_{\widehat{m}}R_n \: d\widehat{m}}.
\end{equation}
Integrating \eqref{eq:WBE_and_source_terms} against $S(\widehat{m},\rho_1^d)$ and $\widehat{m}$ provides the desired relation
\begin{equation}
    \begin{cases}
        \partial_t \Sigma + \bnabla\bcdot(\Sigma\vel)=S_{avg} \int_{\widehat{m}}R_n \: d\widehat{m} = R_{\Sigma},\\
        \partial_t m_1^d + \bnabla\bcdot(m_1^d\vel)=m_{avg} \int_{\widehat{m}}R_n \: d\widehat{m} = -R_{m_1},\\
    \end{cases}
\end{equation}
which implies that
\begin{equation}
    R_{\Sigma}= -\frac{S_{avg}}{m_{avg}} R_{m_1}.
\end{equation}
With the mass transfer process, the system yields
\begin{equation}
    \label{eq:relaxed-system-source-terms}
    \begin{cases}
        \begin{aligned}
            &\partial_t m_1 &+& \bnabla \bcdot (m_1 \vel) &=& R_{m_1},\\
            &\partial_t m_1^d &+& \bnabla \bcdot (m_1^d \vel) &=& -R_{m_1},\\
            &\partial_t m_2 &+& \bnabla \bcdot (m_2 \vel) &=& 0,\\
            &\partial_t \alpha_1^d &+& \bnabla \bcdot (\alpha_1^d\vel) &=& -(\rho_1^d)^{-1}R_{m_1},\\
            &\partial_t \Sigma &+& \bnabla \bcdot (\Sigma\vel) &=& -\frac{S_{avg}}{m_{avg}} R_{m_1},\\
            &\partial_t \overline{\alpha}_1 &+& \vel \bcdot \bnabla \overline{\alpha}_1 &=& R_{\overline{\alpha}_1},\\
        \end{aligned}\\
        \partial_t (\rho\vel) + \bnabla \bcdot \Big(\rho\vel\otimes \vel+(\overline{p}-\sigma\Vert\bnabla\overline{\alpha}_1\Vert)\mathbf{I}+\sigma \frac{\bnabla\overline{\alpha}_1\otimes\bnabla\overline{\alpha}_1}{\Vert\bnabla\overline{\alpha}_1\Vert}\Big)= {R}_{\vel}.
    \end{cases}
\end{equation}
Concerning the dissipation of the model, the mathematical entropy production as defined and computed in Appendix \ref{app:math_entropy_prod}, is 
\begin{equation}
    \label{eq:entropy_prodution_source_terms}
    \varsigma=-((1-\alpha_1^d)(p_1-p_2)-\sigma\overline{H}) R_{\overline{\alpha}_1}-\left(e_1(\rho_1^d)+\frac{\overline{p}}{\rho_1^d}-\left(e_1(\rho_1)+\frac{p_1}{\rho_1}\right)+\sigma \frac{S_{avg}}{m_{avg}}\right)R_{m_1}
    +\vel \bcdot {R}_{\vel}.
\end{equation}
Remark that we recover the same mathematical entropy production as the models \eqref{eq:non-diss-system} and \eqref{eq:relaxed-system} in Section \ref{sec:twoscale} when neither mass transfer nor momentum variation are accounted $R_{m_1}~=~0$ and ${R}_{\vel}~=~\boldsymbol{0}$.
As previously discussed, more information could be recovered about the polydispersity of the mass transfer provided that more geometric variables are accounted for \citep{essadki_adaptive_2016}.
This model can easily be extended to account for these geometric variables and more geometrical parameters would be available in the mass transfer.
To lighten the model under consideration, only the IAD has been kept as it is the only one which impacts the mathematical entropy production through capillarity energies.
\subsection{Large-scale mean curvature limitation \via the pressure relaxation}
\label{sec:curv-limit}
In order to define the mass transfer between scales, we alter the large-scale Laplace equilibrium by introducing a different curvature $H_{lim}$ instead of $\overline{H}$ in \eqref{eq:laplace_eq},
\begin{equation}
    \label{eq:laplace_eq_modified}
    R_{\overline{\alpha}_1} = \frac{1}{\epsilon}\left(p_1-p_2 -\frac{\sigma H_{lim}}{1-\alpha_1^d}\right).
\end{equation}
This change now requires to compensate the mathematical entropy production with the source term of the momentum equation.
As we are interested here in a transfer from large to small scales, we propose to regularize the interface where the local mean curvature is large and positive, as it would happen for a liquid ligament break-up.

The regularization process is then introduced by the definition of $H_{lim}:=\min(\overline{H}, H_{max})$ where $H_{max}$ is a user-specified positive curvature threshold to locally control the deformation of the large-scale interface as suggested by the scheme in Figure \ref{fig:regularization}.

Choosing such a different equilibrium leads to an unsigned term in the mathematical entropy production rate \eqref{eq:entropy_prodution_source_terms} when no mass transfer is accounted for, \textit{i.e.} $R_{m_1}=0$.
Thus, we must determine the right mass and momentum transfer to make the total process dissipative \ie $\varsigma<0$.
With such dynamics, the mathematical entropy production rate is now
\begin{equation}
    \label{eq:math-ent-prod-transfer-1}
    \varsigma=-((1-\alpha_1^d)(p_1-p_2)-\sigma\overline{H})
    \frac{1}{\epsilon}\left(p_1-p_2 -\frac{\sigma H_{lim}}{1-\alpha_1^d}\right)
    -\left(e_1(\rho_1^d)+\frac{\overline{p}}{\rho_1^d}-\left(e_1(\rho_1)+\frac{p_1}{\rho_1}\right)+\sigma \frac{S_{avg}}{m_{avg}}\right)R_{m_1}
    +\vel \bcdot {R}_{\vel}.
\end{equation}
Then, considering that the large-scale and small-scale liquid densities are close, we define $\Delta\rho_1 = \rho_1^d - \rho_1$ and a non-dimensional function $h$ corresponding to the first-order integral remainder of $e_1(\rho_1^d)+\overline{p}/\rho_1^d-(e_1(\rho_1)+p_1/\rho_1)$ such that 
\begin{equation}
    \label{eq:h_def}
    e_1(\rho_1^d)+\frac{\overline{p}}{\rho_1^d}-\left(e_1(\rho_1)+\frac{p_1}{\rho_1}\right) 
    =
    -\frac{\overline{\alpha}_2}{\rho_1}(p_1-p_2)(1+h),
\end{equation}
with $h\rightarrow 0$ when $\Delta\rho_1/\rho_1 \rightarrow 0$.
Then, defining
\begin{equation}
    \Delta H := \overline{H}-H_{lim},
\end{equation}
we reorganize the terms in \eqref{eq:math-ent-prod-transfer-1} to obtain
\begin{equation}
    \label{eq:math_ent_prod_org}
    \begin{aligned}
        \varsigma=&-\frac{1}{\epsilon}(1-\alpha_1^d)\left(p_1-p_2-\frac{\sigma H_{lim}}{1-\alpha_1^d}\right)^2
        +\left(\frac{\overline{\alpha}_2}{\rho_1}
        R_{m_1}+\frac{1}{\epsilon}\sigma \Delta H\right)
        \left(p_1-p_2\right)\\
        &-\frac{1}{\epsilon}\sigma \Delta H
        \left(\frac{\sigma H_{lim}}{1-\alpha_1^d}\right)
        +\left(\frac{\overline{\alpha}_2}{\rho_1}(p_1-p_2)
        h-\sigma \frac{S_{avg}}{m_{avg}}\right) R_{m_1}
        +\vel \bcdot {R}_{\vel}.
    \end{aligned}
\end{equation}
This reorganization suggests nullifying the second unsigned term with the pressure difference by choosing
\begin{equation}
    \label{eq:choice_Rm1}
    R_{m_1} = -\frac{1}{\epsilon}\frac{\rho_1\sigma \Delta H}{\overline{\alpha}_2}.
\end{equation}
This choice notably activates the mass transfer when $\Delta H \neq 0$ \textit{i.e.} when the local curvature $\overline{H}$ is different from the prescribed curvature $H_{lim}$.
Then, the mathematical entropy production rate becomes
\begin{equation}
    \label{eq:math_ent_prod_mass_transfer_lim}
    \varsigma=-\frac{1}{\epsilon}(1-\alpha_1^d)\left(p_1-p_2-\frac{\sigma H_{lim}}{1-\alpha_1^d}\right)^2
    -\frac{1}{\epsilon}\sigma \Delta H
    \left(\frac{\sigma H_{lim}}{1-\alpha_1^d}+\left(p_1-p_2\right)
    h-\sigma \frac{S_{avg}}{m_{avg}}\frac{\rho_1}{\overline{\alpha}_2}\right)
    +\vel \bcdot {R}_{\vel}.
\end{equation}
The first term of the right-hand side is negative for any $\epsilon\ge 0$.
As $h$ is expected to be small, the sign of the second term depends mainly on both $H_{lim}$ and the ratio $S_{avg}/m_{avg}$.
Neglecting $h$, this second term has the sign of 
\begin{equation}
    \label{eq:curvature-comp}
    \frac{S_{avg}}{m_{avg}}\frac{\rho_1}{\overline{\alpha}_2}-\frac{H_{lim}}{1-\alpha_1^d}.
\end{equation}
Let us investigate the sign of this quantity with a dimensional analysis.
Given a mixture-volume of typical length $l$ with a small-scale of typical length $l_{ss}$, large-scale and small-scale densities of liquid are almost the same such that $m_{avg}\sim\rho_1 l_{ss}^3$, and we also have $1-\alpha_1^d\sim 1$ and $\overline{\alpha}_2\sim 1$.
As the regularization of the interface requires setting $H_{lim}^{-1}$ comparable to $l$ and then much larger than $l_{ss}$, the quantity \eqref{eq:curvature-comp} behaves as
\begin{equation}
    \frac{S_{avg}}{m_{avg}}\frac{\rho_1}{\overline{\alpha}_2}-\frac{H_{lim}}{1-\alpha_1^d}
    \approx
    \frac{1}{l_{ss}}-\frac{1}{l},
\end{equation}
which is consequently expected positive.
Then, the dissipative nature of the inter-scale transfer, \ie $\varsigma<0$, must be enforced by choosing a momentum source term that provides a negative contribution through the third term of the mathematical entropy production rate \eqref{eq:math_ent_prod_mass_transfer_lim}.
This also confirms that an energetic transfer from large-scale momentum through a momentum source term in \eqref{eq:model_unclosed_mass_transfer} are necessary.
Indeed, the regularization process creates a small-scale of droplets which generates more interface area and therefore requires more energy due to capillarity energy being proportional to the interface area.

\subsection{Choice of the momentum source term to enforce a dissipative inter-scale transfer}
We look for an expression of the momentum source term such that provides a negatively signed contribution in the mathematical entropy production rate and which is activated similarly as the other source term of the inter-scale mass transfer.
We then propose a momentum source term of the following form
\begin{equation}
    \label{eq:mom-source-term-unclosed}
    \boldsymbol{R}_{\vel} = -\epsilon^{-1}\sigma\Delta H\tilde{R}_{\vel}\vel,
\end{equation}
with $\tilde{R}_{\vel}>0$. 
It particularly enforces that the momentum source term is similarly activated when $\Delta H$ is positive and that both velocity amplitude and the kinetic energy decrease.
Indeed, the negative signing of the mathematical entropy production rate \eqref{eq:math_ent_prod_mass_transfer_lim} is now possible by enforcing a last condition on $\tilde{R}_{\vel}$,
\begin{equation}
    \label{eq:mom-source-term-cond}
    \frac{\sigma H_{lim}}{1-\alpha_1^d}+\left(p_1-p_2\right)
    h-\sigma \frac{S_{avg}}{m_{avg}}\frac{\rho_1}{\overline{\alpha}_2}+\vel^2 \tilde{R}_{\vel}\ge 0.
\end{equation}
Such a source term is only possible for non-zero velocity which is here assumed, but later discussed in Section \ref{sec:mass-transfer-location}.
We propose here to minimize the dissipation of free energy during the inter-scale transfer by choosing $\tilde{R}_{\vel}$ that satisfies the equality case of the above inequality, reading
\begin{equation}
    \label{eq:mom-source-term-partial}
    \tilde{R}_{\vel}=\frac{1}{\vel^2}\left(\sigma\frac{S_{avg}}{m_{avg}}\frac{\rho_1}{\overline{\alpha}_2}-\frac{\sigma H_{lim}}{1-\alpha_1^d}+\left(p_2-p_1\right)
    h\right),
\end{equation}
with the assumption that
\begin{equation}
    \sigma\frac{S_{avg}}{m_{avg}}\frac{\rho_1}{\overline{\alpha}_2}-\frac{\sigma H_{lim}}{1-\alpha_1^d}+\left(p_2-p_1\right)h\ge0,
\end{equation}
as discussed in Section \ref{sec:curv-limit}.
Because of this requirement, the location where the inter-scale mass transfer is activated must be adjusted as further discussed in Section \ref{sec:mass-transfer-location}.
Then, the mathematical entropy production of the model reads
\begin{equation}
    \partial_t \mathcal{H} + \bnabla\bcdot \mathbfcal{G} = \frac{1}{\epsilon}\left(p_1-p_2-\frac{\sigma}{1-\alpha_1^d}H_{lim}\right),
\end{equation}
and the dissipation of the model only comes from pressure relaxation as the model \eqref{eq:relaxed-system}.
The final model reads
\begin{equation}
    \label{eq:final_model}
    \begin{cases}
        \begin{aligned}
            &\partial_t m_1 &+& \bnabla \bcdot (m_1 \vel) &=& -\frac{1}{\epsilon}\frac{\rho_1\sigma }{\overline{\alpha}_2}\Delta H,\\
            &\partial_t m_1^d &+& \bnabla \bcdot (m_1^d \vel) &=& \frac{1}{\epsilon}\frac{\rho_1\sigma }{\overline{\alpha}_2}\Delta H,\\
            &\partial_t m_2 &+& \bnabla \bcdot (m_2 \vel) &=& 0,\\
            &\partial_t \alpha_1^d &+& \bnabla \bcdot (\alpha_1^d\vel) &=& \frac{1}{\epsilon}\frac{\rho_1\sigma }{\overline{\alpha}_2\rho_1^d}\Delta H,\\
            &\partial_t \Sigma &+& \bnabla \bcdot (\Sigma\vel) &=& \frac{1}{\epsilon}\frac{S_{avg}}{m_{avg}} \frac{\rho_1\sigma \Delta H}{\overline{\alpha}_2},\\
            &\partial_t \overline{\alpha}_1 &+& \vel \bcdot \bnabla \overline{\alpha}_1 &=& \frac{1}{\epsilon}(p_1-p_2-\frac{\sigma}{1-\alpha_1^d}H_{lim}),\\
        \end{aligned}\\
        \partial_t (\rho\vel) + \bnabla \bcdot \Big(\rho\vel\otimes \vel+(\overline{p}-\sigma\Vert\bnabla\overline{\alpha}_1\Vert)\mathbf{I}+\sigma \frac{\bnabla\overline{\alpha}_1\otimes\bnabla\overline{\alpha}_1}{\Vert\bnabla\overline{\alpha}_1\Vert}\Big)
        =  -\frac{1}{\epsilon}\sigma\Delta H\left(\sigma\frac{S_{avg}}{m_{avg}}\frac{\rho_1}{\overline{\alpha}_2}-\frac{\sigma H_{lim}}{1-\alpha_1^d}+\left(p_2-p_1\right)
        h\right)\frac{\vel}{\vel^2},
    \end{cases}
\end{equation}
with $\overline{p}= \overline{\alpha}_1 p_1 + \overline{\alpha}_2 p_2$.
In the limit of an instantaneous relaxation, we have 
\begin{equation}
    p_1-p_2=\frac{\sigma H_{lim}}{1-\alpha_1^d},
    \quad
    \text{and}
    \quad
    \overline{H}=H_{lim},
\end{equation}
and the large-scale mean curvature is then expected to be limited and the large-scale interface regularized.

\subsection{Mass transfer location}
\label{sec:mass-transfer-location}
In the model \eqref{eq:final_model}, the mass transfer from large scale to small scale is \apriori triggered everywhere in the domain provided that $\overline{H}\neq H_{lim}$.
However, the discussion of the inter-scale model showed that it can only occur where
\begin{equation}
    \label{eq:non-null-vel-amp}
    \sigma\frac{S_{avg}}{m_{avg}}\frac{\rho_1}{\overline{\alpha}_2}-\frac{\sigma H_{lim}}{1-\alpha_1^d}+\left(p_2-p_1\right)h> 0,
    \qquad
    \text{and,}
    \qquad
    \Vert\vel\Vert
    \neq
    0.
\end{equation}
Moreover, we want to avoid the inter-scale mass transfer where there is not enough volume fraction of the gaseous phase to receive the small-scale inclusions.
This could notably happen in the inner side of the numerical spreading of the large-scale DIM or for small-scale re-impact in the large-scale liquid phase.

We propose then to locate the mass transfer in regions of the flow that avoid these limits of the model.
Therefore, we modify the definition of $H_{lim}$ by setting
\begin{equation}
    H_{lim} = \mathbbm{1}_\mathcal{C}
    \min(\overline{H}, H_{max})
    + (1-\mathbbm{1}_\mathcal{C}) \overline{H},
\end{equation}
where $\mathcal{C}$ is a condition or a set of conditions that enables the mass transfer \via  curvature limitation, only at the location where $\mathcal{C}$ is satisfied.

In order to fulfil the requirements of the regularization and avoid the limitations stated before, we choose
$
\mathcal{C}:=
    \mathcal{C}_1
    \cap
    \mathcal{C}_2
    \cap
    \mathcal{C}_3
$
with
\begin{equation}
    \label{eq:curv-criteria}
    \mathcal{C}_1:=\left(\sigma\frac{S_{avg}}{m_{avg}}\frac{\rho_1}{\overline{\alpha}_2}-\frac{\sigma H_{lim}}{1-\alpha_1^d}+\left(p_2-p_1\right)
    h\ge 0
    \right),
    \qquad
    \mathcal{C}_2:=\left(
        \overline{\alpha}_{I,min}<\overline{\alpha}_1<\overline{\alpha}_{I,max}
    \right),
    \qquad
    \mathcal{C}_3:=\left(
        \bnabla \overline{\alpha}_1\bcdot\vel>0
    \right).
\end{equation}
The condition $\mathcal{C}_1$ corresponds to the decreasing condition of the kinetic energy that also ensures the dissipative nature of the mass transfer.
The condition $\mathcal{C}_2$ ensures that the mass transfer occurs in the outer side  of the large-scale diffuse interface by setting the upper bound $\overline{\alpha}_{I,max}$ such that there is enough gaseous phase in the mixture to receive the small-scale liquid inclusions. Conversely, the lower bound $\overline{\alpha}_{I,min}$ ensures that we indeed are in or very near of the interface.
The condition $\mathcal{C}_3$ avoids re-impact by triggering the mass transfer where the small-scale is advected away from the large-scale interface.
Remark also that $\mathcal{C}_3$ also includes the positivity of velocity amplitude  as required by \eqref{eq:non-null-vel-amp}.

\subsection{Closure of the two-scale model and discussion}
We conclude the modelling part of this work by proposing a specific closure of the two-scale model \eqref{eq:final_model} with expressions of  $S_{avg}/m_{avg}$ and $h$.
With the reduced information about the small-scale geometry $\alpha_1^d$ and $\Sigma$, we propose to make the following assumptions:
\begin{align}
    &\text{\textbullet\: the small-scale and large-scale liquid phases have the same linearized barotropic EOS;} \label{hyp:baro-linearized} \tag{H2a}\\
    &\text{\textbullet\: the inter-scale produces a spray of monodisperse spherical droplets;} \label{hyp:monodisperse} \tag{H2b}\\
    &\text{\textbullet\: the radius of the droplets is smaller than the inverse of the large-scale curvature threshold;}  \label{hyp:smaller-than-curv} \tag{H2c}
\end{align}
With \eqref{hyp:baro-linearized}, we define $p_0$ and $\rho_{0,1}$ a pressure of reference and a density of reference for the liquid such that the EOS reads $p_1(\rho_1) = p_0 + c_1^2 (\rho_1-\rho_{0,1})$.
Integrating the pressure law leads to
\begin{equation}
    e_1(\rho_1)=\frac{(\rho_1 \log(\rho_1))c_1^2-p_0}{\rho_1} + e_c,
\end{equation}
where $e_c$ is an energy constant.
Then, from \eqref{eq:h_def} and denoting $\delta_{\rho_1}=\Delta \rho_1/\rho_1$, we obtain
\begin{equation}
    \label{eq:h-closed}
    h = -\frac{-\delta_{\rho_1}}{1+\delta_{\rho_1}}
    +\frac{\rho_1c_1^2}{(p_2-p_1)\overline{\alpha}_2}
    \left(-\frac{\delta_{\rho_1}}{1+\delta_{\rho_1}}\log(1+\delta_{\rho_1})\right)
    = -\delta_{\rho_1} + O(\delta_{\rho_1}^2).
\end{equation}
Following \eqref{hyp:monodisperse}, let us denote with $r$ the radius of the droplets produced, then $S_{avg}/m_{avg}=3/(r\rho_1^d)$. Then, according to \eqref{hyp:smaller-than-curv}, we have that $r = \kappa H_{lim}^{-1}$, with $\kappa~\le~1$ a scaling factor. 
Remark that $r$ - or more generally $S_{avg}/m_{avg}$ - can be chosen independently of $H_{lim}$.
With the expression proposed, $\kappa$ is an independent parameter, the influence of which is later studied in Section \ref{sec:sim_mass_IAD}.
Using this expression of $S_{avg}/m_{avg}$ in \eqref{eq:mom-source-term-unclosed} and \eqref{eq:mom-source-term-partial}, the source term on the momentum equation becomes
\begin{equation}
    \begin{aligned}
        \boldsymbol{R}_{\vel} &=-\frac{1}{\epsilon}\sigma\Delta H\left(\frac{3}{\kappa\rho_1^d}\frac{\rho_1}{\overline{\alpha}_2}-\frac{1}{1-\alpha_1^d}+\left(p_2-p_1\right)
        \frac{h}{\sigma H_{lim}}\right)\sigma H_{lim}\frac{\vel}{\vel^2}.
    \end{aligned}
\end{equation}
Then, the final two-scale model reads
\begin{equation}
    \label{eq:final_model_closed}
    \begin{cases}
        \begin{aligned}
            &\partial_t m_1 &+& \bnabla \bcdot (m_1 \vel) &=& -\frac{1}{\epsilon}\frac{\rho_1\sigma }{\overline{\alpha}_2}\Delta H,\\
            &\partial_t m_1^d &+& \bnabla \bcdot (m_1^d \vel) &=& \frac{1}{\epsilon}\frac{\rho_1\sigma }{\overline{\alpha}_2}\Delta H,\\
            &\partial_t m_2 &+& \bnabla \bcdot (m_2 \vel) &=& 0,\\
            &\partial_t \alpha_1^d &+& \bnabla \bcdot (\alpha_1^d\vel) &=& \frac{1}{\epsilon}\frac{\rho_1\sigma }{\overline{\alpha}_2\rho_1^d}\Delta H,\\
            &\partial_t \Sigma &+& \bnabla \bcdot (\Sigma\vel) &=& \frac{1}{\epsilon}\frac{3H_{lim}}{\kappa \rho_1^d} \frac{\rho_1\sigma \Delta H}{\overline{\alpha}_2},\\
            &\partial_t \overline{\alpha}_1 &+& \vel \bcdot \bnabla \overline{\alpha}_1 &=& \frac{1}{\epsilon}(p_1-p_2-\frac{\sigma}{1-\alpha_1^d}H_{lim}),\\
        \end{aligned}\\
        \partial_t (\rho\vel) + \bnabla \bcdot \Big(\rho\vel\otimes \vel+(\overline{p}-\sigma\Vert\bnabla\overline{\alpha}_1\Vert)\mathbf{I}+\sigma \frac{\bnabla\overline{\alpha}_1\otimes\bnabla\overline{\alpha}_1}{\Vert\bnabla\overline{\alpha}_1\Vert}\Big)
        =  -\frac{1}{\epsilon}\sigma\Delta H\left(\frac{3}{\kappa\rho_1^d}\frac{\rho_1}{\overline{\alpha}_2}-\frac{1}{1-\alpha_1^d}+\left(p_2-p_1\right)
        \frac{h}{\sigma H_{lim}}\right)\sigma H_{lim}\frac{\vel}{\vel^2},
    \end{cases}
\end{equation}
with $\overline{p}= \overline{\alpha}_1 p_1 + \overline{\alpha}_2 p_2$, and $H_{lim} = \mathbbm{1}_\mathcal{C}
\min(\overline{H}, H_{max})
+ (1-\mathbbm{1}_\mathcal{C}) \overline{H}$.

From this final formulation of the model, one can see that, when there is no curvature limitation, \ie $H_{lim}=\overline{H}$ and $\Delta H =0$, it extends the capillarity model of \cite{chanteperdrix_modelisation_2004} with two additional equations on small-scale variables $\alpha_1^d$ and $\Sigma$ and with a pressure relaxation which accounts for the Laplace pressure jump.
Remark also that the inter-scale mass transfer affects all variables except the effective density of the gaseous phase $m_2$.
Finally, for this specific closure, the inter-scale transfer regularizing process is parametrized by $H_{max}$, which limits the large-scale curvature, $\kappa$ which pilots the amount of small-scale IAD produced by the transfer, and the conditions $\mathcal{C}$ which locates in which regions of the flows the inter-scale transfer occurs.

\section{Numerical strategy}\label{sec:numerics}

We now propose a numerical scheme to solve the model \eqref{eq:final_model_closed} with the definition of $H_{lim}$ given in Section \ref{sec:mass-transfer-location} in the limit $\epsilon\rightarrow 0$. It particularly enforces the local Laplace equilibrium and the limited curvature
\begin{equation}
    p_1-p_2 = \frac{\sigma}{1-\alpha_1^d}H_{lim},
    \qquad
    \overline{H}  = H_{lim}.
\end{equation}
The numerical strategy is based on an adequate decomposition of the model into sub-models, which are then solved with dedicated schemes: a Godunov method \citep{godunov_finite_1959} for the hyperbolic model, an arithmetic solver for the capillarity model \citep{chanteperdrix_modelisation_2004,schmidmayer_model_2017}, and an original implicit-explicit relaxation scheme for the pressure relaxation extending the usual Newton-Raphson methods \citep{chanteperdrix_modelisation_2004,cordesse_diffuse_2020} to avoid a non-local strategy for the solution of the Laplace equilibrium.

\subsection{Splitting, relaxation and time integration}

For the building of the numerical method, we propose to cast the system into the following fully conservative form with a state vector $\bq$, fluxes $\bF$ and source terms $\epsilon^{-1}\boldsymbol{r}(\bq)$,
\begin{equation}
    \label{eq:model-mass-transfer}
    \partial_t \bq + \bnabla\bcdot \bF(\bq)
    = \epsilon^{-1}\boldsymbol{r}(\bq),
\end{equation}
that we will solve in the limit $\epsilon\rightarrow 0$.
The chosen state variable is 
$
\bq:=(
        \overline{\alpha}_1\rho,
        \alpha_1\rho_1,
        \alpha_2\rho_2,
        \alpha_1^d\rho_1^d,
        \alpha_1^d,
        \Sigma,
        \rho \vel
    )^T,
$ switching from the notation $m_k$ to $\alpha_k\rho_k$ for effective densities for further simplifications purposes, and the fluxes $\bF = \bF_{hyp} + \bF_{cap}$ are decomposed following a convective-related part $\bF_{hyp}$ and a capillarity-related part $\bF_{cap}$,
\begin{equation}
    \bF_{hyp}:=
    \begin{pmatrix}
        \overline{\alpha}_1\rho\\
        \alpha_1\rho_1\vel\\
        \alpha_2\rho_2\vel\\
        \alpha_1^d\rho_1^d\vel\\
        \alpha_1^d\vel\\
        \Sigma \vel\\
        \rho \vel \otimes \vel + \overline{p}\Id
    \end{pmatrix},
    \qquad
    \bF_{cap}:=
    \begin{pmatrix}
        0\\
        0\\
        0\\
        0\\
        0\\
        0\\
        \sigma\left(
            \frac{\bnabla\overline{\alpha}_1\otimes\bnabla\overline{\alpha}_1}{\Vert\bnabla\overline{\alpha}_1\Vert}
            -\Vert\bnabla\overline{\alpha}_1\Vert\Id
        \right)
    \end{pmatrix},
\end{equation}
and the source term corresponding to the pressure relaxation and the inter-scale mass transfer process is
\begin{equation}
    \br(\bq) =
    \begin{pmatrix}
        -\frac{\sigma\rho_1}{\overline{\alpha}_2}\Delta H\\
        \frac{\sigma\rho_1}{\overline{\alpha}_2}\Delta H\\
        0\\
        \frac{\sigma\rho_1}{\overline{\alpha}_2\rho_1^d}\Delta H\\
        \frac{3H_{lim}}{\kappa \rho_1^d} \frac{\rho_1\sigma \Delta H}{\overline{\alpha}_2}\\
        \rho(p_1-p_2-\frac{\sigma}{1-\alpha_1^d}H_{lim})\\
        \sigma\Delta H\left(\frac{3}{\kappa\rho_1^d}\frac{\rho_1}{\overline{\alpha}_2}-\frac{1}{1-\alpha_1^d}+\left(p_2-p_1\right)
        \frac{h}{\sigma H_{lim}}\right)\sigma H_{lim}\frac{\vel}{\vel^2}
    \end{pmatrix}.
\end{equation}
In order to develop adapted numerical schemes for each part of the system, we use a splitting procedure for the fluxes and an instantaneous relaxation process for the source term.
It results in solving successively the following three systems,
\begin{subequations}
    \begin{align}
        &\partial_t \bq
        + \bnabla \bcdot \bF^{hyp}(\bq)=0,\label{eq:hyp}\\
        &\partial_t \bq
        + \bnabla \bcdot \bF^{cap}(\bq)=0,\label{eq:cap}\\
        &\br(\bq)=0.\label{eq:relaxation}
    \end{align}
\end{subequations}
Defining a discrete solution $q^n$ at time $t^n$, the discrete operators $\mathbf{L}^{hyp}$, $\mathbf{L}^{cap}$, and $\mathbf{L}^{relax}$ are time integration of systems \eqref{eq:hyp} and \eqref{eq:cap} with explicit Euler scheme and a projection scheme solving \eqref{eq:relaxation}.
We also define $\mathbf{L}^{F}=\mathbf{L}^{cap}\circ\mathbf{L}^{hyp}$.
Then, the state $\bq^n$ computed at the $n-$th time-step results from the following second-order Heun's method with intermediary relaxation steps,
\begin{equation}
    \begin{aligned}
        \bq^{(1)} &= \mathbf{L}^{F}(\bq^{n-1}),\\
        \bq^{(1)}_{rel} &= \mathbf{L}^{relax}(\bq^{(1)}),\\
        \bq^{(2)} &= \mathbf{L}^{F}(\bq^{(1)}_{rel}),\\
        \bq^{(3)} &= \frac{1}{2}(\bq^{n-1}+\bq^{(2)}),\\
        \bq^{n} &= \mathbf{L}^{relax}(\bq^{(3)})
    \end{aligned}
\end{equation}
The stability of this time integration is not well established as the eigenvalues of model \eqref{eq:final_model} are not known.
Therefore, we choose our time step similarly as a CFL condition with a maximal wave speed taken as the upper-bound of the eigenvalues of the augmented model obtained in \eqref{eq:eigenvalues-majorant} such that
\begin{equation}
    \Delta t = \text{CFL} \times \lambda_{max},
\end{equation}
with $\text{CFL}$ the CFL number.
Let us now detail the numerical procedure to solve each of the operators.
\subsection{Hyperbolic fluxes}
We focus here on the numerical method dedicated to the numerical approximation of operator $\boldsymbol{L}^{hyp}: \bq^0\mapsto\bq^{hyp}$.
An adequate strategy to solve this conservative set of equations is to use a Godunov method \citep{godunov_finite_1959,godlewski_hyperbolic_1991,leveque_numerical_1992,toro_riemann_2009} that relies on the conservative form of the equations corresponding to balance equations.
For the sake of readability, we now consider one dimension in space, and we discretize the model into
\begin{equation}
    \label{eq:conservative-schemes}
    \frac{\bq^{hyp}_i-\bq^0_i}{\Delta t}
    = \frac{\mathcal{S}_i}{\mathcal{V}_i}\left(F_{i+1/2}^{hyp}-F_{i-1/2}^{hyp}\right),
\end{equation}
where the subscript indexes the cell in the mesh, the superscript indexes the discretized time.
Then, $\bq^n_i$ is the volume average of the state of the $i-$th cell encompassing the space domain between $x_{i-1/2}=x_i-\Delta x/2$ and $x_{i+1/2}= x_i+\Delta x/2$ at the discretized time $t^n$, $\mathcal{V}_i$ is the volume of the $i-$th cell, $\mathcal{S}_i$ the surface area with the neighbouring cells, and $F_{i\pm1/2}^{hyp}$ are the fluxes at the interface between the $i-$th cell and the $(i+1)-$th cell. Remark that for a one-dimensional regular mesh we have $\mathcal{S}_i/\mathcal{V}_i=1/\Delta x$.

Following the lines proposed by Godunov, we consider the following Riemann problem centred at $x_{i+1/2}$,
\begin{equation}
    \label{eq:riemann-prob}
    \begin{cases}
        \partial_t \bq + \bnabla \bcdot \bF(\bq)=0,\\
        \bq(x,0)=
        \left\{
            \begin{array}{ll}
                \bq_{i} & \quad\text{if }x<0, \\
                \bq_{i+1} &\quad\text{if }x>0.
            \end{array}
        \right.
    \end{cases}
\end{equation}
The solution $\tilde{\bq}$ of this problem is self-similar and for $t>0$, and we note $\tilde{\bq}(x/t)=\bq(x,t)$.
The flux at the interface $F_{i+1/2}^{hyp}$ is computed by evaluating $F^{hyp}(\tilde{\bq}(0))$.
Details of the procedure to solve the Riemann problem \eqref{eq:riemann-prob} is given in Appendix \ref{app:riemann-problem}.
We extend here this approach to a MUSCL scheme \citep{van_leer_towards_1979, toro_riemann_2009} that relies on a linear extrapolation of the conservative set of variables $\bq_{i+1/2}^{\pm}$ of the state within the cell $(x_i,x_{i+1})$ so that the previous method is the same except the initial conditions of the interface Riemann problem that are now
\begin{equation}
    \bq(x,0)=
    \begin{cases}
        \bq_{i+1/2}^{-}, \quad \text{if }x<0, \\
        \bq_{i+1/2}^{+}, \quad \text{if }x>0.
    \end{cases}
\end{equation}
For a regular mesh the linear extrapolation within the $i-$th cell is defined by
\begin{equation}
    \bq^{+} = \bq_i + \frac{\Delta x}{2}\boldsymbol{s}_i,
    \qquad
    \bq^{-} = \bq_i - \frac{\Delta x}{2}\boldsymbol{s}_i,
\end{equation}
where $\boldsymbol{s}_i:=\tfrac{1}{2}(\boldsymbol{s}_{i-1/2}+\boldsymbol{s}_{i+1/2})$ and $\boldsymbol{s}_{i+1/2}:=\tfrac{1}{2}(\bq_{i+1}^n-\bq_{i}^n)$.
Furthermore, the slopes are limited to avoid spurious oscillation using the MINMOD limiter \citep{sweby_high_1984,coquel_entropy_1996,toro_riemann_2009} so that each component $(\boldsymbol{s}_i)_k$ of the slope $\boldsymbol{s}_i$ reads
\begin{equation}
    (\boldsymbol{s}_i)_k=
    \begin{cases}
        \max(0, \min((\boldsymbol{s}_{i-1/2})_k (\boldsymbol{s}_{i+1/2})_k)),\quad\text{if}\quad (\boldsymbol{s}_{i+1/2})_k>0,\\
        \min(0, \max((\boldsymbol{s}_{i-1/2})_k (\boldsymbol{s}_{i+1/2})_k)),\quad\text{if}\quad (\boldsymbol{s}_{i+1/2})_k<0.
    \end{cases}
\end{equation}
\subsection{Capillarity fluxes}
Let us focus now on the building of a numerical scheme for the capillarity fluxes, \ie operator $\boldsymbol{L}^{cap}: \bq^0\mapsto\bq^{cap}$, with an arithmetical-average approach as proposed in \cite{chanteperdrix_modelisation_2004, schmidmayer_model_2017}.
The numerical scheme is here written in two dimensions with $\vel = (u_x,u_y)$, and the cells of a regular mesh space of step sizes $\Delta x$, $\Delta y$ are here discretized with the subscripts $i$ and $j$,
\begin{equation}
    \begin{aligned}
        \frac{(\rho u_x)^{cap}_{i,j}-(\rho u_x)^{0}_{i,j}}{\Delta t}
        &= \sigma \frac{1}{\Delta x}\left[
            \frac{\left(\partial_x \overline{\alpha}_1\right)_{i+1/2,j}\left(\partial_x \overline{\alpha}_1\right)_{i+1/2,j}}
            {\Vert\bnabla\overline{\alpha}_1\Vert_{i+1/2,j}}
            +\frac{\left(\partial_x \overline{\alpha}_1\right)_{i+1/2,j}\left(\partial_y \overline{\alpha}_1\right)_{i+1/2,j}}{\Vert\bnabla\overline{\alpha}_1\Vert_{i+1/2,j}}
            -\Vert\bnabla\overline{\alpha}_1\Vert_{i+1/2,j}\right.\\
            &-
            \left.\left(\frac{\left(\partial_x \overline{\alpha}_1\right)_{i-1/2,j}\left(\partial_x \overline{\alpha}_1\right)_{i-1/2,j}}{\Vert\bnabla\overline{\alpha}_1\Vert_{i-1/2,j}}
            +\frac{\left(\partial_x \overline{\alpha}_1\right)_{i-1/2,j}\left(\partial_y \overline{\alpha}_1\right)_{i-1/2,j}}{\Vert\bnabla\overline{\alpha}_1\Vert_{i-1/2,j}}-\Vert\bnabla\overline{\alpha}_1\Vert_{i-1/2,j}
        \right)\right]\\
        &+ \sigma \frac{1}{\Delta y}\left[
            \frac{\left(\partial_x \overline{\alpha}_1\right)_{i,j+1/2}\left(\partial_x \overline{\alpha}_1\right)_{i,j+1/2}}
            {\Vert\bnabla\overline{\alpha}_1\Vert_{i,+1/2}}
            +\frac{\left(\partial_x \overline{\alpha}_1\right)_{i,j+1/2}\left(\partial_y \overline{\alpha}_1\right)_{i,j+1/2}}{\Vert\bnabla\overline{\alpha}_1\Vert_{i,j+1/2}}
            -\Vert\bnabla\overline{\alpha}_1\Vert_{i,j+1/2}\right.\\
            &-
            \left.\left(\frac{\left(\partial_x \overline{\alpha}_1\right)_{i,j-1/2}\left(\partial_x \overline{\alpha}_1\right)_{i,j-1/2}}{\Vert\bnabla\overline{\alpha}_1\Vert_{i,j-1/2}}
            +\frac{\left(\partial_x \overline{\alpha}_1\right)_{i,j-1/2}\left(\partial_y \overline{\alpha}_1\right)_{i,j-1/2}}{\Vert\bnabla\overline{\alpha}_1\Vert_{i,j-1/2}}-\Vert\bnabla\overline{\alpha}_1\Vert_{i,j-1/2}
        \right)\right],\\
        \frac{(\rho u_y)^{cap}_{i,j}-(\rho u_y)^{0}_{i,j}}{\Delta t}
        &= \sigma \frac{1}{\Delta x}\left[
            \frac{\left(\partial_y \overline{\alpha}_1\right)_{i+1/2,j}\left(\partial_x \overline{\alpha}_1\right)_{i+1/2,j}}
            {\Vert\bnabla\overline{\alpha}_1\Vert_{i+1/2,j}}
            +\frac{\left(\partial_y \overline{\alpha}_1\right)_{i+1/2,j}\left(\partial_y \overline{\alpha}_1\right)_{i+1/2,j}}{\Vert\bnabla\overline{\alpha}_1\Vert_{i+1/2,j}}
            -\Vert\bnabla\overline{\alpha}_1\Vert_{i+1/2,j}\right.\\
            &-
            \left.\left(\frac{\left(\partial_y \overline{\alpha}_1\right)_{i-1/2,j}\left(\partial_x \overline{\alpha}_1\right)_{i-1/2,j}}{\Vert\bnabla\overline{\alpha}_1\Vert_{i-1/2,j}}
            +\frac{\left(\partial_y \overline{\alpha}_1\right)_{i-1/2,j}\left(\partial_y \overline{\alpha}_1\right)_{i-1/2,j}}{\Vert\bnabla\overline{\alpha}_1\Vert_{i-1/2,j}}-\Vert\bnabla\overline{\alpha}_1\Vert_{i-1/2,j}
        \right)\right]\\
        &+ \sigma \frac{1}{\Delta y}\left[
            \frac{\left(\partial_y \overline{\alpha}_1\right)_{i,j+1/2}\left(\partial_x \overline{\alpha}_1\right)_{i,j+1/2}}
            {\Vert\bnabla\overline{\alpha}_1\Vert_{i,+1/2}}
            +\frac{\left(\partial_y \overline{\alpha}_1\right)_{i,j+1/2}\left(\partial_y \overline{\alpha}_1\right)_{i,j+1/2}}{\Vert\bnabla\overline{\alpha}_1\Vert_{i,j+1/2}}
            -\Vert\bnabla\overline{\alpha}_1\Vert_{i,j+1/2}\right.\\
            &-
            \left.\left(\frac{\left(\partial_y \overline{\alpha}_1\right)_{i,j-1/2}\left(\partial_x \overline{\alpha}_1\right)_{i,j-1/2}}{\Vert\bnabla\overline{\alpha}_1\Vert_{i,j-1/2}}
            +\frac{\left(\partial_y \overline{\alpha}_1\right)_{i,j-1/2}\left(\partial_y \overline{\alpha}_1\right)_{i,j-1/2}}{\Vert\bnabla\overline{\alpha}_1\Vert_{i,j-1/2}}-\Vert\bnabla\overline{\alpha}_1\Vert_{i,j-1/2}
        \right)\right],
    \end{aligned}
\end{equation}
where the gradients are discretized following
\begin{equation}
    (\partial_x \overline{\alpha}_1)_{i+1/2,j} = \frac{(\overline{\alpha}_1)_{i+1,j}-(\overline{\alpha}_1)_{i,j}}{\Delta x},
    \qquad
    (\partial_y \overline{\alpha}_1)_{i+1/2,j} = \frac{1}{2}\left(\frac{(\overline{\alpha}_1)_{i,j+1}-(\overline{\alpha}_1)_{i,j-1}}{2\Delta x}+\frac{(\overline{\alpha}_1)_{i+1,j+1}-(\overline{\alpha}_1)_{i+1,j-1}}{2\Delta x}\right),
\end{equation}
and $(\partial_y \overline{\alpha}_1)_{i,j+1/2}$, $(\partial_x \overline{\alpha}_1)_{i,j+1/2}$ are obtained by symmetrically inverting the role of the $x$-axis and $y$-axis.
Finally, the norm of the gradient is obtained with
\begin{equation}
    \Vert\bnabla\overline{\alpha}_1\Vert_{i+1/2,j}
    =
    \sqrt{
        (\partial_x \overline{\alpha}_1)_{i+1/2,j}^2
        +
        (\partial_y \overline{\alpha}_1)_{i+1/2,j}^2
    }.
\end{equation}
\subsection{Relaxation}
The relaxation operator $\boldsymbol{L}^{rel}: \bq^0\mapsto\bq^{rel}$ projects the state $\bq^0$ towards a state $\bq^{rel}$ satisfying both
\begin{equation}
    (1-\alpha_1^d)\left(
        p_1\left(
            \frac{(\alpha_1\rho_1)^{cap}}{\overline{\alpha}_1(1-\alpha_1^d)}
        \right)
        -p_2\left(
            \frac{(\alpha_2\rho_2)^{cap}}{(1-\overline{\alpha}_1)(1-\alpha_1^d)}
        \right)
        - \sigma H_{lim}(\overline{\alpha}_1)
    \right)
    =0,
\end{equation}
and
\begin{equation}
    H_{max} = H_{lim} := \max(H_{max},\overline{H}(\overline{\alpha}_1)),
\end{equation}
at fixed $m_1^{0}+(m_1^d)^{0}$, $m_2^{0}$ and $H_{max}$.
As the definition of $\overline{H}$ involves space derivatives of $\overline{\alpha}_1$, a numerical solution \via a Newton-Raphson procedure would require to couple all the cells of the mesh because of the non-local discretization of $\overline{H}$.
It is not desirable for computational reasons, and we introduce a local strategy instead.

\subsubsection{Implicit-explicit integration of the instantaneous relaxation process to account for mass transfer}
Sharing similar ideas as dual time stepping \citep{jameson_time_1991}, we propose to use an integration procedure for a fictitious time $\tau$ by introducing the dynamical system related to the relaxation with inter-scale mass transfer
\begin{equation}
    \label{eq:relax-fictitious}
    \begin{cases}
        \begin{aligned}
            &\partial_{\tau} (\alpha_1\rho_1) &=& -\frac{1}{\epsilon}\frac{\rho_1\sigma }{\overline{\alpha}_2}\Delta H,\\
            &\partial_{\tau} (\alpha_1^d\rho_1^d) &=& \frac{1}{\epsilon}\frac{\rho_1\sigma }{\overline{\alpha}_2}\Delta H,\\
            &\partial_{\tau} (\alpha_2\rho_2) &=& 0,\\
            &\partial_{\tau} \alpha_1^d &=& \frac{1}{\epsilon}\frac{\rho_1\sigma }{\overline{\alpha}_2\rho_1^d}\Delta H,\\
            &\partial_{\tau} \Sigma &=& \frac{1}{\epsilon}\frac{S_{avg}}{m_{avg}} \frac{\rho_1\sigma \Delta H}{\overline{\alpha}_2},\\
            &\partial_{\tau} \overline{\alpha}_1 &=& \frac{1}{\epsilon}(p_1-p_2-\frac{\sigma}{1-\alpha_1^d}H_{lim}),\\
            &\partial_{\tau} (\rho\vel) &=& -\frac{1}{\epsilon}\left(\frac{3}{\kappa\rho_1^d}\frac{\rho_1}{\overline{\alpha}_2}-\frac{1}{1-\alpha_1^d}+\left(p_2-p_1\right)
            \frac{h}{\sigma H_{lim}}\right)\sigma H_{lim}\frac{\vel}{\vel^2}\sigma\Delta H.
        \end{aligned}
    \end{cases}
\end{equation}
The relaxed state $\bq_{rel}$ is then asymptotically reached for $\tau\rightarrow+\infty$ in \eqref{eq:relax-fictitious} and initial state $\bq^0$.
We propose to  integrate implicitly the source terms on $\overline{\alpha}_1$ and explicitly the other ones involving $\overline{H}$.
This notably enables to recover a method similar to the classical Newton-Raphson method where there is no mass transfer.
Following the chosen implicit-explicit time integration, and defining the function of the Laplace pressure equilibrium
\begin{equation}
    \mathcal{F}(\overline{\alpha}_1,\alpha_1^d):=
    (1-\alpha_1^d)\left(
        p_1\left(
            \frac{m_1}{\overline{\alpha}_1(1-\alpha_1^d)}
        \right)
        -p_2\left(
            \frac{m_2}{(1-\overline{\alpha}_1)(1-\alpha_1^d)}
        \right)
        - \sigma \overline{H}(\overline{\alpha}_1)
    \right),
\end{equation}
the discretization in fictitious time reads
\begin{equation}
    \label{eq:relax-discretization-1}
    \left\{
    \begin{aligned}
        (\alpha_1\rho_1)^{k+1}-(\alpha_1\rho_1)^{k}&=-\frac{\Delta \tau}{\epsilon}\frac{\rho_1^k\sigma }{\overline{\alpha}_2^k}(\Delta H)^k,\\
        (\alpha_1^d\rho_1^d)^{k+1}-(\alpha_1^d\rho_1^d)^{k}&=\frac{\Delta \tau}{\epsilon}\frac{\rho_1\sigma }{\overline{\alpha}_2}(\Delta H)^k,\\
        (\alpha_2\rho_2)^{k}&=(\alpha_2\rho_2)^{0},\\
        (\alpha_1^d)^{k+1}-(\alpha_1^d)^{k}&=\frac{\Delta \tau}{\epsilon}\frac{\rho_1\sigma }{\overline{\alpha}_2\rho_1^d}(\Delta H)^k,\\
        \Sigma^{k+1}-\Sigma^{k}&=\frac{\Delta \tau}{\epsilon}\frac{S_{avg}}{m_{avg}}\frac{\rho_1\sigma }{\overline{\alpha}_2}(\Delta H)^k,\\
        \overline{\alpha}_1^{k+1}-\overline{\alpha}_1^{k} &= \frac{\Delta \tau}{\epsilon}\frac{1}{1-(\alpha_1^d)^{k}}\Big(
            \mathcal{F}^k
            +(\overline{\alpha}_1^{k+1}-\overline{\alpha}_1^{k})(\partial_{\overline{\alpha}_1}\mathcal{F})^k\\
            &+((\alpha_1\rho_1)^{k+1}-(\alpha_1\rho_1)^{k})(\partial_{(\alpha_1\rho_1)}\mathcal{F})^k
            +((\alpha_1^d)^{k+1}-(\alpha_1^d)^{k})(\partial_{(\alpha_1^d)}\mathcal{F})^k
            \Big),\\
        (\rho\vel)^{k+1}-(\rho\vel)^k &= -\frac{\Delta \tau}{\epsilon}\left(\frac{3}{\kappa\rho_1^d}\frac{\rho_1^k}{\overline{\alpha}_2^k}-\frac{1}{1-(\alpha_1^d)^{k}}+\left(p_2^k-p_1^k\right)
        \frac{h^k}{\sigma H_{lim}^k}\right)\sigma H_{lim}^k\frac{(\vel^k)}{(\vel^k)^2} \sigma(\Delta H)^k,
    \end{aligned}
    \right.
\end{equation}
where $\rho_1^k:=(\alpha_1\rho_1)^{k}/(\overline{\alpha}_1^{k})/(1-(\alpha_1^d)^{k})$, $\overline{\alpha}_2^{k}:=1-\overline{\alpha}_1^{k}$, and $(\vel^k):=(\rho\vel)^k/((\alpha_1\rho_1)^{k}+(\alpha_2\rho_2)^{k}+(\alpha_1^d\rho_1^d)^{k})$ and with the initial condition $\bq^{0} = \bq^{cap}$.
After some manipulations for the update of $\overline{\alpha}_1$, we obtain 

\begin{equation}
    \label{eq:relax-discretization-2}
    \begin{aligned}
        %
        \overline{\alpha}_1^{k+1}-\overline{\alpha}_1^{k}
        &= \frac{
            \frac{\Delta \tau}{\epsilon}\frac{1}{1-(\alpha_1^d)^{k}}
        }{
            1-\frac{\Delta \tau}{\epsilon}\frac{1}{1-(\alpha_1^d)^{k}}(\partial_{\overline{\alpha}_1}\mathcal{F})^k
        }
        \left(
            \mathcal{F}^k
            -\frac{\Delta \tau}{\epsilon}\frac{\rho_1^k\sigma }{\overline{\alpha}_2^k}(\Delta H)^k
            \left(
                (\partial_{(\alpha_1\rho_1)}\mathcal{F})^k
                +\frac{\Delta \tau}{\epsilon}\frac{1}{\rho_1^d}(\partial_{(\alpha_1^d)}\mathcal{F})^k
            \right)
        \right),\\
    \end{aligned}
\end{equation}
Now the choice of the fictitious time step ratio $\frac{\Delta \tau}{\epsilon}$ remains.
If no mass transfer were accounted for, one would choose an arbitrarily large value to recover a Newton-Raphson method, but the explicit scheme together with admissibility conditions enforces a finite choice that is now discussed.

\subsubsection{Relaxation restricted to admissible states}
In order to keep the integration scheme providing valid states during the relaxation process, we want to enforce ``stability conditions'' for $\rho\vel$, $\alpha_1\rho_1$, $\alpha_1^d$ and $\overline{\alpha}_1$, by keeping the updated values in their admissible sets.
These set are $(0,1)$ for $\overline{\alpha}_1$ and $\alpha_1^d$, $\mathbb{R}^{+}$ for $\alpha_1\rho_1$ and such that the velocity amplitude decreases.
For $\overline{\alpha}_1$, a parameter  $0<\lambda<1$ close to $1$ to ensure that the next iteration $\overline{\alpha}_1^{k+1}$ does not get closer than a fraction $\lambda$ of the distance that separated $\overline{\alpha}_1^k$ from the boundaries of $(0,1)$.

Starting with the stability condition on $\rho\vel$, a decreasing amplitude is equivalent as
\begin{equation}
    \begin{aligned}
        &(\rho\vel)^{k+1} \bcdot \vel^k\ge 0\\
        \iff&
        \frac{\Delta \tau}{\epsilon}\le\left(\sigma(\Delta H)^k\left(\frac{3}{\kappa\rho_1^d}\frac{\rho_1^k}{\overline{\alpha}_2^k}-\frac{1}{1-(\alpha_1^d)^k}+\left(p_2^k-p_1^k\right)
        \frac{h^k}{\sigma H_{lim}^k}\right)\sigma H_{lim}\right)^{-1}(\rho\vel)^k \bcdot \vel^k=:\left(\frac{\Delta\tau}{\epsilon}\right)_{\rho\vel, max}.
    \end{aligned}
\end{equation}
For the stability on $\alpha_1\rho_1$, we have the following condition ensuring positivity
\begin{equation}
    (\alpha_1\rho_1)^{k+1}\ge 0 \iff 
    \frac{\Delta \tau}{\epsilon}\le (\alpha_1\rho_1)^k \left(\frac{\rho_1^k\sigma }{\overline{\alpha}_2^k}(\Delta H)^k\right)^{-1}=:\left(\frac{\Delta\tau}{\epsilon}\right)_{\alpha_1\rho_1, max}.
\end{equation}
For the stability on $\alpha_1^d$ that is only increasing during the process, we ensure that it does not go beyond $1$ even if we actually expect it to remain small compared to $1$.
It yields
\begin{equation}
    \begin{aligned}
        (\alpha_1^d)^{k+1}-(\alpha_1^d)^{k}\le \lambda(1-(\alpha_1^d)^{k})
        &\iff \frac{\Delta \tau}{\epsilon}\le \lambda(1-(\alpha_1^d)^{k})\left(\frac{\rho_1^k\sigma }{\overline{\alpha}_2^k\rho_1^d}(\Delta H)^k\right)^{-1}=:\left(\frac{\Delta\tau}{\epsilon}\right)_{\alpha_1^d, max}.
    \end{aligned}
\end{equation}
Finally, for the stability of $\overline{\alpha}_1$, we have the following condition,
\begin{equation}
    \label{eq:stability-cond-alpha-transfer}
    -\lambda \overline{\alpha}_1^{k}
    \le
    \overline{\alpha}_1^{k+1}-\overline{\alpha}_1^{k}
    \le
    \lambda (1-\overline{\alpha}_1^{k})
    \iff
    \Big(
    \mathcal{P}_1\left(
        \frac{\Delta \tau}{\epsilon}
    \right)\le0
    \quad
    \text{and}
    \quad
    \mathcal{P}_2\left(
        \frac{\Delta \tau}{\epsilon}
    \right)\ge0
    \Big),
\end{equation}
with
\begin{equation}
    \mathcal{P}_1(X):=a X^2+b_1X+c_1,
    \quad
    \mathcal{P}_2(X):=a X^2+b_2X+c_2,
\end{equation}
and
\begin{equation}
    \begin{gathered}
        a := \frac{\rho_1^k\sigma}{\overline{\alpha}_2^k(1-(\alpha_1^d)^k)}(\Delta H)^k((\rho_1^d)^{-1}(\partial_{\alpha_1^d}\mathcal{F})^k-(\partial_{\alpha_1\rho_1}\mathcal{F})^k),
        \quad
        b_1 :=(1-(\alpha_1^d)^k)^{-1}(\mathcal{F}^k+\lambda\overline{\alpha}_2^k(\partial_{\overline{\alpha}_1}\mathcal{F})^k),\\
        b_2 :=(1-(\alpha_1^d)^k)^{-1}(\mathcal{F}^k-\lambda\overline{\alpha}_1^{k}(\partial_{\overline{\alpha}_1}\mathcal{F})^k),
        \quad
        c_1 :=-\lambda(1-\overline{\alpha}_1^{k}),
        \quad
        c_2 :=\lambda\overline{\alpha}_1^{k}.
    \end{gathered}
\end{equation}
For the first condition $\mathcal{P}_1(\Delta \tau/\epsilon)\le0$, its validity depends on the sign of the discriminant $\Delta_1$ of the second-order polynomial $\mathcal{P}_1$.
\begin{itemize}
    \item If $\Delta_1<0$, the condition is always satisfied as $c_1<0$ and the stability of $\overline{\alpha}_1$ does not introduce any restriction on $\Delta \tau/\epsilon$;
    \item If $\Delta_1>0$, $\Delta \tau/\epsilon$ is restricted by either the first root of $\mathcal{P}_1$ when $a<0$ or the second root when $a>0$.
\end{itemize}
A similar discussion can be conducted for the second condition
$\mathcal{P}_2(\Delta \tau/\epsilon)\ge0$
such that one can define a maximal step ration $(\Delta \tau/\epsilon)_{\overline{\alpha}_1,max}$ defined by the minimal bound enforced on $\Delta \tau/\epsilon$ by the two conditions of \eqref{eq:stability-cond-alpha-transfer}.

In the end, the final step ratio $\Delta \tau/\epsilon$ to be used for the integration in fictitious time \eqref{eq:relax-discretization-2} is 
\begin{equation}
    \frac{\Delta \tau}{\epsilon}
    =\min\left(
        \left(\frac{\Delta \tau}{\epsilon}\right)_{\rho\vel,max},
        \left(\frac{\Delta \tau}{\epsilon}\right)_{\alpha_1\rho_1,max},
        \left(\frac{\Delta \tau}{\epsilon}\right)_{\alpha_1^d,max},
        \left(\frac{\Delta \tau}{\epsilon}\right)_{\overline{\alpha}_1,max}
    \right),
\end{equation}
such that the next iteration $\bq^{k+1}$ fulfils all the admissibility conditions.

\section{Simulations}
\label{sec:simulations}
We propose now to illustrate the inter-scale transfer mechanism and its regularization property with the study of a two-dimensional test-case where a liquid column is deformed by an incident gaseous flow such that filaments appear on each side of the deformed column.
Such a test-case provides us with a curved interface which deforms down to the smallest scales, and eventually breaks up. We propose then to use the regularization of the interface to separate this phenomenon into two different scales: a large-scale interface corresponding to the shape of the regularized column core and a small scale resulting from the regularization of the filaments.
As this numerical experiment has only an illustrative purpose, the parameters of the model were purposely chosen such that the time needed for the deformation of the column is comparable with the one of advection throughout the domain.
Therefore, only the mass density ratio and capillarity coefficient representative of a liquid water/air configuration were retained, but the sound velocities were decreased to mitigate the computational cost of the fast propagation of shock waves.
Nevertheless, this test-case presents all the ingredients to illustrate the inter-scale mass transfer and its curvature-limitation properties.
The numerical method is implemented and the test-case is available in the open-source finite-volume solver [dataset] \texttt{Josiepy}.

After a qualitative presentation of the interface dynamics without the regularizing inter-scale transfer in Section \ref{sec:no-transfer}, we successively assess in Section \ref{sec:transfer}: the regularizing properties of the model with different curvature thresholds in Section \ref{sec:regul}, the repartition of mass and IAD between scales with respect to the choice of the small-scale droplets' size in Section \ref{sec:sim_mass_IAD} and the influence of mesh refinement on the global dynamics with and without regularization in Section \ref{sec:mesh_refinement}.
\subsection{Description of the test-case and simulation without inter-scale transfer}
\label{sec:no-transfer}
We consider a two-dimensional $4\times2$ m domain $\mathcal{D}$ filled with a liquid column filled with water (denoted by the subscript $1$) of circular section of radius $R=0.15$ m and located at the position $C=(1,1)$ m, immersed in a gaseous phase filled with air (denoted by the subscript $2$).
The fluids are given a linearized barotropic EOS: $p(\rho)=p_0 + c_0^2 (\rho-\rho_0)$, the parameters of which are listed in Table \ref{tab:EOS_param}.
The capillarity coefficient is set at $10^{-2}$ N.m${}^{-1}$.
\begin{table}
    \centering
    \begin{tabular}[c]{c|ccc}
        Phase & $p_0$ & $c_0$ & $\rho_0$\\
        \hline
        $1$ & $10^5$ Pa & $10$ m.s${}^{-1}$ & $10^3$ kg.m${}^{-3}$\\
        $2$ & $10^5$ Pa & $10$ m.s${}^{-1}$ & $1$ kg.m${}^{-3}$\\
    \end{tabular}
    \caption{Parameters of the fluids' barotropic linearized EOS.}
    \label{tab:EOS_param}
\end{table}
We distinguish then three areas : the gaseous area (G), the liquid area (L) and the mixture area (M) resulting from a smoothening of the large-scale volume fraction field over a thickness of $R/5$. The location of these areas along with the initialization parameters are summarized in Table \ref{tab:init_state}.
\begin{table}
    \centering
    \begin{tabular}[c]{c|c|ccccccc}
        Area & Location & $\overline{\alpha}_1$ & $p_1$ & $p_2$  & $u_x$ & $u_y$ & $\alpha_1^d$ & $\Sigma$\\
        \hline
        (G) & $\mathcal{D}\setminus \mathcal{B}_C(R+R/5)$ & $0$ & NaN & $p_0$  & $6.66$ m.s${}^{-1}$ & $0$ m.s${}^{-1}$ & $0$ & $0$ m${}^{-1}$\\
        (L) & $\mathcal{B}_C(R)$ & $1$ & $p_0+\sigma/R$ & NaN  & $0$ m.s${}^{-1}$ & $0$ m.s${}^{-1}$ & $0$ & $0$ m${}^{-1}$\\
        (M) & $\mathcal{B}_C(R+R/5)\setminus \mathcal{B}_C(R)$ & $h_{\overline{\alpha}_1}(\boldsymbol{x})$ & $p_0+\sigma \overline{H}$ & $p_0$  & $Y_1 u_{x,\text{(L)}}+Y_2 u_{x,\text{(G)}}$ & $0$ m.s${}^{-1}$ & $0$ & $0$ m${}^{-1}$
    \end{tabular}
    \caption{Initialization state for each area. $\mathcal{B}_C(r)$ denotes the ball of radius $r$ centred in $C$, and $h_{\overline{\alpha}_1}$ is a smoothening function defined by $h_{\overline{\alpha}_1}:\boldsymbol{x}\mapsto\tilde{h}_{\overline{\alpha}_1}(\Vert\boldsymbol{x}-\boldsymbol{x}_C\Vert)$ with $\tilde{h}_{\overline{\alpha}_1}:\boldsymbol{x}\mapsto \exp(2x^2(x^2-3)/(x^2-1)^2)$, and the mass fractions are denoted $Y_k=\alpha_k\rho_k/\rho$.}
    \label{tab:init_state}
\end{table}
An inlet boundary condition is enforced on the left side of the domain with Dirichlet conditions on $\overline{\alpha}_1$, $\vel$, $\alpha_1^d$ and $\Sigma$ to keep the boundary at the initial state, while a homogeneous Neumann condition is set on phase pressures.
An outlet boundary condition is set on the right side with a Neumann condition for all components.
Top and bottom boundaries are periodic.
The simulations are then performed over a time period of $3$ s on  $400\times200$ cells with a CFL condition set to $0.4$.
In this first simulation of reference, the regularizing mass transfer is deactivated by choosing $H_{max}=10^3$~m${}^{-1}$ which is an order of magnitude larger than the inverse of the discretization length $\Delta x^{-1}=10^2$~m${}^{-1}$.

The overall dynamics is showed in Fig. \ref{fig:dynamics-no-transfer} and described in three successive stages:
\begin{itemize}
    \item Stage 1: The liquid column deforms as it undergoes the upstream pressure of the incident flow between $t=0$ s and $t=0.25$ s.
    The interface is well resolved as the iso-line $\overline{\alpha}_1=0.5$ and the interface area estimator maxima are superposed.
    \item Stage 2: We observe the growth of two filaments on both the top and  bottom sides of the liquid column between $t=0.25$ s and $t=1.25$ s.
    The interface is less and less well-located as we go further to the filament's extremity and the IAD estimator shows an opening at its end.
    This shows that the simulation is not converged enough in space discretization and the capillarity phenomena are lost at these small scales.
    \item Stage 3: The water column breaks in two and gets out of the simulation domain between $t=1.25$ s and $t=2$ s.
    The interface has numerically spread too much such that the liquid core of the column does not reach a volume fraction of $1$.
\end{itemize}
These numerical difficulties can also be quantified through the evolution of $H_{lig}~:=~\underset{\mathcal{D}}{\max}(\overline{H}\mathbbm{1}_{\mathcal{C}})$ defined with the criteria \eqref{eq:curv-criteria} that is always located at the end of the ligaments.
Fig. \ref{fig:H-evo-no-transfer} shows that $H_{lig}$ quickly rises from $1/R$ as the ligaments start to grow, and it saturates at approximately $150$ m${}^{-1}$ which corresponds to the scales of the space discretization length.

\subsection{Comparison with the activated mass-transfer}
\label{sec:transfer}
In order to circumvent the challenging resolution of the filaments' growth at large scale, we introduce now the inter-scale transfer to both regularize the large-scale interface, and model the primary atomization in the under-resolved mixed-regime region.
We consider then the same initial setup as the one described in Section \ref{sec:no-transfer}.
However, we change the settings dedicated to the inter-scale mass transfer by choosing $H_{max}=40$ m${}^{-1}$, $\rho_1^d=10^3$ kg.m${}^{-3}$ and $\kappa=1$.
We expect that the curvature threshold $H_{max}$ limits the mean curvature $H_{lig}$ while the latter parameter $\kappa$ pilots the amount of IAD created when mass is transferred form large to small scales.

We propose to discuss the dynamics of this system by highlighting the impact of the two main effects of the inter-scale mass transfer:
(i) the large-scale regularizing properties of the inter-scale process,
(ii) the quantitative repartition of both the liquid mass and IAD between large and small scales.
The following two sets of figures address each of these effects:
\begin{itemize}
    \item[(i)] In Figs. \ref{fig:compare-alpha}-\ref{fig:compare-IAD}, we compare the dynamics at large scale of the two cases by plotting respectively the large-scale volume fraction $\overline{\alpha}_1$ and the large-scale IAD estimator $\Vert\bnabla\overline{\alpha}_1\Vert$.
    In Fig. \ref{fig:H-evo-compare}, we compare the evolution in time of the curvature $H_{lig}$ to measure the regularizing impact of the inter-scale transfer.
    \item[(ii)] In Fig. \ref{fig:compare-mass}, effective densities at large scale $\alpha_1\rho_1$ and small scale $\alpha_1^d\rho_1^d$ are compared and their repartition between the two scales is plotted in time in Fig. \ref{fig:compare-mass-evo}.
    The same discussion is proposed for the IAD in Fig. \ref{fig:compare-IAD_2_scale} along with its evolution in time in Fig. \ref{fig:compare-IAD-evo}.
\end{itemize}
\subsubsection{Regularizing properties}
\label{sec:regul}
Let us first observe from Figs. \ref{fig:compare-alpha}-\ref{fig:compare-IAD} that the dynamics is similar during stage 1 as the inter-scale transfer has not started yet.
When the filaments begin to grow during stage 2, we see that the growth is stopped when the mass transfer is activated \via a curvature threshold set to $H_{max}=40$ m${}^{-1}$.
The interface is locally regularized in the sense that the under-resolved filaments, appearing when there is no mass transfer, have been transferred to the small-scale part of the model.
As showed in Fig. \ref{fig:H-evo-compare}, the curvature $H_{lig}$ is indeed limited starting from stage 2 and is almost always kept below the threshold $H_{max}$.
We observe that the curvature goes over the limit for some snapshots which correspond to situations where the condition $C_1$ is not satisfied, and then, mass transfer cannot occur despite the mean curvature higher than the threshold.

This regularization also allows to ``close'' the interface through a non-negligible amount of IAD $\Vert\bnabla\overline{\alpha}_1\Vert$ all around the iso-line $\overline{\alpha}_1=0.5$ at large scale which makes the capillarity fluxes more effective.
This consequently impacts the overall dynamics, and we particularly observe that the core of the liquid column has a more compact shape.
Given the mesh resolution considered, we have a better resolution of the large-scale capillarity phenomena with the inter-scale transfer, while the previously under-resolved interface dynamics previously observed is now purposely modelled in the small-scale model with geometrical quantities.

Finally, we further investigate the large-scale interface dynamics for several values of $H_{max}\in \{ 30,35,40,60 \}$ comprised into the spectrum of curvature length-scales ranging from the curvature of the initial sphere with $R^{-1}=6.66$ to the highest curvatures measurable on a $200\times400$-cell mesh \ie of order of magnitude of $\Delta x^{-1}=100$.
The selected threshold values were chosen as they all allow sufficient deformation of the interface to let filaments grow while preventing the bending of the diffuse interface reaching the numerical cell size. The impact of such choice of values on the global dynamics is now studied.
In Fig. \ref{fig:H-evo-compare-2} the maximal positive mean curvature measured during the dynamics at the tip of the filaments.
We observe that the regularizing mass transfer is robust for each threshold value as the measured mean curvature does not exceed the preset threshold or only temporarily if the regularization conditions are not fulfilled.
In Fig. \ref{fig:H-compare-alpha}, we can appreciate the impact of the regularization process on the global dynamics as the large-scale liquid phase is subject to capillarity forces but not the small-scale one. 
When the regularization threshold is set by a high curvature $H_{max}\in\{40,60\}$, the large-scale dynamics is not affected in comparison with the case without regularization.
In these cases, the inter-scale transfer only crops the extremities of filaments where the interface is under-resolved and the diffuse interface capillarity forces are weak.
For the smallest curvature threshold $H_{max}\in\{30\}$, the regularization process totally prevents the development of the filament and strongly changes the large-scale interface dynamics.
In the end, this inter-scale regularization keeps the usage of the diffuse interface capillarity model where the interface is resolved enough, but it can also affect the dynamics of the resolved interface depending on the threshold setting as demonstrated here.
\subsubsection{Repartition of mass and IAD between scales}
\label{sec:sim_mass_IAD}
Now let us discuss the repartition of the liquid mass and the IAD between both scales.
As expected one can observe in Fig. \ref{fig:compare-mass-evo} that some large-scale liquid mass is transferred to the small-scale model while conserving the total liquid mass during stage 2.
The superposition of the effective densities at both scales shows that the mass transfer has happened at the extremities of the large-scale ligaments, and the small-scale liquid phase is then advected by the flow.
We can again measure the overall impact on the dynamics as the liquid mass is not spatially distributed at the same location, whether the inter-scale transfer is activated or not, by summing the contributions of both the large- and small-scale components. 

Regarding the IAD, one can see in Fig. \ref{fig:compare-IAD-evo} that the regularization tends to decrease the total large-scale IAD when mass transfer is activated, in accordance with the more compact shape of the liquid core.
However, the sum of the IAD from both scales resulting from the regularization largely exceeds the large-scale IAD when the inter-scale transfer is deactivated.
Indeed, the IAD production associated with small-scale droplets outweighs the reduction of IAD of the large-scale interface induced by the regularization.
In this sense, the regularization process can be interpreted as a primary break-up model for under-resolved interface instabilities of length-scales smaller than the diffuse interface thickness.
Parameters of such process could then be chosen to reproduce experimental heuristics by adding flow-dependent condition to activate at some specific interface location or by choosing specific droplet sizes produced at small scale. 
For illustration purposes, we propose a set of simulations with varying $\kappa$ which involves the production of smaller and smaller droplets in comparison with the regularization length-scale threshold $H_{max}$. 
We display in Fig. \ref{fig:compare-kappa} the amount of IAD, large-scale and small-scale combined, for $\kappa\in\{0.1,0.2,0.5,1\}$ and for no inter-scale transfer.
We particularly observe an additional amount of small-scale IAD approximately proportional to the inverse of $\kappa$ according to the dependency of the source term of IAD (see \eqref{eq:final_model_closed}).
Moreover, the evolution of the IAD also exhibits more clearly that the regularization process is mostly performed gradually with punctually short period of intense regularization.
These moments correspond to developments of new structures at the large-scale which eventually trigger the regularization.
For instance, two new filaments grow downstream of the liquid inclusion from $t=1.2$ s to $t=1.6$ s as showed notably in Fig. \ref{fig:H-compare-alpha} and their regularization explains the change of IAD production rate at the same period of time in Fig. \ref{fig:compare-kappa}.

\section{Conclusion}
\label{sec:conclusion}
%
Following the two-scale modelling approach, we have successfully introduced a proper way of combining two scales in a unified manner, 
accounting for capillarity at both scales, and proposed an innovative local regularization of the large-scale interface through the definition of a dissipative mass transfer between scales.
%
With a multi-fluid CSS approach for the large-scale capillarity model, we have added a supplementary potential energy depending on the gradient of a colour function, that is chosen to be the large-scale volume fraction.
The resulting model includes both capillarity fluxes along with a local Laplace equilibrium.
A modification of this local equilibrium is then used to build a mass transfer model between scales, that induces a regularization technique of the large-scale interface.
More specifically, we enforce a relaxation evolution towards a Laplace equilibrium with a preset maximal curvature which sets an upper limit for the large-scale interface curvatures.
This modified large-scale Laplace equation then triggers the transfer to a small-scale kinetic-based model based on at least two geometric quantities the small-scale IAD and volume fraction.
Finally, a numerical scheme along with simulations allows to confirm the expected behaviour of the model on a first demonstrating case.
The parameters of the model need to be further investigated with DNS comparisons where smaller scales are resolved.
Nevertheless, the approach lays the foundations of a key feature for unified two-scale models including the mixed zone, while controlling the range of scales to be resolved in the numerical simulations at the modelling level.


For further studies and models, the introduction of a different velocity at small scale would improve the description of the disperse regime by including key phenomena such as drag, added mass or secondary break-up.
Then, it would allow more complex numerical test-cases to evaluate the performance of the interface-regularizing property of the model along with heuristic tuning of the parameters of the source terms.
The model could then be tested against classic numerical benchmarks including phase inversion problem \citep{estivalezes_phase_2022} or primary atomization of a liquid jet \citep{shinjo_recent_2018}.
However, such numerical setup requires a much higher computational effort which are not available in the current implementation of the model.
The use of more advanced solvers, including adaptive mesh refinement is currently in progress to propose validation test-cases.
Such a model could also be extended to both an improved small-scale description in the disperse regime and the mixed regime, and would rely on an extended small-scale model accounting for deformed and polydisperse inclusions as proposed in \cite{loison_two-scale_2023}.
Another perspective enabled by our two-scale approach lies in the modelling of a small-scale gaseous phase in the liquid bulk phase.
A similar derivation could then hold with a symmetric regularizing mass transfer active on zones where mean curvature amplitude is large but negative.
Furthermore, coalescence models must be considered to obtain reliable droplet distribution out of the dilute regime.
Pressure and temperature relaxations in the context of fluids with full EOS and thermodynamical mass exchange such as proposed by \cite{pelanti_arbitrary-rate_2022} are also under consideration for further developments.


\section*{Declaration of competing interest}

The authors declare that they have no known competing financial interests or personal relationships that could have appeared to influence the work reported in this paper.

\section*{Data availability}

Source code of the finite-volume solver \texttt{Josiepy}, used in this study and developed by the HPC@Maths team, is available on the GitHub repository \texttt{github.com/hpc-maths/josiepy}.
Simulation data will be made available on request.

\section*{Acknowledgments}

This work has been funded by the French Department of Defence (Defence Innovation Agency) through the MMEED project. The PhD of A.L. is funded by a grant of the French Department of Higher Education and Research and the French Department of Defence (Defence Innovation Agency).

\appendix
\section{Hamilton's stationary action principle}
\label{app:SAP}

Hamilton's SAP consists in the minimization of the action $\Action:=\int_\Omega \lag$ on the space-time domain $\Omega$.
Consider a small parameter $\lambda$ in the vicinity of $0$ and a family of trajectories $\phi^{\lambda}(\boldsymbol{X},t,\lambda)$ that maps a position $\boldsymbol{X}$ of the referential domain $\Omega_{x}(0)$ to its position $\boldsymbol{x}\in \Omega_{x}(t)$ at instant $t$.
This enables the definition of an infinitesimal Eulerian displacement
\begin{equation}
    \boldsymbol{\VarTraj}(\boldsymbol{x},t) := \left(\partial_{\lambda}\boldsymbol{\phi}^{\lambda}\right)_{\boldsymbol{X},t}((\boldsymbol{\phi}^{\lambda})^{-1}(\boldsymbol{x},t,\lambda=0),t, \lambda=0).
\end{equation}
Introducing a corresponding family of Eulerian fields  $b^{\lambda}(\boldsymbol{x},t, \lambda)$, one can then define a variational operator  $\delta(\cdot)$ which acts on Eulerian fields $b$ following
%
\begin{equation}
    \delta b(\boldsymbol{x},t) := \left(\partial_{\lambda} b^{\lambda}\right)_{\boldsymbol{x},t}(\boldsymbol{x},t,\lambda=0).
\end{equation}
We assume that these families of Lagrangian mappings and Eulerian fields satisfy the following properties:
\begin{itemize}[leftmargin=*]
    \item The mapping $\boldsymbol{\phi}$ and Eulerian fields $b$ of the solution are included in the families for $\lambda=0$ \textit{i.e.} for all $\boldsymbol{X}\in\Omega_{x}(0)$ and $(\boldsymbol{x},t)\in \Omega$,
    \begin{equation}
        \boldsymbol{\phi}^{\lambda}(\boldsymbol{X},t,\lambda=0)=\boldsymbol{\phi}(\boldsymbol{X},t),
        \qquad
        b^{\lambda}(\boldsymbol{x},t,\lambda=0)=b(\boldsymbol{x},t).
    \end{equation}
    \item All the mappings and Eulerian fields preserve the constraints. Denote the conserved Eulerian field $b_{c}\in\{m_1, m_2, m_1^d, \alpha_1^d\}$, and the advected Eulerian fields $b_{a}\in\{z, \rho_1^d\}$, then for all $(\boldsymbol{x},t)\in \Omega$,
    \begin{equation}
        \partial_t b_c^{\lambda} + \bnabla\bcdot(b_c^{\lambda}\vel)=0,
        \qquad
        \partial_t b_a^{\lambda} + \vel\bcdot\bnabla b_a^{\lambda}=0.
    \end{equation}
    \item All the mappings and families of Eulerian fields $b$ preserve the values at the boundaries of the space-time domain \textit{i.e.} for all $(\boldsymbol{x},t)\in \partial\Omega$,
    \begin{equation}
        b^{\lambda}(\boldsymbol{x},t,\lambda)=b(\boldsymbol{x},t).
    \end{equation}
\end{itemize}
With this variational operator, Hamilton's SAP writes
\begin{equation}
    \delta\mathcal{A}=0.
\end{equation}
Following \cite{gavrilyuk_hyperbolic_1998,gavrilyuk_mathematical_2002}, the variations of the conserved fields $b_c$, the advected fields $b_a$ and $\vel$ are related to $\boldsymbol{\VarTraj}$ through relations
\begin{equation}
    \label{eq:variation_constraints}
    \delta b_c = - \bnabla \bcdot (b_c\boldsymbol{\VarTraj}),
    \qquad
    \delta b_a = -(\boldsymbol{\VarTraj}\bcdot\bnabla)b_a,
    \delta\vel = D_t \boldsymbol{\VarTraj}-(\boldsymbol{\VarTraj}\bcdot\bnabla)\vel.
\end{equation}
The variation of $\overline{\alpha}_2$ is $\delta\overline{\alpha}_2=-\delta\overline{\alpha}_1$ because of the volume occupation relation $\overline{\alpha}_1+\overline{\alpha}_2=1$.
As the variational operator $\delta$ commutes with space derivatives, we also have that $\delta(\bnabla \overline{\alpha}_1)=\bnabla(\delta\overline{\alpha}_1)$.
We also change the dependencies of the Lagrangian \eqref{eq:lagrangian} by writing $\alpha_k= \overline{\alpha}_k(1-\alpha_1^d)$ for $k=1,2$ and $\Sigma=m_1^d(\rho_1^d)^{-2/3}z$.
Then, the Lagrangian \eqref{eq:lagrangian} solely depends on the conserved quantities $b_c\in\{m_1, m_2, m_1^d, \alpha_1^d\}$, the advected quantities $b_a\in\{z, \rho_1^d\}$, and $\vel$, $\overline{\alpha}_1$, $\overline{\alpha}_2$, $\bnabla\overline{\alpha}_1$.
\begin{equation}
    \lag(m_1,m_2,m_1^d,\rho_1^d,z,\overline{\alpha}_1,\alpha_1^d,\vel)
    =
    \lag_1(m_1,\overline{\alpha}_1,\alpha_1^d,\vel)
    +\lag_2(m_2,\overline{\alpha}_1,\alpha_1^d,\vel)
    +\lag_{cap}(\Vert\bnabla\overline{\alpha}_1\Vert)
    +\lag_1^d(m_1^d,\rho_1^d,z,\vel).
\end{equation}
For concision purposes, we denote some partial derivatives of the Lagrangian with $\lag_k^*:=m_k\partial_{m_k}\lag_k - \lag_k$,  for $k=1,2,1^d$, $\boldsymbol{D}^T:=\partial_{\bnabla\overline{\alpha}_1}\lag_{cap}$, and $\boldsymbol{K}^T:=\partial_{\vel}\lag$.
We also write the divergence of a matrix $\boldsymbol{A}$, $\bnabla\bcdot\boldsymbol{A}=\partial_{x_j}A_{ij}$ with summation on repeated indexes.
Using integration by parts, the variation of the action reads
\begin{equation}
    \begin{aligned}
        \delta\Action = \int_{\Omega}
        &-\left\{
            \partial_t \boldsymbol{K} + \bnabla\bcdot\Big[\boldsymbol{K}\otimes\vel-(\lag_1^*+\lag_2^*+\lag_1^{d,*}-\lag_{cap}
            +\alpha_1^d\partial_{\alpha_1^d}(\lag_1+\lag_2))\boldsymbol{I}
        -\bnabla\overline{\alpha}_1\otimes\boldsymbol{D}\Big]\right.\\
        &\hspace{20pt}\left.-(\partial_{\overline{\alpha}_1}\lag_1-\partial_{\overline{\alpha}_2}\lag_2-\bnabla\bcdot\boldsymbol{D})\bnabla\overline{\alpha}_1\right\}\bcdot\VarTraj\\
        &+\Big(
            \partial_{\overline{\alpha}_1}\lag_1-\partial_{\overline{\alpha}_2}\lag_2-\bnabla\bcdot\boldsymbol{D}
        \Big)\delta\overline{\alpha}_1,
    \end{aligned}
\end{equation}
where $\boldsymbol{I}$ is the identity matrix.
Then, Hamilton's SAP, \textit{i.e.} $\delta\Action=0$ for any variation of the trajectories $\VarTraj$ and large-scale volume fraction variation $\delta\overline{\alpha}_1$, yields
\begin{equation}
    \begin{cases}
        \partial_t \boldsymbol{K} + \bnabla\bcdot\Big[\boldsymbol{K}\otimes\vel-(\lag_1^*+\lag_2^*+\lag_1^{d,*}-\lag_{cap}
        +\alpha_1^d\partial_{\alpha_1^d}(\lag_1+\lag_2))\boldsymbol{I}-\bnabla\overline{\alpha}_1\otimes\boldsymbol{D}\Big]=\boldsymbol{0},\\
        \partial_{\overline{\alpha}_1}\lag_1-\partial_{\overline{\alpha}_2}\lag_2-\bnabla \bcdot \boldsymbol{D}=0.
    \end{cases}
\end{equation}
Evaluating the derivatives of the Lagrangian gives
\begin{equation}
    \boldsymbol{K}=\rho\vel,
    \quad
    \boldsymbol{D}=-\sigma \frac{\bnabla\overline{\alpha}_1}{\Vert\bnabla\overline{\alpha}_1\Vert},
    \quad
    \lag_1^*=-\alpha_1p_1,
    \quad
    \lag_2^*=-\alpha_2p_2,
    \quad
    \lag_1^{d,*}=0,
\end{equation}
and for $k=1,2$,
\begin{equation}
    \partial_{\alpha_1^d}\lag_k = -\overline{\alpha}_kp_k,
    \quad
    \partial_{\overline{\alpha}_k}\lag_k=(1-\alpha_1^d)p_k.
\end{equation}
Using the conservative variables $(m_1,m_2,m_1^d,\alpha_1^d,\Sigma,\rho\vel)$, the full system including constraints \eqref{eq:mass_conservation}-\eqref{eq:incomp}-\eqref{eq:Sigma_cons} reads
\begin{equation}
    \label{eq:non-diss-system}
    \begin{cases}
        \begin{aligned}
            &\partial_t m_k &+& \bnabla \bcdot (m_k \vel) &=& 0, \qquad k=1,2,1^d,\\
            &\partial_t \alpha_1^d &+& \bnabla \bcdot (\alpha_1^d\vel) &=& 0,\\
            &\partial_t \Sigma &+& \bnabla \bcdot (\Sigma\vel) &=& 0,\\
        \end{aligned}\\
        \partial_t (\rho\vel) + \bnabla \bcdot \Big(\rho\vel\otimes \vel
        +(\overline{p}-\sigma\Vert\bnabla\overline{\alpha}_1\Vert)\mathbf{I}+\sigma \frac{\bnabla\overline{\alpha}_1\otimes\bnabla\overline{\alpha}_1}{\Vert\bnabla\overline{\alpha}_1\Vert}\Big)= \boldsymbol{0},
    \end{cases}
\end{equation}
with
\begin{equation}
    \overline{p}
    :=
    \overline{\alpha}_1p_1\left(\frac{m_1}{\overline{\alpha}_1(1-\alpha_1^d)}\right)
    +
    \overline{\alpha}_2p_2
    \left(\frac{m_2}{(1-\overline{\alpha}_1)(1-\alpha_1^d)}\right),
\end{equation}
and $\overline{\alpha}_1$ defined by the implicit Laplace equilibrium
\begin{equation}
    \label{eq:laplace_eq}
    p_1\left(\frac{m_1}{\overline{\alpha}_1(1-\alpha_1^d)}\right)
    -p_2\left(\frac{m_2}{(1-\overline{\alpha}_1)(1-\alpha_1^d)}\right)
    =\frac{\sigma}{1-\alpha_1^d}\overline{H}(\overline{\alpha}_1),
\end{equation}
where $\overline{H}(\overline{\alpha}_1)$ is defined by \eqref{eq:implicit-mean-curvature}.

\section{Mathematical entropy production of the two-scale capillarity model}
\label{app:math_entropy_prod}
For calculation purposes, we consider a transport equation on the variable $z=(\rho_1^d)^{2/3}\Sigma/m_1^d$ similarly to \cite{di_battista_towards_2021} instead of the conservation equation on $\Sigma$.
We introduce then the source term $R_z$ such that $D_t z = R_z$, and $R_z = R_{\Sigma}(\rho_1^d)^{2/3}/m_1^d+ z R_{m_1}/m_1^d$.
We do not prescribe the dynamics of $\overline{\alpha}_1$, and we consider then the following system of equations,
\begin{equation}
    \label{eq:general_system}
    \begin{cases}
        \begin{aligned}
            &\partial_t m_1 &+& \bnabla\bcdot(m_1\vel)&=&R_{m_1},\\
            &\partial_t m_2 &+& \bnabla\bcdot(m_2\vel)&=&0,\\
            &\partial_t m_1^d &+& \bnabla\bcdot(m_1^d\vel)&=&-R_{m_1},\\
            &\partial_t \alpha_1^d &+& \bnabla\bcdot(\alpha_1^d\vel)&=&-(\rho_1^d)^{-1}R_{m_1},\\
            &\partial_t z &+&  \vel \cdot \bnabla z &=& R_z,\\
            &\partial_t (\rho \vel) &+& \bnabla\bcdot (\rho \vel\otimes\vel+\mathbf{P})&=&\boldsymbol{R}_{\vel},
        \end{aligned}
    \end{cases}
\end{equation}
where the dynamics of $\overline{\alpha}_1$ is not specified and $\mathbf{P}$ is a general pressure tensor.
Remark also that the fourth and fifth equations are equivalent to $D_t \rho_1^d=0$ and $D_t z = R_z$.
We look for a supplementary conservation equation for an entropy-entropy flux pair $(\mathcal{H},\mathbfcal{G})$ such that the entropy production rate $\varsigma$ is negatively signed,
\begin{equation}
    \varsigma:=\partial_t \mathcal{H} + \bnabla\bcdot \mathbf{\mathbfcal{G}} \le 0.
\end{equation}
With the summation convention on repeated indexes, the divergence of a matrix $\boldsymbol{A}$ is $\bnabla\bcdot\boldsymbol{A}=\partial_{x_j}A_{ij}$, and the double scalar product of two matrices $\boldsymbol{A}$ and $\boldsymbol{B}$ is $\boldsymbol{A}:\boldsymbol{B}=A_{ij}B_{ji}$.
Furthermore, the gradient of a vector $\boldsymbol{a}$ is $\bnabla\boldsymbol{a}=\partial_{x_j}a_i$.
With $\mathcal{H}:= \boldsymbol{K}\bcdot\vel -\lag$, multiplying the momentum equation of \eqref{eq:general_system} by $\vel$ gives
\begin{equation}
    \begin{aligned}
        0=&\vel\bcdot\partial_t \boldsymbol{K} + \vel \bcdot \left[ \bnabla \bcdot (\boldsymbol{K}\otimes \vel+\boldsymbol{P})\right]-\vel \bcdot {R}_{\vel}\\
        =&\partial_t (\boldsymbol{K}\bcdot \vel)-\boldsymbol{K}\bcdot \partial_t \vel + \bnabla \bcdot\left[(\boldsymbol{K}\bcdot \vel)\otimes \vel+\boldsymbol{P}^T\vel\right]
        -(\boldsymbol{K}\otimes\vel):\bnabla \vel-\boldsymbol{P}:\bnabla\vel-\vel \bcdot {R}_{\vel}\\
        =&\partial_t \mathcal{H}+ \bnabla \bcdot (\mathcal{H}\vel+\boldsymbol{P}^T\vel)
        +\partial_t \lag
        +\bnabla \bcdot (\lag \vel)
        -\boldsymbol{K}\bcdot\partial_t \vel 
        -(\boldsymbol{K}\otimes\vel):\bnabla \vel-\boldsymbol{P}:\bnabla\vel-\vel \bcdot {R}_{\vel}.
    \end{aligned}
\end{equation}
Developing the derivatives of the Lagrangian and accounting for the dynamics given by \eqref{eq:general_system} yields
\begin{align*}
    0=&\partial_t \mathcal{H}+ \bnabla \bcdot \left[\mathcal{H}\vel+\boldsymbol{P}^T\vel\right]\\
    &+\partial_{m_1}\lag_1D_t m_1
    +\partial_{\overline{\alpha}_1}\lag_1 D_t \overline{\alpha}_1
    +\partial_{\alpha_1^d}\lag_1D_t\alpha_1^d\\
    &+\partial_{m_2}\lag_2D_t m_2
    +\partial_{\overline{\alpha}_2}\lag_2D_t \overline{\alpha}_2
    +\partial_{\alpha_1^d}\lag_2D_t\alpha_1^d\\
    &+\partial_{m_1^d}\lag_1^dD_t m_1^d
    +\partial_{\rho_1^d}\lag_1^dD_t \rho_1^d
    +\partial_z\lag_1^dD_t z\\
    &+\boldsymbol{D}\bcdot D_t (\bnabla\overline{\alpha}_1)+\lag\bnabla\bcdot \vel-\boldsymbol{P}:\bnabla\vel-\vel \bcdot {R}_{\vel}\\
    =&\partial_t \mathcal{H}+ \bnabla\bcdot \left[\mathcal{H}\vel+\boldsymbol{P}^T\vel+(D_t\overline{\alpha}_1)\boldsymbol{D}\right]
    +(\partial_{\overline{\alpha}_1}\lag_1-\partial_{\overline{\alpha}_2}\lag_2-\bnabla\bcdot \boldsymbol{D}) D_t \overline{\alpha}_1\\
    &+\left[\partial_{m_1}\lag_1-\partial_{m_1^d}\lag_1^d-\frac{1}{\rho_1^d}\left(\partial_{\alpha_1^d}\lag_1+\partial_{\alpha_1^d}\lag_2\right)\right]R_{m_1}
    +\partial_z\lag_1^d R_z\\
    &-\left\{\boldsymbol{P}+\left[\lag_1^*+\lag_2^*+\lag_1^{d,*}-\lag_{cap}+\alpha_1^d\left(\partial_{\alpha_1^d}\lag_1+\partial_{\alpha_1^d}\lag_2\right)\right]\boldsymbol{I}+\boldsymbol{D}\otimes\bnabla\overline{\alpha}_1\right\}:\bnabla\vel-\vel \bcdot {R}_{\vel}.
\end{align*}
Evaluating the Lagrangian leads to
\begin{align*}
    0=&\partial_t \mathcal{H}+ \bnabla\bcdot \left(\mathcal{H}\vel+\mathbf{P}^T\vel-\sigma \frac{\bnabla\overline{\alpha}_1}{\Vert\bnabla\overline{\alpha}_1\Vert}D_t\overline{\alpha}_1\right)
    +\left[(1-\alpha_1^d)(p_1-p_2)-\sigma\overline{H}\right] D_t \overline{\alpha}_1\\
    &+\left(e_1^d-e_1-\frac{p_1}{\rho_1}+\sigma z (\rho_1^d)^{-2/3}+\frac{\overline{\alpha}_1p_1+\overline{\alpha}_2p_2}{\rho_1^d}\right)R_{m_1}\\
    &-\left(\mathbf{P}-\left((\overline{\alpha}_1p_1+\overline{\alpha}_2p_2-\sigma\Vert\bnabla\overline{\alpha}_1\Vert)\mathbf{I}+\sigma\frac{\bnabla\overline{\alpha}_1\otimes\bnabla\overline{\alpha}_1}{\Vert\bnabla\overline{\alpha}_1\Vert}\right)\right):\bnabla\vel\\
    &-\sigma m_1^d (\rho_1^d)^{-2/3} R_z-\vel \bcdot {R}_{\vel},
\end{align*}
with $e_1^d:=e_1(\rho_1^d)$ and $e_1:=e_1(\rho_1)$.
We choose the entropy flux $\mathbfcal{G}$ by setting
\begin{equation}
    \begin{aligned}
        \mathbf{P}&:=\left(\overline{\alpha}_1p_1+\overline{\alpha}_2p_2-\sigma\Vert\bnabla\overline{\alpha}_1\Vert\right)\mathbf{I}+\sigma\frac{\bnabla\overline{\alpha}_1\otimes\bnabla\overline{\alpha}_1}{\Vert\bnabla\overline{\alpha}_1\Vert},\\
        \mathbfcal{G}&:=\mathcal{H}\vel+\mathbf{P}^T\vel-\sigma \frac{\bnabla\overline{\alpha}_1}{\Vert\bnabla\overline{\alpha}_1\Vert}D_t\overline{\alpha}_1.
    \end{aligned}
\end{equation}
With the expression of Lagrangian \eqref{eq:lagrangian}, the mathematical entropy production rate finally evaluates to
\begin{equation}
    \begin{aligned}
        \varsigma=&-\left[(1-\alpha_1^d)(p_1-p_2)-\sigma\overline{H}\right] D_t \overline{\alpha}_1
        -\left(e_1^d-e_1-\frac{p_1}{\rho_1}+\sigma z (\rho_1^d)^{-2/3}+\frac{\overline{\alpha}_1p_1+\overline{\alpha}_2p_2}{\rho_1^d}\right)R_{m_1}\\
        &+\sigma m_1^d (\rho_1^d)^{-2/3} R_z+\vel \bcdot {R}_{\vel}.
    \end{aligned}
\end{equation}
Then, the sign of the mathematical entropy production rate $\varsigma$ depends on the assumptions on the dynamics of $\overline{\alpha}_1$ and the source terms $R_{m_1}$ and $R_z$.
\section{Hyperbolicity of the augmented two-scale model with capillarity}
\label{app:hyperbolicity}
The system of conservation equations modelling our two-scale two-phase flow with capillarity (\ref{eq:relaxed-system}) involves fluxes, which not only depend on the set of conserved variables, but also on their gradients, in particular for $\overline{\alpha}_1$. 
A possible mean to study the mathematical properties of the system consists in considering
an augmented system of equation including a new conserved variable  $\boldsymbol{w}:=\bnabla\overline{\alpha}_1$.
Depending on capillarity fluxes model, the system may still involve derivative of the conservative variables, and it is possible to resort to a symmetrization of the system using entropy variables in order to study the structure of the resulting system of Partial Differential Equations (PDEs) \cite{gavrilyuk_new_1999,giovangigli_symmetrization_2022}.

Nevertheless, within the framework of our model, a study of hyperbolicity for \eqref{eq:relaxed-system} can be led under the following assumptions along the same lines as \cite{schmidmayer_model_2017}:
1- we consider an augmented system of conservation equations, where the new variable  $\boldsymbol{w}$ is introduced and satisfies an independent conservation equation. The link between $\bnabla\overline{\alpha}_1$ and $\boldsymbol{w}$ is then a result of initial conditions and of the dynamics of the system of PDEs. 
2- Even if we rely on this augmented variable, we are still in the presence of gradients of the conserved variables in the sources terms, where the mean curvature involves the derivative of  $\boldsymbol{w}$. These terms are still considered as source terms and are supposed to be local fields, in the sense that they are not taken into account in the convective part of the system.

We then consider an augmented model with $\boldsymbol{w}$ as an independent variable.
Taking the gradient of the equation on $\overline{\alpha}_1$ leads to
\begin{equation}
    \partial_t \boldsymbol{w}+\bnabla (\vel\bcdot \boldsymbol{w}) = \boldsymbol{S},
\end{equation}
where $\boldsymbol{S}$ is a source term which does not impact the hyperbolicity study.
We then consider the following first-order homogeneous system
\begin{equation}
    \label{eq:relaxed-system-homo-app}
    \begin{cases}
        \setlength\arraycolsep{2pt}
        \begin{array}{lclcl}
            \partial_t m_k &+& \bnabla \bcdot (m_k \vel) &=& 0, \qquad k=1,2,1^d,\\
            \partial_t \alpha_1^d &+& \bnabla \bcdot (\alpha_1^d\vel) &=& 0,\\
            \partial_t \Sigma &+& \bnabla \bcdot (\Sigma\vel) &=& 0,\\
            \partial_t \overline{\alpha}_1 &+& \vel \bcdot \bnabla \overline{\alpha}_1 &=& 0,\\
            \partial_t \boldsymbol{w}&+&\bnabla (\vel\bcdot \boldsymbol{w}) &=& 0,
        \end{array}\\
        \partial_t (\rho\vel) + \bnabla \bcdot \left[\rho\vel\otimes \vel+(\overline{p}-\sigma\Vert\boldsymbol{w}\Vert)\mathbf{I}+\sigma \frac{\boldsymbol{w}\otimes\boldsymbol{w}}{\Vert\boldsymbol{w}\Vert}\right]= 0.
    \end{cases}
\end{equation}
Remark that the above system is not rotational invariant as the equation on $\boldsymbol{w}$ is not an equation of conservation and that hyperbolicity must be studied for each direction $\boldsymbol{\omega}$ with $\Vert\boldsymbol{\omega}\Vert=1$.
Denote the primitive set of variables $\boldsymbol{q}=(m_1,m_2,m_1^d,\alpha_1^d,\Sigma, \overline{\alpha}_1,w_x,w_y,w_z,u_x,u_y,u_z)$.
We consider a smooth solution such that we look for a quasi-linear form 
\begin{equation}
    \partial_t \boldsymbol{q}
    + \mathbf{A}_x(\boldsymbol{q})\partial_x\boldsymbol{q}
    + \mathbf{A}_y(\boldsymbol{q})\partial_y\boldsymbol{q}
    + \mathbf{A}_z(\boldsymbol{q})\partial_z\boldsymbol{q}
    =\boldsymbol{0},
\end{equation}
with $\mathbf{A}_i$ are the Jacobian matrices in the direction $i$.
Denote $\boldsymbol{n}:=\boldsymbol{w}/\Vert\boldsymbol{w}\Vert$ and $\Delta p:=p_1-p_2$, then \eqref{eq:relaxed-system-homo-app} admits a linearized form with the matrices $\mathbf{A}_i$ given by \eqref{eq:hyp_matrices}.
\tiny
\begin{equation}
  \label{eq:hyp_matrices}
    \begin{aligned}
        &{\scriptstyle\mathbf{A}_x(\boldsymbol{q})=}
        {
        \left(
\begin{array}{cccccccccccc}
 u_x & 0 & 0 & 0 & 0 & 0 & 0 & 0 & 0 & m_1 & 0 & 0 \\
 0 & u_x & 0 & 0 & 0 & 0 & 0 & 0 & 0 & m_2 & 0 & 0 \\
 0 & 0 & u_x & 0 & 0 & 0 & 0 & 0 & 0 & m_1^d & 0 & 0 \\
 0 & 0 & 0 & u_x & 0 & 0 & 0 & 0 & 0 & \alpha_1^d & 0 & 0 \\
 0 & 0 & 0 & 0 & u_x & 0 & 0 & 0 & 0 & \Sigma  & 0 & 0 \\
 0 & 0 & 0 & 0 & 0 & u_x & 0 & 0 & 0 & 0 & 0 & 0 \\
 0 & 0 & 0 & 0 & 0 & 0 & u_x & 0 & 0 & w_1 & w_2 & w_3 \\
 0 & 0 & 0 & 0 & 0 & 0 & 0 & u_x & 0 & 0 & 0 & 0 \\
 0 & 0 & 0 & 0 & 0 & 0 & 0 & 0 & u_x & 0 & 0 & 0 \\
 \frac{c_1^2}{\rho \left(1-\alpha_1^d\right)} & \frac{c_2^2}{\rho \left(1-\alpha_1^d\right)} &
   0 & (c_F^d)^2 & 0 & \frac{c_2^2 \rho_2-c_1^2
   \rho_1+\Delta p}{\rho} & \frac{\sigma}{\rho}n_x(n_y^2+n_z^2) & -\frac{\sigma}{\rho}n_yn_x^2 & -\frac{\sigma}{\rho}n_zn_x^2 & u_x & 0 & 0 \\
 0 & 0 & 0 & 0 & 0 & 0 & \frac{\sigma}{\rho}n_y(n_y^2+n_z^2) & \frac{\sigma}{\rho}n_x(n_x^2+n_z^2)& -\frac{\sigma}{\rho}n_x
 n_y n_z & 0 & u_x & 0 \\
 0 & 0 & 0 & 0 & 0 & 0 & \frac{\sigma}{\rho}n_z(n_y^2+n_z^2) & -\frac{\sigma  n_x n_y n_z}{
   \rho} & \frac{\sigma}{\rho}n_x(n_x^2+n_y^2) & 0 & 0 & u_x \\
\end{array}
\right)}\\
        &{\scriptstyle\mathbf{A}_y(\boldsymbol{q})}=
        {
        \left(
\begin{array}{cccccccccccc}
 u_y & 0 & 0 & 0 & 0 & 0 & 0 & 0 & 0 & 0 & m_1 & 0 \\
 0 & u_y & 0 & 0 & 0 & 0 & 0 & 0 & 0 & 0 & m_2 & 0 \\
 0 & 0 & u_y & 0 & 0 & 0 & 0 & 0 & 0 & 0 & m_1^d & 0 \\
 0 & 0 & 0 & u_y & 0 & 0 & 0 & 0 & 0 & 0 & \alpha _1^d & 0 \\
 0 & 0 & 0 & 0 & u_y & 0 & 0 & 0 & 0 & 0 & \Sigma  & 0 \\
 0 & 0 & 0 & 0 & 0 & u_y & 0 & 0 & 0 & 0 & 0 & 0 \\
 0 & 0 & 0 & 0 & 0 & 0 & u_y & 0 & 0 & 0 & 0 & 0 \\
 0 & 0 & 0 & 0 & 0 & 0 & 0 & u_y & 0 & w_1 & w_2 & w_3 \\
 0 & 0 & 0 & 0 & 0 & 0 & 0 & 0 & u_y & 0 & 0 & 0 \\
 0 & 0 & 0 & 0 & 0 & 0 & \frac{\sigma}{\rho}n_y (n_y^2+n_z^2)&
   \frac{\sigma}{\rho}n_x (n_x^2+n_z^2) & -\frac{\sigma}{\rho}n_x n_y
   n_z & u_y & 0 & 0 \\
 \frac{c_1^2}{\rho \left(1-\alpha_1^d\right)} & \frac{c_2^2}{\rho \left(1-\alpha_1^d\right)}
   & 0 & (c_F^d)^2 & 0 & \frac{
    \rho_2c_2^2-\rho_1c_1^2+\Delta p}{\rho}& -\frac{\sigma}{\rho} n_xn_y^2& \frac{\sigma}{\rho}n_y
    (n_x^2+n_z^2) & -\frac{\sigma}{\rho}n_z n_y^2 & 0 & u_y & 0 \\
 0 & 0 & 0 & 0 & 0 & 0 & -\frac{\sigma}{\rho} n_x n_y n_z & \frac{\sigma}{\rho}n_z
(n_x^2+n_z^2) & \frac{\sigma}{\rho}n_y
 (n_x^2+n_y^2) & 0 & 0 & u_y \\
\end{array}
\right)}\\
        &{\scriptstyle\mathbf{A}_z(\boldsymbol{q})}=
        {
            \left(
\begin{array}{cccccccccccc}
 u_z & 0 & 0 & 0 & 0 & 0 & 0 & 0 & 0 & 0 & 0 & m_1 \\
 0 & u_z & 0 & 0 & 0 & 0 & 0 & 0 & 0 & 0 & 0 & m_2 \\
 0 & 0 & u_z & 0 & 0 & 0 & 0 & 0 & 0 & 0 & 0 & m_1^d \\
 0 & 0 & 0 & u_z & 0 & 0 & 0 & 0 & 0 & 0 & 0 & \alpha _1^d \\
 0 & 0 & 0 & 0 & u_z & 0 & 0 & 0 & 0 & 0 & 0 & \Sigma  \\
 0 & 0 & 0 & 0 & 0 & u_z & 0 & 0 & 0 & 0 & 0 & 0 \\
 0 & 0 & 0 & 0 & 0 & 0 & u_z & 0 & 0 & 0 & 0 & 0 \\
 0 & 0 & 0 & 0 & 0 & 0 & 0 & u_z & 0 & 0 & 0 & 0 \\
 0 & 0 & 0 & 0 & 0 & 0 & 0 & 0 & u_z & w_1 & w_2 & w_3 \\
 0 & 0 & 0 & 0 & 0 & 0 & \frac{\sigma}{\rho}n_z (n_y^2+n_z^2) &
   -\frac{\sigma}{\rho}n_x n_y n_z & \frac{\sigma}{\rho}n_x
   (n_x^2+n_y^2) & u_z & 0 & 0 \\
 0 & 0 & 0 & 0 & 0 & 0 & -\frac{\sigma}{\rho}n_x n_y n_z & \frac{\sigma}{\rho}n_z
 (n_x^2+n_z^2) & \frac{\sigma}{\rho}n_y
 (n_x^2+n_y^2) & 0 & u_z & 0 \\
 \frac{c_1^2}{\rho \left(1-\alpha_1^d\right)} & \frac{c_2^2}{\rho \left(1-\alpha_1^d\right)}
   & 0 & (c_F^d)^2 & 0 & \frac{
    \rho_2c_2^2-\rho_1c_1^2+\Delta p}{\rho} & -\frac{\sigma}{\rho}n_x
    n_z^2 & -\frac{\sigma}{\rho}n_y n_z^2 & \frac{\sigma}{\rho}n_z(n_x^2+n_y^2) & 0 & 0 & u_z \\
\end{array}
\right)
        }
    \end{aligned}
\end{equation}
\normalsize
As the system is not rotational invariant, consider then the direction $\boldsymbol{\omega}$ with $\Vert\boldsymbol{\omega}\Vert=1$.
Let us study then the eigenvalues of the Jacobian matrix $\mathbf{A}_{\boldsymbol{\omega}}:=\omega_x\mathbf{A}_x+\omega_y\mathbf{A}_y+\omega_z\mathbf{A}_z$ associated to this direction.
The characteristic polynomial $P_{\boldsymbol{\omega}}$ of $\mathbf{A}_{\boldsymbol{\omega}}$ reads
\begin{equation}
    \begin{aligned}
        P_{\boldsymbol{\omega}}(\lambda)
        =
        (\lambda-u_{\boldsymbol{\omega}})^8
        \left[(\lambda-u_{\boldsymbol{\omega}})^4
        + (\lambda-u_{\boldsymbol{\omega}})^2(- 
           (c_F^d)^2 - \frac{\sigma}{\rho} \Vert\boldsymbol{w}\Vert (1 - (\boldsymbol{\omega} \bcdot \boldsymbol{n})^2))
           + 
           (c_F^d)^2 \frac{\sigma}{\rho}  \Vert\boldsymbol{w}\Vert (1 - (\boldsymbol{\omega} \bcdot \boldsymbol{n})^2 )\right],
    \end{aligned}
\end{equation}
with $u_{\boldsymbol{\omega}}=\vel\bcdot\boldsymbol{\omega}$.
Denote $\We_{\boldsymbol{\omega}}:=\rho u_{\boldsymbol{\omega}}^2 /(\sigma\Vert\bnabla\overline{\alpha}_1\Vert)$ and $Ma_{\boldsymbol{\omega}} = u_{\boldsymbol{\omega}}/c_F^d$, the roots of $P_{\boldsymbol{\omega}}$ gives the following eigenvalues
\begin{gather}
    \label{eq:eigenvalues}
    \lambda_{1,2,3,4,5,6,7,8} = u_{\boldsymbol{\omega}},\\
    \lambda_{9,10} = 
    {
        \scriptstyle
        u_{\boldsymbol{\omega}} \pm c_F^d \sqrt{
        \frac{1}{2}\left[1+\frac{\Ma_{\boldsymbol{\omega}}^2}{\We_{\boldsymbol{\omega}}}(1-(\boldsymbol{\omega}\bcdot \boldsymbol{n})^2)\right]
        +
        \frac{1}{2}
        \sqrt{
            \left[1 - \frac{\Ma_{\boldsymbol{\omega}}^2}{\We_{\boldsymbol{\omega}}} (1 - (\boldsymbol{\omega}\bcdot \boldsymbol{n})^2)\right]^2 + 
            4 \frac{\Ma_{\boldsymbol{\omega}}^2}{\We_{\boldsymbol{\omega}}} (1 - (\boldsymbol{\omega}\bcdot \boldsymbol{n})^2) (\boldsymbol{\omega}\bcdot \boldsymbol{n})^2
        }
        }
    },\\
    \lambda_{11,12} = 
    {
        \scriptstyle
        u_{\boldsymbol{\omega}} \pm c_F^d \sqrt{
        \frac{1}{2}\left[1+\frac{\Ma_{\boldsymbol{\omega}}^2}{\We_{\boldsymbol{\omega}}}(1-(\boldsymbol{\omega}\bcdot \boldsymbol{n})^2)\right]
        -
        \frac{1}{2}
        \sqrt{
            \left[1 - \frac{\Ma_{\boldsymbol{\omega}}^2}{\We_{\boldsymbol{\omega}}} (1 - (\boldsymbol{\omega}\bcdot \boldsymbol{n})^2)\right]^2 + 
            4 \frac{\Ma_{\boldsymbol{\omega}}^2}{\We_{\boldsymbol{\omega}}} (1 - (\boldsymbol{\omega}\bcdot \boldsymbol{n})^2) (\boldsymbol{\omega}\bcdot \boldsymbol{n})^2
        }
        }.
    }
\end{gather}
As $u_{\boldsymbol{\omega}}$ is a multiple eigenvalue, we are particularly interested in whether there are as many independent eigenvectors associated to $u_{\boldsymbol{\omega}}$ as the degree of multiplicity which is here $8$.
Denoting $\boldsymbol{r}=(r_i)_{i=1,...,10}$, finding the eigenvectors of $u_{\boldsymbol{\omega}}$ are obtained by solving
\begin{equation}
    (\mathbf{A}_{\boldsymbol{\omega}}-u_{\boldsymbol{\omega}}\mathbf{I})\boldsymbol{r}=0.
\end{equation}
Using \cite{wolfram_research_mathematica_2023}, we obtain the following eigenvectors
\begin{equation}
    \begin{aligned}
        \boldsymbol{r}_1^T&=
        \Big(
            \begin{array}{ccc wc{65pt} cc *{3}{wc{41pt}} *{3}{wc{11pt}}}
            1 & 0 & 0 & -\frac{c_1^2}{\rho  (c_F^d)^2 (1-\alpha_1^d)} & 0 & 0 & 0 & 0 & 0 &0 & 0 & 0
            \end{array}
        \Big),\\
        \boldsymbol{r}_2^T&=
        \Big(
            \begin{array}{ccc wc{65pt} cc *{3}{wc{41pt}} *{3}{wc{11pt}}}
            0 & 1 & 0 & -\frac{c_2^2}{\rho  (c_F^d)^2 (1-\alpha_1^d)} & 0 & 0 & 0 & 0 & 0 & 0 & 0 & 0
            \end{array}
        \Big),\\
        \boldsymbol{r}_3^T&=
        \Big(
            \begin{array}{ccc wc{65pt} cc *{3}{wc{41pt}} *{3}{wc{11pt}}}
            0 & 0 & 1 & 0 & 0 & 0 & 0 & 0 & 0 & 0 & 0 & 0
            \end{array}
        \Big),\\
        \boldsymbol{r}_4^T&=
        \Big(
            \begin{array}{ccc wc{65pt} cc *{3}{wc{41pt}} *{3}{wc{11pt}}}
            0 & 0 & 0 & 0 & 1 & 0 & 0 & 0 & 0 & 0 & 0 & 0
            \end{array}
        \Big),\\
        \boldsymbol{r}_5^T&=
        \Big(
            \begin{array}{ccc wc{65pt} cc *{3}{wc{41pt}} *{3}{wc{11pt}}}
            0 & 0 & 0 & \frac{\rho_1c_1^2 -\rho_2c_2^2+p_2-p_1}{\rho(c_F^d)^2} & 0 & 1 & 0 & 0 & 0 & 0 & 0 & 0
            \end{array}
        \Big),\\
        \boldsymbol{r}_6^T&=
        \Big(
            \begin{array}{ccc wc{65pt} cc *{3}{wc{41pt}} *{3}{wc{11pt}}}
            0 & 0 & 0 & 0 & 0 & 0 & 0 & 0 & 0 & \multicolumn{3}{c}{\hspace{10pt}(\boldsymbol{n}\times\boldsymbol{\omega})^T}
            \end{array}
        \Big),\\
        \boldsymbol{r}_7^T&=
        \Big(
            \begin{array}{ccc wc{65pt} cc *{3}{wc{41pt}} *{3}{wc{11pt}}}
            0 & 0 & 0 & \frac{\sigma(1-(\boldsymbol{\omega}\bcdot\boldsymbol{n})^2)}{\rho (c_F^d)^2} & 0 & 0 & \multicolumn{3}{c}{\hspace{15pt}\left[(1-(\boldsymbol{\omega}\bcdot\boldsymbol{n})^2)\boldsymbol{n} - (\boldsymbol{\omega}\bcdot\boldsymbol{n})\boldsymbol{\omega}\right]^T\hspace{45pt}} & 0 & 0 & 0
            \end{array}
        \Big).
    \end{aligned}
\end{equation}
Remark that the eigenvectors are independent and span a subspace of dimension $7$ when $\boldsymbol{n}$ and $\boldsymbol{\omega}$ are not collinear, and a subspace of dimension $6$ when they are collinear as $\boldsymbol{r}_6=\boldsymbol{0}$.
In either case, the system \eqref{eq:relaxed-system} is weakly hyperbolic.
\section{Solution of the Riemann problem}
\label{app:riemann-problem}

We detail here the computational method to evaluate the flux at the interface between two cells, arbitrarily called "left" and "right" and denoted with the indexes L and R.
We consider the $x$-axis oriented in the direction of the interface.
For the considered Godunov method, we recall that the fluxes at the interface are evaluated using the solution $\bq$ of the Riemann problem
\begin{equation}
    \label{eq:riemann-problem-app}
    \begin{cases}
        \partial_t \bq + \bnabla\bcdot\boldsymbol{F}(\bq)=0,\\
        \bq(x,0) = 
        \begin{cases}
            \bq_L\quad\text{if}\: x<0,\\
            \bq_R\quad\text{if}\: x>0,\\
        \end{cases}
    \end{cases}
\end{equation}
with $\bq=(\rho\overline{\alpha}_1,\alpha_1\rho_1,\alpha_2\rho_2,\alpha_1^d\rho_1^d, \alpha_1^d,\rho u_x,\rho u_y)$.
Given the self-similar nature of the solution, we denote $\tilde{\bq}(x/t)=\bq(x,t)$ for $t>0$, and the interface flux is evaluated as $\boldsymbol{F}(\tilde{\bq}(0))$.
The solution of this Riemann problem with linearized barotropic EOS is an extension of the work proposed by \citep{chanteperdrix_compressible_2002}.
Indeed, the model presented in their work is recovered when $\alpha_1^d\rightarrow 0$, and the structure of the eigenvalues is the same with two truly non-linear waves of velocity $u_x\pm c_F^d$, and additional linearly degenerate fields to the material velocity $u_x$.

Given the structure of the eigenvalues and eigenvectors, the solution of this problem is self-similar with three waves denoted from left to right in the usual $x-t$ plane as the $1$-wave, the discontinuity wave, and the $3$-wave.
They separate the $x-t$ plane in four regions:
\begin{itemize}
    \item the left state $\bq_L$ at the left of the $1$-wave,
    \item the left star-state $\bq_L^*$ between the $1$-wave and the discontinuity wave,
    \item the left star-state $\bq_R^*$ between the discontinuity wave and the $3$-wave,
    \item the right state $\bq_R$ at the right of the $3$-wave.
\end{itemize}
From the Rankine-Hugoniot conditions, one can demonstrate that the normal velocity $u_x$ and the pressure $\overline{p}$ are constant across the discontinuity wave.
For either shocks or rarefaction waves, left and right states are both linked to their respective star regions of same velocity $u_x^*$ and $\overline{p}^*$.
We express that relation with functions $f_L$ and $f_R$ giving respectively the velocity of the star region from the left/right state and the pressure of the star region.
The common normal velocity within the star region gives
\begin{equation}
    f_L(p^*,\bq_L)=(u_x)^*=(u_x)_R^*=f_R(p^*,\bq_R).
\end{equation}
For concision purposes, only the main computational procedure along with the differences are highlighted here, and the reader is referred to their work for an exhaustive discussion.
We propose here to establish the expression of $f_L$ for the $1$-wave only, as the expression of $f_R$ is similarly obtained.

\subsection{Expression of $f_L$ for a $1$-shock}
Let us write the Rankine-Hugoniot conditions for a $1$-shock of velocity $s$,
\begin{equation}
    s\left(\bq_L^*-\bq_L\right)
    = \boldsymbol{F}^{hyp}(\bq_L^*)-\boldsymbol{F}^{hyp}(\bq_L).
\end{equation}
Such a shock is only valid if the Lax inequality $(u_x)_L>s>(u_x)^*$ holds.
We develop and reorganize this set of equation to obtain for $q\in(\rho\overline{\alpha}_1,\alpha_1\rho_1,\alpha_2\rho_2,\alpha_1^d\rho_1^d, \alpha_1^d,\rho u_y)$
\begin{equation}
    \label{eq:RH}
    \left\{
    \begin{array}{lcll}
        q_L^* &=& q_L & \frac{(u_x)_L-s}{(u_x)^*-s},\\
        (\rho u_x)_L^* &=& (\rho_L u_x)_L&\frac{(u_x)_L-s}{(u_x)^*-s}
        +\frac{\overline{p}_L-\overline{p}_L^*}{(u_x)^*-s}.
    \end{array}
    \right.
\end{equation}
From these equations, we particularly obtain that
\begin{equation}
    \label{eq:RH-extra-rel}
    (\overline{\alpha}_1)_L^* = (\overline{\alpha}_1)_L,
    \qquad
    \rho_L^* = \rho_L \frac{(u_x)_L-s}{(u_x)^*-s},
    \qquad
    s=(u_x)_L+\frac{\overline{p}_L-\overline{p}^*}{\rho_L\left[(u_x)_L-(u_x)^*\right]}.
\end{equation}
In order to get the expression of $f_L$ for a shock, we need to express $s$ as a function of $\overline{p}^*$ and $(u_x)^*$.
We do so by using the linearized barotropic EOS and the first relation of \eqref{eq:RH} in the last relation of \eqref{eq:RH-extra-rel} to express $\overline{p}^*$ with $s$ and $(u_x)^*$.
Then, isolating $s$ yields
\begin{equation}
    s=\frac{1-(\alpha_1^d)_L\frac{(u_x)_L}{(u_x)^*}}{1-(\alpha_1^d)_L}(u_x)^*+\rho_L (c_F^d)_L^2\frac{(u_x)_L-(u_x)^*}{\overline{p}_L-\overline{p}^*}.
\end{equation}
Using this relation with the last relation of \eqref{eq:RH-extra-rel} finally gives
\begin{equation}
    (u_x)^*=(u_x)_L
    -\sqrt{1-(\alpha_1^d)_L}
    \frac{
        \overline{p}^*-\overline{p}_L
    }{
        \sqrt{\rho_L( \overline{p}^*-\overline{p}_L+(1-(\alpha_1^d)_L)\rho_L (c_F^d)_L^2)}
    }
    =:f_L^{shock}(\overline{p}^*,\bq_L).
\end{equation}
According to the Lax inequality, this last relation is only valid for $\overline{p}^*>\overline{p}_L$. 
\subsection{Expression of $f_L$ for a $1$-rarefaction}
Consider now a rarefaction wave connecting the state $\bq_L$ and $\bq_L^*$.
From the Riemann invariants associated with $u_x-c_F^d$ for the barotropic linearized EOS,
\begin{equation}
    \overline{\alpha}_1,
    \quad
    \frac{\alpha_1\rho_1}{\alpha_2\rho_2},
    \quad
    \frac{\alpha_1^d\rho_1^d}{\alpha_1\rho_1},
    \quad
    \frac{\Sigma}{\alpha_1\rho_1},
    \quad
    \rho_1^d,
    \quad
    c_F^d(1-\alpha_1^d),
    \quad
    u_x+\frac{1}{2}c_F^d(1-\alpha_1^d)
    \log\left(\frac{(\alpha_1\rho_1)(\alpha_2\rho_2)}{(1-\alpha_1^d)^2\overline{\alpha}_1(1-\overline{\alpha}_18)\rho_{0,1}\rho_{0,2}}\right).
\end{equation}
As these invariants are equal in state $\bq_L$ and $\bq_L^*$, some calculations provide for $q\in(\rho\overline{\alpha}_1,\alpha_1\rho_1,\alpha_2\rho_2,\alpha_1^d\rho_1^d, \alpha_1^d,\rho u_y)$
\begin{equation}
    \label{eq:rel-invariants}
    \left\{
    \begin{array}{lcll}
        q_L^* &=&q_L &\frac{1-\alpha^1_{d,*}}{(1-\alpha^1_{d,g})}\exp\left(\frac{u_g-u^*}{c_g(1-\alpha^1_{d,g})}\right),\\
        (u_x)_L^* &=&(u_x)_L&+(c_F^d)_L(1-(\alpha_1^d)_L)\log\left(\frac{\rho_L (c_F^d)_L^2(1-(\alpha_1^d)_L)}{\overline{p}^*-\overline{p}_L+\rho_L (c_F^d)_L^2(1-(\alpha_1^d)_L)}\right)=:f_L^{raref}(\overline{p}^*,\bq_L),
    \end{array}
    \right.
\end{equation}
where the last relation defines the function $f_L$ for $\overline{p}^*<\overline{p}_L$ such that $(u_x)_L<(u_x)_L^*$
Remark that we start computing the state in the star region with the component $\alpha^1_{d,*}$ thanks to the first relation of \eqref{eq:rel-invariants} with $q=\alpha_1^d$ and a Newton-Raphson method.

We finally define the function $f_L$ with 
\begin{equation}
    f_L(\overline{p}^*,\bq_L)=
    \begin{cases}
        f_L^{raref}(\overline{p}^*,\bq_L)
        \quad
        \text{if}\:
        \overline{p}^*<\overline{p}_L,\\
        f_L^{shock}(\overline{p}^*,\bq_L)
        \quad
        \text{if}\:
        \overline{p}^*>\overline{p}_L.
    \end{cases}
\end{equation}

\subsection{Solution algorithm}
Given the definition of $f_L$ and assuming that we have obtained $f_R$ similarly, we obtain the solution of the Riemann problem \eqref{eq:riemann-problem-app} by proceeding as follows:
\begin{itemize}
    \item[(i)] Identifying the nature of the $1$-wave and $3$ wave by solving in $\overline{p}$ the invariance of velocity $(u_x)$ in the star region with a Newton-Raphson method,
    \begin{equation}
        f_L(\overline{p},\bq_L) - f_R(\overline{p},\bq_R) = 0.
    \end{equation}
    \item[(ii)] Identifying the region where the cell interface stationary wave $x/t=0$ belongs,
    \item[(iii)] Computing the state $\tilde{\bq}(0)$ and the flux $\boldsymbol{F}(\tilde{\bq}(0))$ with the set of relations \eqref{eq:RH} or \eqref{eq:rel-invariants}.
\end{itemize}
%

\bibliographystyle{abbrvnat}

\bibliography{biblio}

\begin{thebibliography}{87}
\providecommand{\natexlab}[1]{#1}
\expandafter\ifx\csname urlstyle\endcsname\relax
  \providecommand{\doi}[1]{\url{#1}}\else
  \providecommand{\doi}{\begingroup \Url}\fi

\bibitem[Baer and Nunziato(1986)Baer, M.~R. and Nunziato,
  J.~W.]{baer_two-phase_1986}
Baer, M.~R. and Nunziato, J.~W.
\newblock A two-phase mixture theory for the deflagration-to-detonation
  transition ({DDT}) in reactive granular materials.
\newblock \emph{International Journal of Multiphase Flow}, 12\penalty0
  (6):\penalty0 861--889, 1986.

\bibitem[Bedford(1985)Bedford, A.]{bedford_hamiltons_1985}
Bedford, A.
\newblock \emph{Hamilton's {Principle} in {Continuum} {Mechanics}}.
\newblock Pitman Publishing Ltd., 1985.
\newblock \doi{10.13140/2.1.1603.4887}.

\bibitem[Berdichevsky(2009)Berdichevsky, V.]{berdichevsky_variational_2009}
Berdichevsky, V.
\newblock \emph{Variational {Principles} of {Continuum} {Mechanics}: {I}.
  {Fundamentals}}, volume~5 of \emph{Interaction of {Mechanics} and
  {Mathematics}}.
\newblock Springer, 2009.
\newblock \doi{10.1007/978-3-540-88467-5}.

\bibitem[Bode et~al.(2014)Bode, M., Diewald, F., Broll, D.~O., Heyse, J.~F.,
  Le~Chenadec, V., and Pitsch, H.]{bode_influence_2014}
Bode, M., Diewald, F., Broll, D.~O., Heyse, J.~F., Le~Chenadec, V., and Pitsch,
  H.
\newblock Influence of the {Injector} {Geometry} on {Primary} {Breakup} in
  {Diesel} {Injector} {Systems}.
\newblock pp. 2014--01--1427, 2014.
\newblock \doi{10.4271/2014-01-1427}.

\bibitem[Bonometti and Magnaudet(2007)Bonometti, T. and Magnaudet,
  J.]{bonometti_interface-capturing_2007}
Bonometti, T. and Magnaudet, J.
\newblock An interface-capturing method for incompressible two-phase flows.
  {Validation} and application to bubble dynamics.
\newblock \emph{International Journal of Multiphase Flow}, 33\penalty0
  (2):\penalty0 109--133, 2007.
\newblock \doi{10.1016/j.ijmultiphaseflow.2006.07.003}.

\bibitem[Brackbill et~al.(1992)Brackbill, J., Kothe, D., and Zemach,
  C.]{brackbill_continuum_1992}
Brackbill, J., Kothe, D., and Zemach, C.
\newblock A continuum method for modeling surface tension.
\newblock \emph{Journal of Computational Physics}, 100\penalty0 (2):\penalty0
  335--354, 1992.
\newblock \doi{10.1016/0021-9991(92)90240-Y}.

\bibitem[Bueno and Gomez(2016)Bueno, J. and Gomez, H.]{bueno_liquid-vapor_2016}
Bueno, J. and Gomez, H.
\newblock Liquid-vapor transformations with surfactants. {Phase}-field model
  and {Isogeometric} {Analysis}.
\newblock \emph{Journal of Computational Physics}, 321:\penalty0 797--818,
  2016.
\newblock \doi{10.1016/j.jcp.2016.06.008}.
\newblock Publisher: Elsevier BV.

\bibitem[Burtea et~al.(2021)Burtea, C., Gavrilyuk, S., and Perrin,
  C.]{burtea_hamiltons_2021}
Burtea, C., Gavrilyuk, S., and Perrin, C.
\newblock Hamilton's principle of stationary action in multiphase flow
  modeling.
\newblock 2021.

\bibitem[Cahn and Hilliard(1958)Cahn, J.~W. and Hilliard,
  J.~E.]{cahn_free_1958}
Cahn, J.~W. and Hilliard, J.~E.
\newblock Free {Energy} of a {Nonuniform} {System}. {I}. {Interfacial} {Free}
  {Energy}.
\newblock \emph{The Journal of Chemical Physics}, 28\penalty0 (2):\penalty0
  258--267, 1958.
\newblock \doi{10.1063/1.1744102}.

\bibitem[Caro et~al.(2005)Caro, F., Coquel, F., Jamet, D., and Kokh,
  S.]{caro_dinmod_2005}
Caro, F., Coquel, F., Jamet, D., and Kokh, S.
\newblock {DINMOD}: {A} diffuse interface model for two-phase flows modelling.
\newblock In \emph{{IRMA} {Lectures} in {Mathematics} and {Theoretical}
  {Physics}}, pp. 209--237. EMS Press, 2005.
\newblock \doi{10.4171/012-1/10}.

\bibitem[Chanteperdrix(2004)Chanteperdrix, G.]{chanteperdrix_modelisation_2004}
Chanteperdrix, G.
\newblock \emph{Modélisation et simulation numérique d’écoulements
  diphasiques à interface libre. {Application} à l’étude des mouvements de
  liquides dans les réservoirs de véhicules spatiaux.}
\newblock PhD thesis, 2004.

\bibitem[Chanteperdrix et~al.(2002)Chanteperdrix, G., Villedieu, P., and Vila,
  J.-P.]{chanteperdrix_compressible_2002}
Chanteperdrix, G., Villedieu, P., and Vila, J.-P.
\newblock A {Compressible} {Model} for {Separated} {Two}-{Phase} {Flows}
  {Computations}.
\newblock In \emph{{FEDSM2002}}, pp. 809--816, Volume 1: Fora, Parts A and B,
  2002.
\newblock \doi{10.1115/FEDSM2002-31141}.

\bibitem[Coquel and LeFloch(1996)Coquel, F. and LeFloch,
  P.~G.]{coquel_entropy_1996}
Coquel, F. and LeFloch, P.~G.
\newblock An entropy satisfying {MUSCL} scheme for systems of conservation
  laws.
\newblock \emph{Numerische Mathematik}, 74\penalty0 (1):\penalty0 1--33, 1996.
\newblock \doi{10.1007/s002110050205}.

\bibitem[Cordesse(2020)Cordesse, P.]{cordesse_contribution_2020}
Cordesse, P.
\newblock \emph{Contribution to the study of combustion instabilities in
  cryotechnic rocket engines: coupling diffuse interface models with
  kinetic-based moment methods for primary atomization simulations}.
\newblock PhD thesis, École polytechnique, 2020.

\bibitem[Cordesse et~al.(2019)Cordesse, P., Kokh, S., Battista, R.~D., and
  Massot, M.]{cordesse_derivation_2019}
Cordesse, P., Kokh, S., Battista, R.~D., and Massot, M.
\newblock Derivation of a two-phase flow model with two-scale kinematics and
  surface tension by means of variational calculus.
\newblock p.~7, Rio de Janeiro, Brazil, 2019.

\bibitem[Cordesse et~al.(2020)Cordesse, P., Di~Battista, R., Chevalier, Q.,
  Matuszewski, L., Ménard, T., Kokh, S., and Massot,
  M.]{cordesse_diffuse_2020}
Cordesse, P., Di~Battista, R., Chevalier, Q., Matuszewski, L., Ménard, T.,
  Kokh, S., and Massot, M.
\newblock A diffuse interface approach for disperse two-phase flows involving
  dual-scale kinematics of droplet deformation based on geometrical variables.
\newblock \emph{ESAIM: Proceedings and Surveys}, 69:\penalty0 24--46, 2020.
\newblock \doi{10.1051/proc/202069024}.

\bibitem[Desjardins et~al.(2008)Desjardins, O., Moureau, V., and Pitsch,
  H.]{desjardins_accurate_2008}
Desjardins, O., Moureau, V., and Pitsch, H.
\newblock An accurate conservative level set/ghost fluid method for simulating
  turbulent atomization.
\newblock \emph{Journal of Computational Physics}, 227\penalty0 (18):\penalty0
  8395--8416, 2008.
\newblock \doi{10.1016/j.jcp.2008.05.027}.

\bibitem[Devassy et~al.(2015)Devassy, B.~M., Habchi, C., and Daniel,
  E.]{devassy_atomization_2015}
Devassy, B.~M., Habchi, C., and Daniel, E.
\newblock Atomization {Modelling} of {Liquid} {Jets} using a {Two}-{Surface}
  {Density} {Approach}.
\newblock \emph{Atomization and Sprays}, 25\penalty0 (1):\penalty0 47--80,
  2015.

\bibitem[Di~Battista(2021)Di~Battista, R.]{di_battista_towards_2021}
Di~Battista, R.
\newblock \emph{Towards a unified eulerian modeling framework for two-phase
  flow: geometrical subscale phenomena and associated highly-scalable numerical
  methods}.
\newblock PhD thesis, Institut Polytechnique de Paris, 2021.

\bibitem[Drew(1983)Drew, D.~A.]{drew_mathematical_1983}
Drew, D.~A.
\newblock Mathematical {Modeling} of {Two}-{Phase} {Flow}.
\newblock \emph{Annual Review of Fluid Mechanics}, p.~31, 1983.

\bibitem[Drui et~al.(2019)Drui, F., Larat, A., Kokh, S., and Massot,
  M.]{drui_small-scale_2019}
Drui, F., Larat, A., Kokh, S., and Massot, M.
\newblock Small-scale kinematics of two-phase flows: identifying relaxation
  processes in separated- and disperse-phase flow models.
\newblock \emph{Journal of Fluid Mechanics}, 876:\penalty0 326--355, 2019.
\newblock \doi{10.1017/jfm.2019.538}.

\bibitem[Dunn(1986)Dunn, J.~E.]{dunn_interstitial_1986}
Dunn, J.~E.
\newblock Interstitial {Working} and a {Nonclassical} {Continuum}
  {Thermodynamics}.
\newblock In \emph{New {Perspectives} in {Thermodynamics}}, pp. 187--222.
  Springer Berlin Heidelberg, 1986.
\newblock \doi{10.1007/978-3-642-70803-9_11}.

\bibitem[Dunn and Serrin(1986)Dunn, J.~E. and Serrin,
  J.]{dunn_thermomechanics_1986}
Dunn, J.~E. and Serrin, J.
\newblock On the {Thermomechanics} of {Interstitial} {Working}.
\newblock In \emph{The {Breadth} and {Depth} of {Continuum} {Mechanics}}, pp.
  705--743. Springer Berlin Heidelberg, 1986.
\newblock \doi{10.1007/978-3-642-61634-1_33}.

\bibitem[Essadki(2016)Essadki, M.]{essadki_new_2016}
Essadki, M.
\newblock A new high order moment method for polydisperse evaporating sprays
  dedicated to the coupling with separated two-phase flows in automotive
  engine.
\newblock Florence, 2016.

\bibitem[Essadki et~al.(2016)Essadki, M., de~Chaisemartin, S., Massot, M.,
  Laurent, F., Larat, A., and Jay, S.]{essadki_adaptive_2016}
Essadki, M., de~Chaisemartin, S., Massot, M., Laurent, F., Larat, A., and Jay,
  S.
\newblock Adaptive {Mesh} {Refinement} and {High} {Order} {Geometrical}
  {Moment} {Method} for the {Simulation} of {Polydisperse} {Evaporating}
  {Sprays}.
\newblock \emph{Oil Gas Sci. Technol. – Rev. IFP Energies nouvelles},
  71\penalty0 (5), 2016.
\newblock \doi{10.2516/ogst/2016012}.

\bibitem[Essadki et~al.(2018)Essadki, M., de~Chaisemartin, S., Laurent, F., and
  Massot, M.]{essadki_high_2018}
Essadki, M., de~Chaisemartin, S., Laurent, F., and Massot, M.
\newblock High {Order} {Moment} {Model} for {Polydisperse} {Evaporating}
  {Sprays} towards {Interfacial} {Geometry} {Description}.
\newblock \emph{SIAM Journal on Applied Mathematics}, 78\penalty0 (4):\penalty0
  2003--2027, 2018.
\newblock \doi{10.1137/16M1108364}.

\bibitem[Fiorina et~al.(2016)Fiorina, B., Vie, A., Franzelli, B., Darabiha, N.,
  Massot, M., Dayma, G., Dagaut, P., Moureau, V., Vervisch, L., Berlemont, A.,
  Sabelnikov, V., Riber, E., and Cuenot, B.]{fiorina_modeling_2016}
Fiorina, B., Vie, A., Franzelli, B., Darabiha, N., Massot, M., Dayma, G.,
  Dagaut, P., Moureau, V., Vervisch, L., Berlemont, A., Sabelnikov, V., Riber,
  E., and Cuenot, B.
\newblock Modeling {Challenges} in {Computing} {Aeronautical} {Combustion}
  {Chambers}.
\newblock \emph{AerospaceLab Journal}, Issue 11:\penalty0 19 pages, 2016.
\newblock \doi{10.12762/2016.AL11-05}.
\newblock Artwork Size: 19 pages Medium: PDF Publisher: ONERA.

\bibitem[Fox and Marchisio(2007)Fox, R.~O. and Marchisio,
  D.~L.]{fox_multiphase_2007}
\emph{Multiphase reacting flows: modelling and simulation}.
\newblock Number no. 492 in Courses and lectures. Springer, Wien ; New York,
  2007.
\newblock OCLC: ocn145453832.

\bibitem[Gavrilyuk and Gouin(1999)Gavrilyuk, S. and Gouin,
  H.]{gavrilyuk_new_1999}
Gavrilyuk, S. and Gouin, H.
\newblock A new form of governing equations of fluids arising from {Hamilton}'s
  principle.
\newblock \emph{International Journal of Engineering Science}, p.~26, 1999.

\bibitem[Gavrilyuk and Saurel(2002)Gavrilyuk, S. and Saurel,
  R.]{gavrilyuk_mathematical_2002}
Gavrilyuk, S. and Saurel, R.
\newblock Mathematical and {Numerical} {Modeling} of {Two}-{Phase}
  {Compressible} {Flows} with {Micro}-{Inertia}.
\newblock \emph{Journal of Computational Physics}, 175\penalty0 (1):\penalty0
  326--360, 2002.
\newblock \doi{10.1006/jcph.2001.6951}.

\bibitem[Gavrilyuk et~al.(1998)Gavrilyuk, S., Gouin, H., and Perepechko,
  Y.~V.]{gavrilyuk_hyperbolic_1998}
Gavrilyuk, S., Gouin, H., and Perepechko, Y.~V.
\newblock Hyperbolic {Models} of {Homogeneous} {Two}-{Fluid} {Mixtures}.
\newblock \emph{Meccanica}, 33\penalty0 (2):\penalty0 161--175, 1998.
\newblock \doi{10.1023/A:1004354528016}.

\bibitem[Giovangigli(2021)Giovangigli, V.]{giovangigli_kinetic_2021}
Giovangigli, V.
\newblock Kinetic derivation of {Cahn}-{Hilliard} fluid models.
\newblock \emph{Physical Review E}, 104\penalty0 (5):\penalty0 054109, 2021.
\newblock \doi{10.1103/PhysRevE.104.054109}.

\bibitem[Giovangigli et~al.(2022)Giovangigli, V., Le~Calvez, Y., and Nabet,
  F.]{giovangigli_symmetrization_2022}
Giovangigli, V., Le~Calvez, Y., and Nabet, F.
\newblock Symmetrization and local existence of strong solutions for diffuse
  interface fluid models.
\newblock 2022.

\bibitem[Godlewski and Raviart(1991)Godlewski, E. and Raviart,
  P.-A.]{godlewski_hyperbolic_1991}
Godlewski, E. and Raviart, P.-A.
\newblock \emph{Hyperbolic {Systems} {Of} {Conservation} {Laws}}.
\newblock Ellipses, 1991.

\bibitem[Godunov and Bohachevsky(1959)Godunov, S.~K. and Bohachevsky,
  I.]{godunov_finite_1959}
Godunov, S.~K. and Bohachevsky, I.
\newblock Finite difference method for numerical computation of discontinuous
  solutions of the equations of fluid dynamics.
\newblock \emph{Matematičeskij sbornik}, 47(89)\penalty0 (3):\penalty0
  271--306, 1959.
\newblock Publisher: Steklov Mathematical Institute of Russian Academy of
  Sciences.

\bibitem[Goldman(2005)Goldman, R.]{goldman_curvature_2005}
Goldman, R.
\newblock Curvature formulas for implicit curves and surfaces.
\newblock \emph{Computer Aided Geometric Design}, 22\penalty0 (7):\penalty0
  632--658, 2005.
\newblock \doi{10.1016/j.cagd.2005.06.005}.

\bibitem[Gouin(1996)Gouin, H.]{gouin_second_1996}
Gouin, H.
\newblock The second gradient theory applied to interfaces: {Models} of
  continuum mechanics for fluid interfaces.
\newblock In \emph{Dynamics of {Multiphase} {Flows} {Across} {Interfaces}}, pp.
  8--13. Springer Berlin Heidelberg, 1996.
\newblock \doi{10.1007/bfb0102656}.

\bibitem[Gouin(2020)Gouin, H.]{gouin_introduction_2020}
Gouin, H.
\newblock \emph{Introduction to {Mathematical} {Methods} of {Analytical}
  {Mechanics}}.
\newblock ISTE Press/Elsevier, 2020.

\bibitem[Gouin and Gavrilyuk(1999)Gouin, H. and Gavrilyuk,
  S.]{gouin_hamiltons_1999}
Gouin, H. and Gavrilyuk, S.
\newblock Hamilton’s {Principle} and {Rankine}-{Hugoniot} {Conditions} for
  {General} {Motions} of {Mixtures}.
\newblock \emph{Meccanica}, 34\penalty0 (1):\penalty0 39--47, 1999.
\newblock \doi{10.1023/A:1004370127958}.

\bibitem[Gouin and Ruggeri(2009)Gouin, H. and Ruggeri, T.]{gouin_hamilton_2009}
Gouin, H. and Ruggeri, T.
\newblock The {Hamilton} {Principle} for {Fluid} {Binary} {Mixtures} with two
  {Temperatures}.
\newblock \emph{Bollettino dell'Unione Matematica Italiana}, 2\penalty0
  (2):\penalty0 403--422, 2009.
\newblock Publisher: Unione Matematica Italiana.

\bibitem[Grenier et~al.(2013)Grenier, N., Vila, J.-P., and Villedieu,
  P.]{grenier_accurate_2013}
Grenier, N., Vila, J.-P., and Villedieu, P.
\newblock An accurate low-{Mach} scheme for a compressible two-fluid model
  applied to free-surface flows.
\newblock \emph{Journal of Computational Physics}, 252:\penalty0 1--19, 2013.
\newblock \doi{10.1016/j.jcp.2013.06.008}.

\bibitem[Gueyffier et~al.(1999)Gueyffier, D., Li, J., Nadim, A., Scardovelli,
  R., and Zaleski, S.]{gueyffier_volume--fluid_1999}
Gueyffier, D., Li, J., Nadim, A., Scardovelli, R., and Zaleski, S.
\newblock Volume-of-{Fluid} {Interface} {Tracking} with {Smoothed} {Surface}
  {Stress} {Methods} for {Three}-{Dimensional} {Flows}.
\newblock \emph{Journal of Computational Physics}, 152\penalty0 (2):\penalty0
  423--456, 1999.
\newblock \doi{10.1006/jcph.1998.6168}.

\bibitem[Herivel(1955)Herivel, J.~W.]{herivel_derivation_1955}
Herivel, J.~W.
\newblock The derivation of the equations of motion of an ideal fluid by
  {Hamilton}'s principle.
\newblock \emph{Mathematical Proceedings of the Cambridge Philosophical
  Society}, 51\penalty0 (2):\penalty0 344--349, 1955.
\newblock \doi{10.1017/s0305004100030267}.
\newblock Publisher: Cambridge University Press (CUP).

\bibitem[Herrmann(2009)Herrmann, M.]{herrmann_detailed_2009}
Herrmann, M.
\newblock Detailed simulations of the breakup processes of turbulent liquid
  jets in subsonic crossflows.
\newblock In \emph{11th {International} {Annual} {Conference} on {Liquid}
  {Atomization} and {Spray} {Systems} 2009, {ICLASS} 2009}, 11th
  {International} {Annual} {Conference} on {Liquid} {Atomization} and {Spray}
  {Systems} 2009, {ICLASS} 2009. ILASS Americas/Professor Scott Samuelsen UCI
  Combustion Laboratory University of California Irvine, CA 92697-3550, 2009.

\bibitem[Herrmann(2010)Herrmann, M.]{herrmann_parallel_2010}
Herrmann, M.
\newblock A parallel {Eulerian} interface tracking/{Lagrangian} point particle
  multi-scale coupling procedure.
\newblock \emph{Journal of Computational Physics}, 229\penalty0 (3):\penalty0
  745--759, 2010.
\newblock \doi{10.1016/j.jcp.2009.10.009}.

\bibitem[Hoarau et~al.(2023)Hoarau, J.-C., Dorey, L.-H., Zuzio, D., Granger,
  F., and Estivalezes, J.-L.]{hoarau_direct_2023}
Hoarau, J.-C., Dorey, L.-H., Zuzio, D., Granger, F., and Estivalezes, J.-L.
\newblock Direct numerical simulation of a subcritical coaxial injection in
  fiber regime using sharp interface reconstruction.
\newblock 2023.

\bibitem[Ishii and Hibiki(1975)Ishii, M. and Hibiki,
  T.]{ishii_thermo-fluid_1975}
Ishii, M. and Hibiki, T.
\newblock \emph{Thermo-fluid dynamic theory of two-phase flow}.
\newblock Eyrolles, France, 1975.

\bibitem[Jacqmin(1999)Jacqmin, D.]{jacqmin_calculation_1999}
Jacqmin, D.
\newblock Calculation of {Two}-{Phase} {Navier}–{Stokes} {Flows} {Using}
  {Phase}-{Field} {Modeling}.
\newblock \emph{Journal of Computational Physics}, 155\penalty0 (1):\penalty0
  96--127, 1999.
\newblock \doi{10.1006/jcph.1999.6332}.
\newblock Publisher: Elsevier BV.

\bibitem[Jameson(1991)Jameson, A.]{jameson_time_1991}
Jameson, A.
\newblock Time dependent calculations using multigrid, with applications to
  unsteady flows past airfoils and wings.
\newblock In \emph{10th {Computational} {Fluid} {Dynamics} {Conference}}. 1991.
\newblock \doi{10.2514/6.1991-1596}.
\newblock \_eprint: https://arc.aiaa.org/doi/pdf/10.2514/6.1991-1596.

\bibitem[Jamet et~al.(2001)Jamet, D., Lebaigue, O., Coutris, N., and Delhaye,
  J.]{jamet_second_2001}
Jamet, D., Lebaigue, O., Coutris, N., and Delhaye, J.
\newblock The {Second} {Gradient} {Method} for the {Direct} {Numerical}
  {Simulation} of {Liquid}–{Vapor} {Flows} with {Phase} {Change}.
\newblock \emph{Journal of Computational Physics}, 169\penalty0 (2):\penalty0
  624--651, 2001.
\newblock \doi{10.1006/jcph.2000.6692}.

\bibitem[Janodet et~al.(2022)Janodet, R., Guillamón, C., Moureau, V., Mercier,
  R., Lartigue, G., Bénard, P., Ménard, T., and Berlemont,
  A.]{janodet_massively_2022}
Janodet, R., Guillamón, C., Moureau, V., Mercier, R., Lartigue, G., Bénard,
  P., Ménard, T., and Berlemont, A.
\newblock A massively parallel accurate conservative level set algorithm for
  simulating turbulent atomization on adaptive unstructured grids.
\newblock \emph{Journal of Computational Physics}, 458:\penalty0 111075, 2022.
\newblock \doi{10.1016/j.jcp.2022.111075}.

\bibitem[Korteweg(1901)Korteweg, D.~J.]{korteweg_sur_1901}
Korteweg, D.~J.
\newblock Sur la forme que prennent les équations du mouvements des fluides si
  l'on tient compte des forces capillaires causées par des variations de
  densité considérables mais connues et sur la théorie de la capillarité
  dans l'hypothèse d'une variation continue de la densité.
\newblock \emph{Archives Néerlandaises des Sciences exactes et naturelles},
  6:\penalty0 1--24, 1901.

\bibitem[Lafaurie et~al.(1994)Lafaurie, B., Nardone, C., Scardovelli, R.,
  Zaleski, S., and Zanetti, G.]{lafaurie_modelling_1994}
Lafaurie, B., Nardone, C., Scardovelli, R., Zaleski, S., and Zanetti, G.
\newblock Modelling {Merging} and {Fragmentation} in {Multiphase} {Flows} with
  {SURFER}.
\newblock \emph{Journal of Computational Physics}, 113\penalty0 (1):\penalty0
  134--147, 1994.
\newblock \doi{10.1006/jcph.1994.1123}.

\bibitem[Laurent and Massot(2001)Laurent, F. and Massot,
  M.]{laurent_multi-fluid_2001}
Laurent, F. and Massot, M.
\newblock Multi-fluid modelling of laminar polydisperse spray flames: origin,
  assumptions and comparison of sectional and sampling methods.
\newblock \emph{Combustion Theory and Modelling}, 5\penalty0 (4):\penalty0
  537--572, 2001.
\newblock \doi{10.1088/1364-7830/5/4/303}.

\bibitem[Le~Martelot et~al.(2014)Le~Martelot, S., Saurel, R., and Nkonga,
  B.]{le_martelot_towards_2014}
Le~Martelot, S., Saurel, R., and Nkonga, B.
\newblock Towards the direct numerical simulation of nucleate boiling flows.
\newblock \emph{International Journal of Multiphase Flow}, 66:\penalty0 62--78,
  2014.
\newblock \doi{10.1016/j.ijmultiphaseflow.2014.06.010}.

\bibitem[Le~Touze et~al.(2020)Le~Touze, C., Dorey, L.-H., Rutard, N., and
  Murrone, A.]{le_touze_compressible_2020}
Le~Touze, C., Dorey, L.-H., Rutard, N., and Murrone, A.
\newblock A compressible two-phase flow framework for {Large} {Eddy}
  {Simulations} of liquid-propellant rocket engines.
\newblock \emph{Applied Mathematical Modelling}, 84:\penalty0 265--286, 2020.
\newblock \doi{10.1016/j.apm.2020.03.028}.

\bibitem[Lebas et~al.(2009)Lebas, R., Menard, T., Beau, P.-A., Berlemont, A.,
  and Demoulin, F.-X.]{lebas_numerical_2009}
Lebas, R., Menard, T., Beau, P.-A., Berlemont, A., and Demoulin, F.-X.
\newblock Numerical simulation of primary break-up and atomization: {DNS} and
  modelling study.
\newblock \emph{International Journal of Multiphase Flow}, 35\penalty0
  (3):\penalty0 247--260, 2009.
\newblock \doi{10.1016/j.ijmultiphaseflow.2008.11.005}.

\bibitem[LeVeque and Leveque(1992)LeVeque, R.~J. and Leveque,
  R.~J.]{leveque_numerical_1992}
LeVeque, R.~J. and Leveque, R.~J.
\newblock \emph{Numerical methods for conservation laws}, volume 214.
\newblock Springer, 1992.

\bibitem[Lhuillier(2004)Lhuillier, D.]{lhuillier_evolution_2004}
Lhuillier, D.
\newblock Evolution of the volumetric interfacial area in two-phase mixtures.
\newblock \emph{Comptes Rendus Mécanique}, 332\penalty0 (2):\penalty0
  103--108, 2004.
\newblock \doi{10.1016/j.crme.2003.12.004}.

\bibitem[Ling et~al.(2017)Ling, Y., Fuster, D., Zaleski, S., and Tryggvason,
  G.]{ling_spray_2017}
Ling, Y., Fuster, D., Zaleski, S., and Tryggvason, G.
\newblock Spray formation in a quasiplanar gas-liquid mixing layer at moderate
  density ratios: {A} numerical closeup.
\newblock \emph{Physical Review Fluids}, 2\penalty0 (1):\penalty0 014005, 2017.
\newblock \doi{10.1103/PhysRevFluids.2.014005}.

\bibitem[Loison et~al.(2023)Loison, A., Pichard, T., Kokh, S., and Massot,
  M.]{loison_two-scale_2023}
Loison, A., Pichard, T., Kokh, S., and Massot, M.
\newblock Two-scale modelling of two-phase flows based on the {Stationary}
  {Action} {Principle} and a {Geometric} {Method} {Of} {Moments}, 2023.

\bibitem[Massot(2007)Massot, M.]{massot_eulerian_2007}
Massot, M.
\newblock Eulerian {Multi}-{Fluid} {Models} for {Polydisperse} {Evaporating}
  {Sprays}.
\newblock In , \emph{Multiphase {Reacting} {Flows}: {Modelling} and
  {Simulation}}, pp. 79--123. Springer Vienna, Vienna, 2007.
\newblock \doi{10.1007/978-3-211-72464-4_3}.

\bibitem[Massot et~al.(1998)Massot, M., Kumar, M., Gomez, A., and Smooke,
  M.]{massot_counterflow_1998}
Massot, M., Kumar, M., Gomez, A., and Smooke, M.
\newblock Counterflow spray diffusion flames of heptane: computations and
  experiments.
\newblock \emph{In Proceedings of the 27th Symposium International on
  Combustion, The Comb. Institute}, pp. 1975--1983, 1998.

\bibitem[Pelanti(2022)Pelanti, M.]{pelanti_arbitrary-rate_2022}
Pelanti, M.
\newblock Arbitrary-rate relaxation techniques for the numerical modeling of
  compressible two-phase flows with heat and mass transfer.
\newblock \emph{International Journal of Multiphase Flow}, 153:\penalty0
  104097, 2022.
\newblock \doi{10.1016/j.ijmultiphaseflow.2022.104097}.

\bibitem[Perigaud and Saurel(2005)Perigaud, G. and Saurel,
  R.]{perigaud_compressible_2005}
Perigaud, G. and Saurel, R.
\newblock A compressible flow model with capillary effects.
\newblock \emph{Journal of Computational Physics}, 209\penalty0 (1):\penalty0
  139--178, 2005.
\newblock \doi{10.1016/j.jcp.2005.03.018}.

\bibitem[Providakis et~al.(2012)Providakis, T., Zimmer, L., Scouflaire, P., and
  Ducruix, S.]{providakis_characterization_2012}
Providakis, T., Zimmer, L., Scouflaire, P., and Ducruix, S.
\newblock Characterization of the {Acoustic} {Interactions} in a {Two}-{Stage}
  {Multi}-{Injection} {Combustor} {Fed} {With} {Liquid} {Fuel}.
\newblock \emph{Journal of Engineering for Gas Turbines and Power},
  134\penalty0 (11):\penalty0 111503, 2012.
\newblock \doi{10.1115/1.4007200}.

\bibitem[Raviart and Sainsaulieu(1995)Raviart, P.-A. and Sainsaulieu,
  L.]{raviart_non-conservative_1995}
Raviart, P.-A. and Sainsaulieu, L.
\newblock A {Non}-{Conservative} hyperbolic system modeling spray dynamics.
  {Part} {I}. {Solution} of the {Riemann} {Problem}.
\newblock \emph{Mathematical Models and Methods in Applied Sciences},
  5\penalty0 (3):\penalty0 297--333, 1995.

\bibitem[Reitz and Bracco(1979)Reitz, R.~D. and Bracco,
  F.~B.]{reitz_dependence_1979}
Reitz, R.~D. and Bracco, F.~B.
\newblock On the {Dependence} of {Spray} {Angle} and {Other} {Spray}
  {Parameters} on {Nozzle} {Design} and {Operating} {Conditions}.
\newblock p. 790494, 1979.
\newblock \doi{10.4271/790494}.

\bibitem[Sakano et~al.(2022)Sakano, Y., Nambu, T., Mizobuchi, Y., and Sato,
  T.]{sakano_evaluation_2022}
Sakano, Y., Nambu, T., Mizobuchi, Y., and Sato, T.
\newblock Evaluation of three-dimensional droplet shape for analysis of the
  crossflow-type atomization.
\newblock \emph{Mechanical Engineering Journal}, 9\penalty0 (1):\penalty0
  21--00378--21--00378, 2022.
\newblock \doi{10.1299/mej.21-00378}.

\bibitem[Salmon(1983)Salmon, R.]{salmon_practical_1983}
Salmon, R.
\newblock Practical use of {Hamilton}'s principle.
\newblock \emph{Journal of Fluid Mechanics}, 132:\penalty0 431--444, 1983.
\newblock \doi{10.1017/S0022112083001706}.
\newblock Publisher: Cambridge University Press.

\bibitem[Saurel et~al.(2017)Saurel, R., Chinnayya, A., and Carmouze,
  Q.]{saurel_modelling_2017}
Saurel, R., Chinnayya, A., and Carmouze, Q.
\newblock Modelling compressible dense and dilute two-phase flows.
\newblock \emph{Physics of Fluids}, 29\penalty0 (6):\penalty0 063301, 2017.
\newblock \doi{10.1063/1.4985289}.

\bibitem[Schmidmayer et~al.(2017)Schmidmayer, K., Petitpas, F., Daniel, E.,
  Favrie, N., and Gavrilyuk, S.]{schmidmayer_model_2017}
Schmidmayer, K., Petitpas, F., Daniel, E., Favrie, N., and Gavrilyuk, S.
\newblock A model and numerical method for compressible flows with capillary
  effects.
\newblock \emph{Journal of Computational Physics}, 334:\penalty0 468--496,
  2017.
\newblock \doi{10.1016/j.jcp.2017.01.001}.

\bibitem[Seppecher(2002)Seppecher, P.]{seppecher_second-gradient_2002}
Seppecher, P.
\newblock Second-gradient theory: {Application} to {Cahn}-{Hilliard} fluids.
\newblock In \emph{Solid {Mechanics} and {Its} {Applications}}, pp. 379--388.
  Kluwer Academic Publishers, 2002.
\newblock \doi{10.1007/0-306-46946-4_29}.

\bibitem[Serre(2010)Serre, D.]{serre_structure_2010}
Serre, D.
\newblock The structure of dissipative viscous system of conservation laws.
\newblock \emph{Physica D: Nonlinear Phenomena}, 239\penalty0 (15):\penalty0
  1381--1386, 2010.
\newblock \doi{10.1016/j.physd.2009.03.014}.

\bibitem[Serrin(1959)Serrin, J.]{serrin_mathematical_1959}
Serrin, J.
\newblock Mathematical {Principles} of {Classical} {Fluid} {Mechanics}.
\newblock In , \emph{Fluid {Dynamics} {I} / {Strömungsmechanik} {I}}, pp.
  125--263. Springer Berlin Heidelberg, Berlin, Heidelberg, 1959.
\newblock \doi{10.1007/978-3-642-45914-6_2}.

\bibitem[Shinjo(2018)Shinjo, J.]{shinjo_recent_2018}
Shinjo, J.
\newblock Recent {Advances} in {Computational} {Modeling} of {Primary}
  {Atomization} of {Liquid} {Fuel} {Sprays}.
\newblock \emph{Energies}, 11\penalty0 (11):\penalty0 2971, 2018.
\newblock \doi{10.3390/en11112971}.

\bibitem[Shinjo and Umemura(2010)Shinjo, J. and Umemura,
  A.]{shinjo_simulation_2010}
Shinjo, J. and Umemura, A.
\newblock Simulation of liquid jet primary breakup: {Dynamics} of ligament and
  droplet formation.
\newblock \emph{International Journal of Multiphase Flow}, 36\penalty0
  (7):\penalty0 513--532, 2010.
\newblock \doi{10.1016/j.ijmultiphaseflow.2010.03.008}.

\bibitem[Shukla et~al.(2010)Shukla, R.~K., Pantano, C., and Freund,
  J.~B.]{shukla_interface_2010}
Shukla, R.~K., Pantano, C., and Freund, J.~B.
\newblock An interface capturing method for the simulation of multi-phase
  compressible flows.
\newblock \emph{Journal of Computational Physics}, 229\penalty0 (19):\penalty0
  7411--7439, 2010.
\newblock \doi{10.1016/j.jcp.2010.06.025}.

\bibitem[Sussman et~al.(1994)Sussman, M., Smereka, P., and Osher,
  S.]{sussman_level_1994}
Sussman, M., Smereka, P., and Osher, S.
\newblock A {Level} {Set} {Approach} for {Computing} {Solutions} to
  {Incompressible} {Two}-{Phase} {Flow}.
\newblock \emph{Journal of Computational Physics}, 114\penalty0 (1):\penalty0
  146--159, 1994.
\newblock \doi{10.1006/jcph.1994.1155}.

\bibitem[Sweby(1984)Sweby, P.~K.]{sweby_high_1984}
Sweby, P.~K.
\newblock High {Resolution} {Schemes} {Using} {Flux} {Limiters} for
  {Hyperbolic} {Conservation} {Laws}.
\newblock \emph{SIAM Journal on Numerical Analysis}, 21\penalty0 (5):\penalty0
  995--1011, 1984.
\newblock \doi{10.1137/0721062}.

\bibitem[Tomar et~al.(2010)Tomar, G., Fuster, D., Zaleski, S., and Popinet,
  S.]{tomar_multiscale_2010}
Tomar, G., Fuster, D., Zaleski, S., and Popinet, S.
\newblock Multiscale simulations of primary atomization.
\newblock \emph{Computers \& Fluids}, 39\penalty0 (10):\penalty0 1864--1874,
  2010.
\newblock \doi{10.1016/j.compfluid.2010.06.018}.

\bibitem[Toro(2009)Toro, E.~F.]{toro_riemann_2009}
Toro, E.~F.
\newblock \emph{Riemann solvers and numerical methods for fluid dynamics: a
  practical introduction}.
\newblock Springer, Dordrecht ; New York, 3rd ed edition, 2009.
\newblock OCLC: ocn401321914.

\bibitem[Truskinovsky(1991)Truskinovsky, L.]{fosdick_kinks_1991}
Truskinovsky, L.
\newblock Kinks versus shocks.
\newblock In , \emph{Shock induced transitions and phase structures in general
  media}, volume~52 of \emph{The {IMA} {Volumes} in {Mathematics} and its
  {Applications}}. Springer Verlag, 1991.

\bibitem[Van~Leer(1979)Van~Leer, B.]{van_leer_towards_1979}
Van~Leer, B.
\newblock Towards the ultimate conservative difference scheme. {V}. {A}
  second-order sequel to {Godunov}'s method.
\newblock \emph{Journal of Computational Physics}, 32\penalty0 (1):\penalty0
  101--136, 1979.
\newblock \doi{10.1016/0021-9991(79)90145-1}.

\bibitem[Vaudor et~al.(2017)Vaudor, G., Ménard, T., Aniszewski, W., Doring,
  M., and Berlemont, A.]{vaudor_consistent_2017}
Vaudor, G., Ménard, T., Aniszewski, W., Doring, M., and Berlemont, A.
\newblock A consistent mass and momentum flux computation method for two phase
  flows. {Application} to atomization process.
\newblock \emph{Computers \& Fluids}, 152:\penalty0 204--216, 2017.
\newblock \doi{10.1016/j.compfluid.2017.04.023}.

\bibitem[Williams(1958)Williams, F.~A.]{williams_spray_1958}
Williams, F.~A.
\newblock Spray {Combustion} and {Atomization}.
\newblock \emph{The Physics of Fluids}, 1\penalty0 (6):\penalty0 6, 1958.

\bibitem[Wolfram~Research(2023)Wolfram~Research,
  I.]{wolfram_research_mathematica_2023}
Wolfram~Research, I.
\newblock Mathematica, {Version} 13.3, 2023.

\end{thebibliography}


\begin{figure}
    \centering
    \includegraphics[width=.4\linewidth]{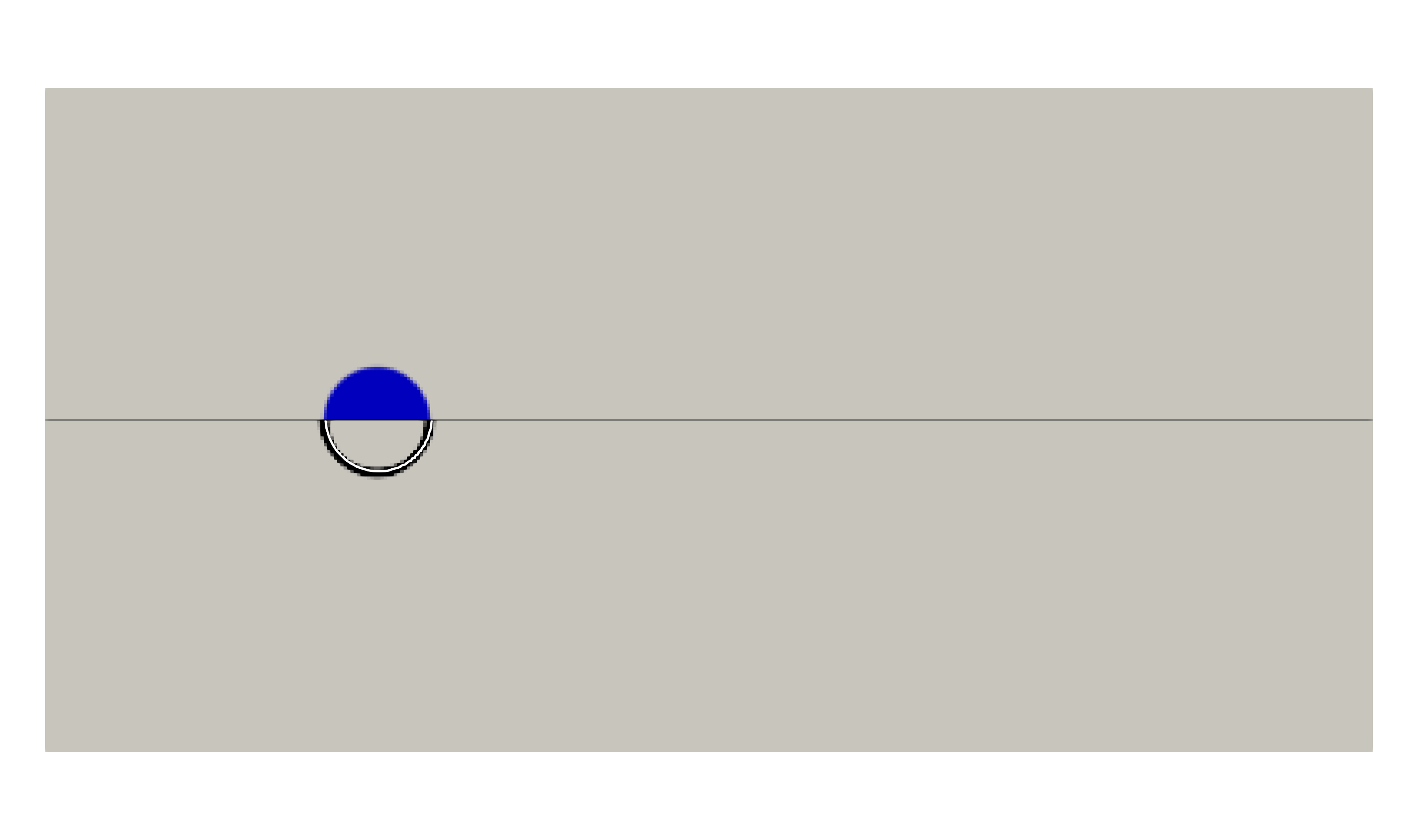}
    \includegraphics[width=.4\linewidth]{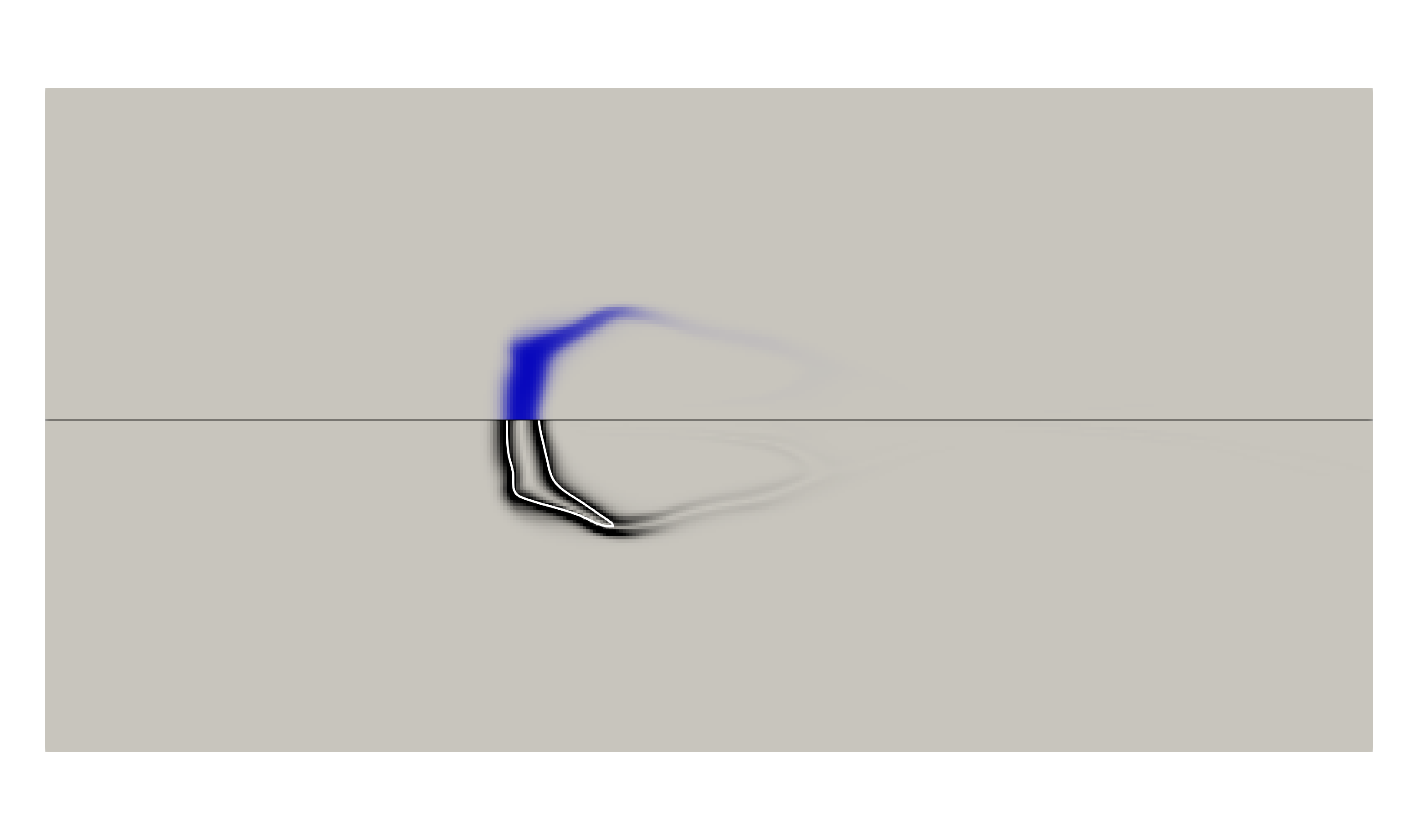}\\
    \includegraphics[width=.4\linewidth]{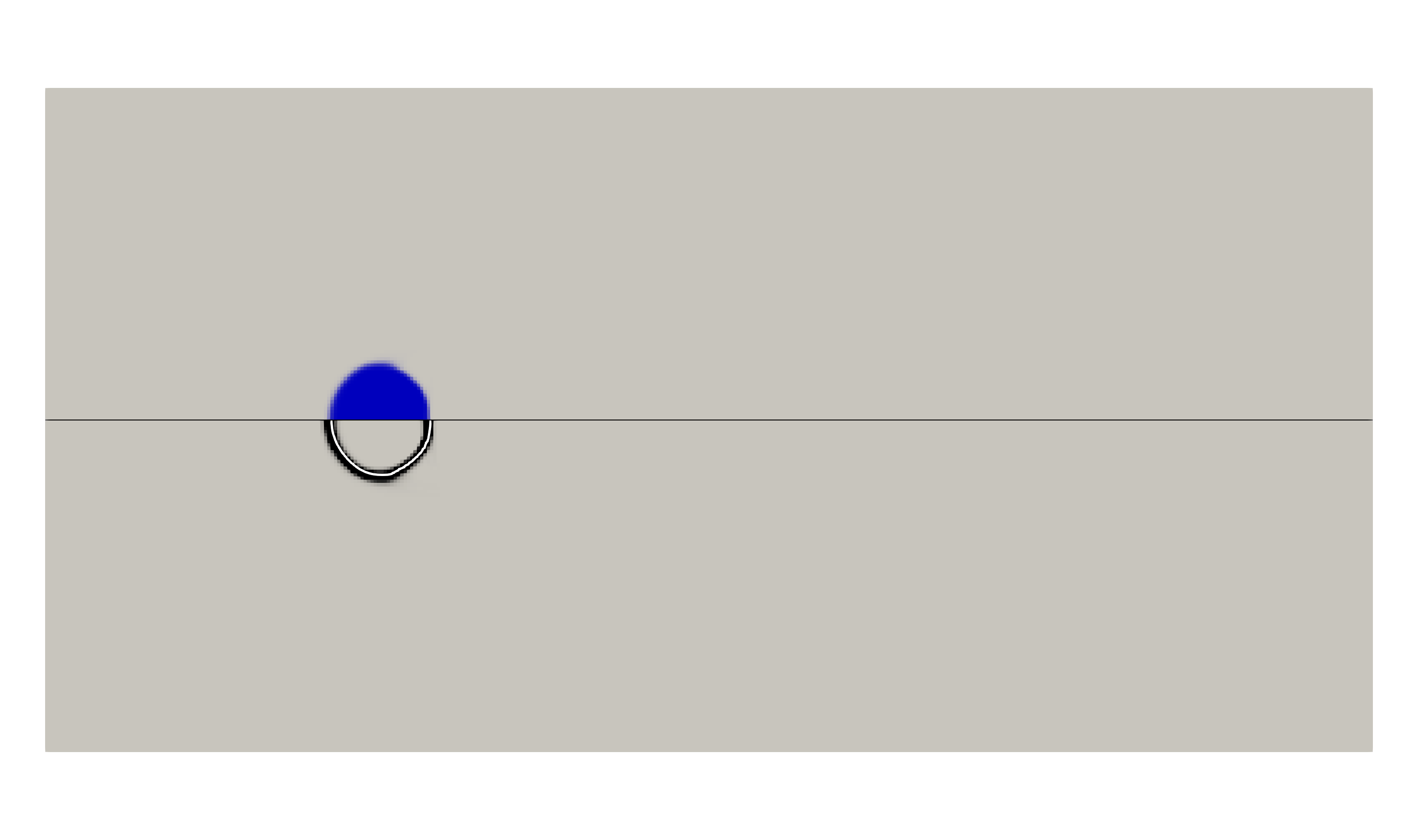}
    \includegraphics[width=.4\linewidth]{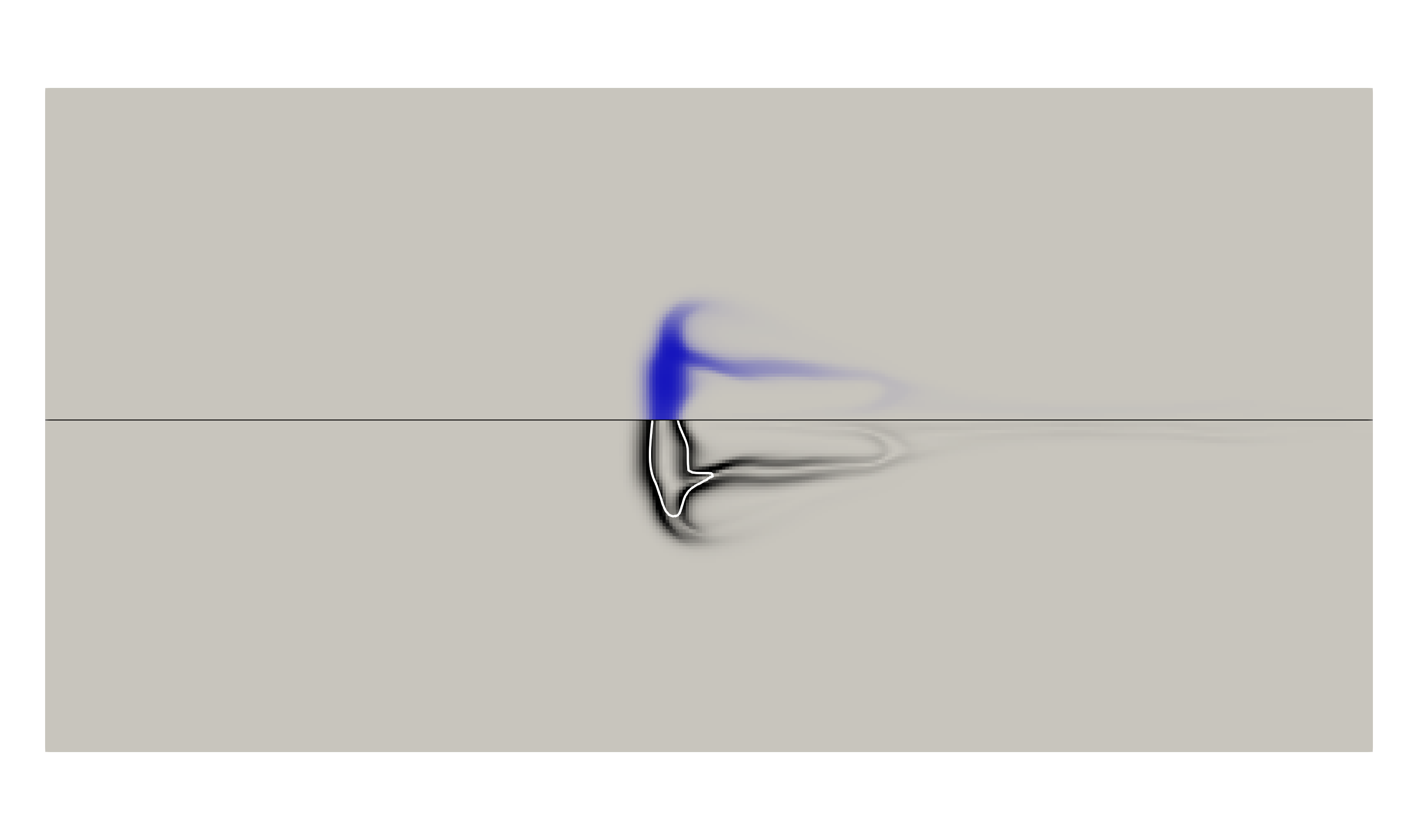}\\
    \includegraphics[width=.4\linewidth]{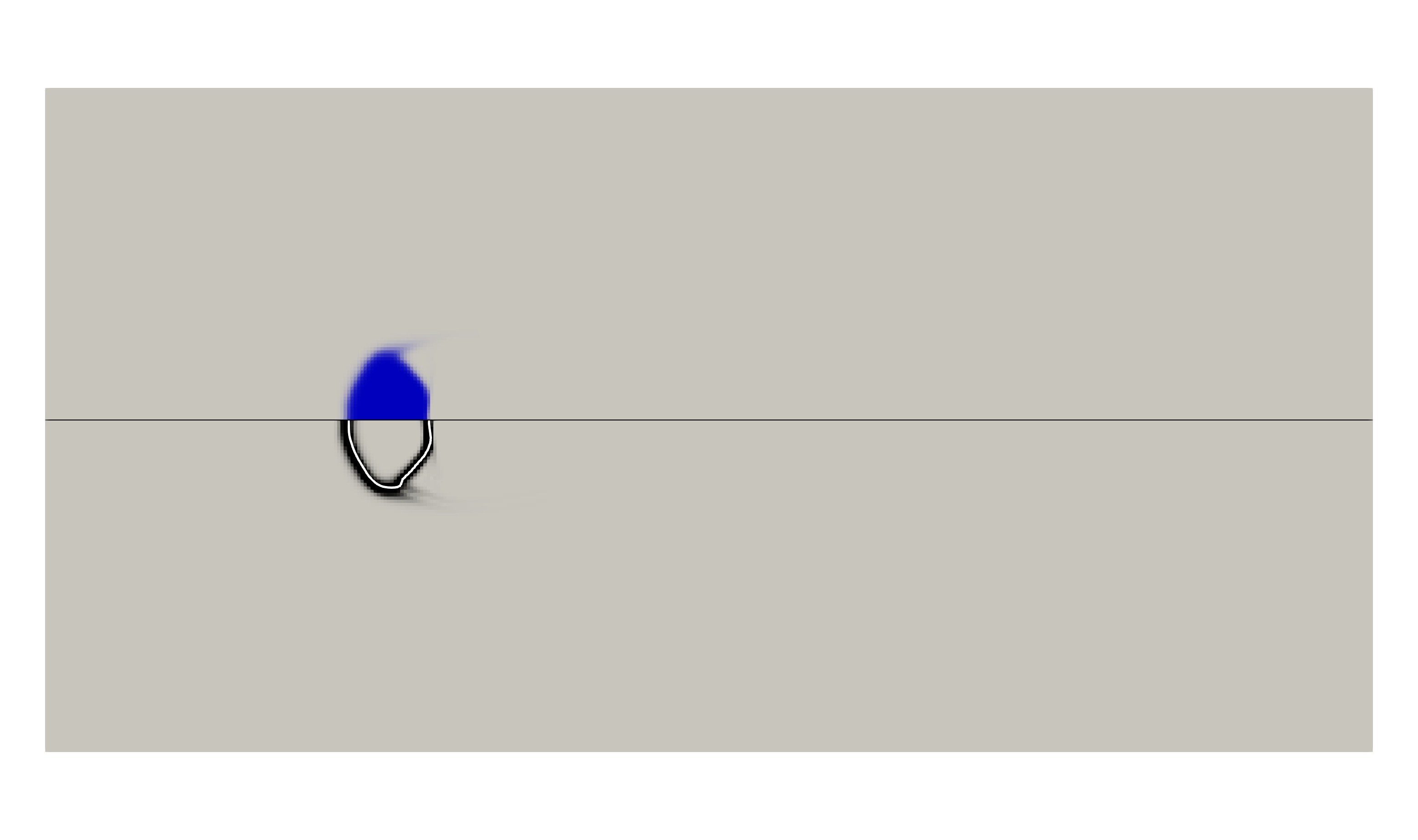}
    \includegraphics[width=.4\linewidth]{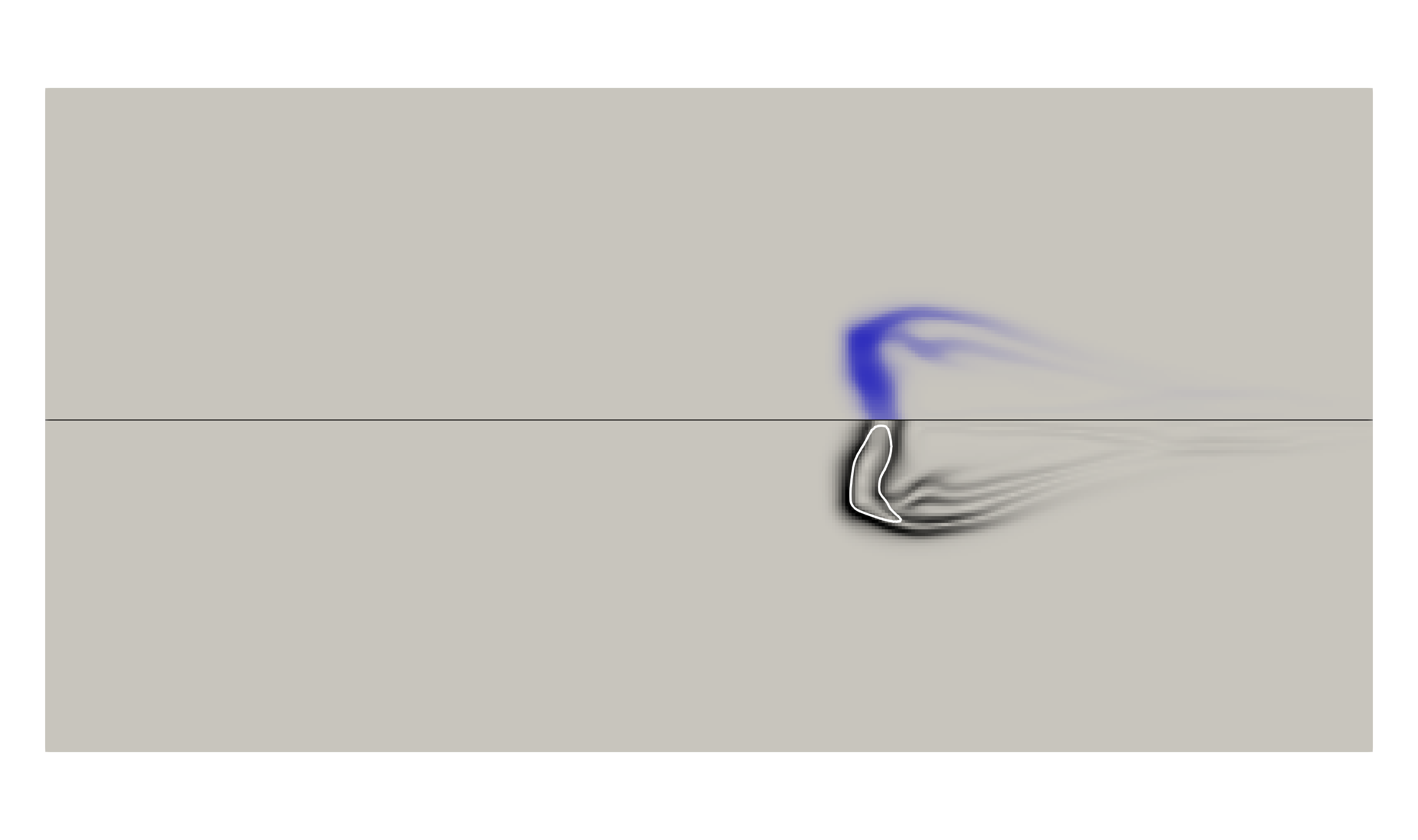}\\
    \includegraphics[width=.4\linewidth]{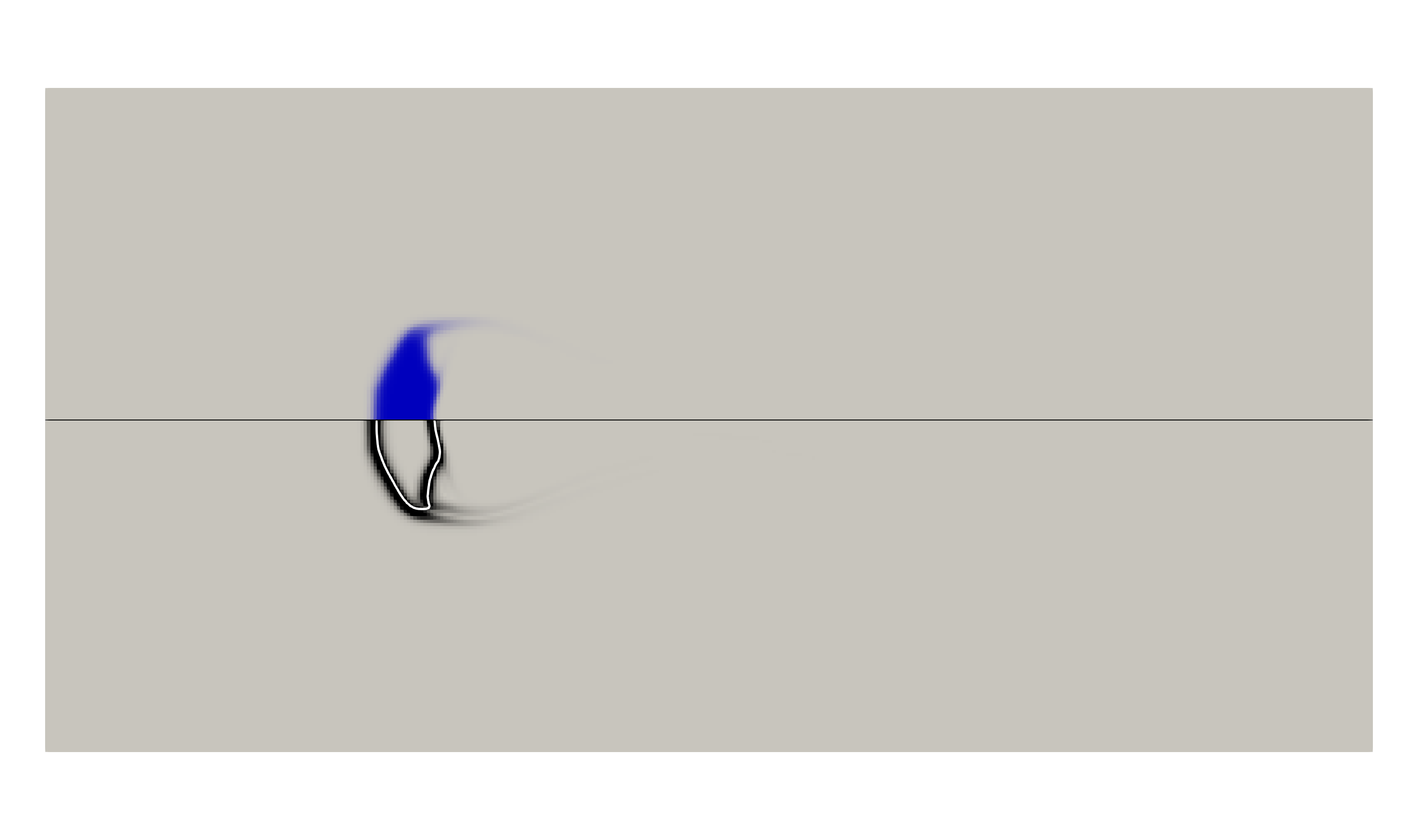}
    \includegraphics[width=.4\linewidth]{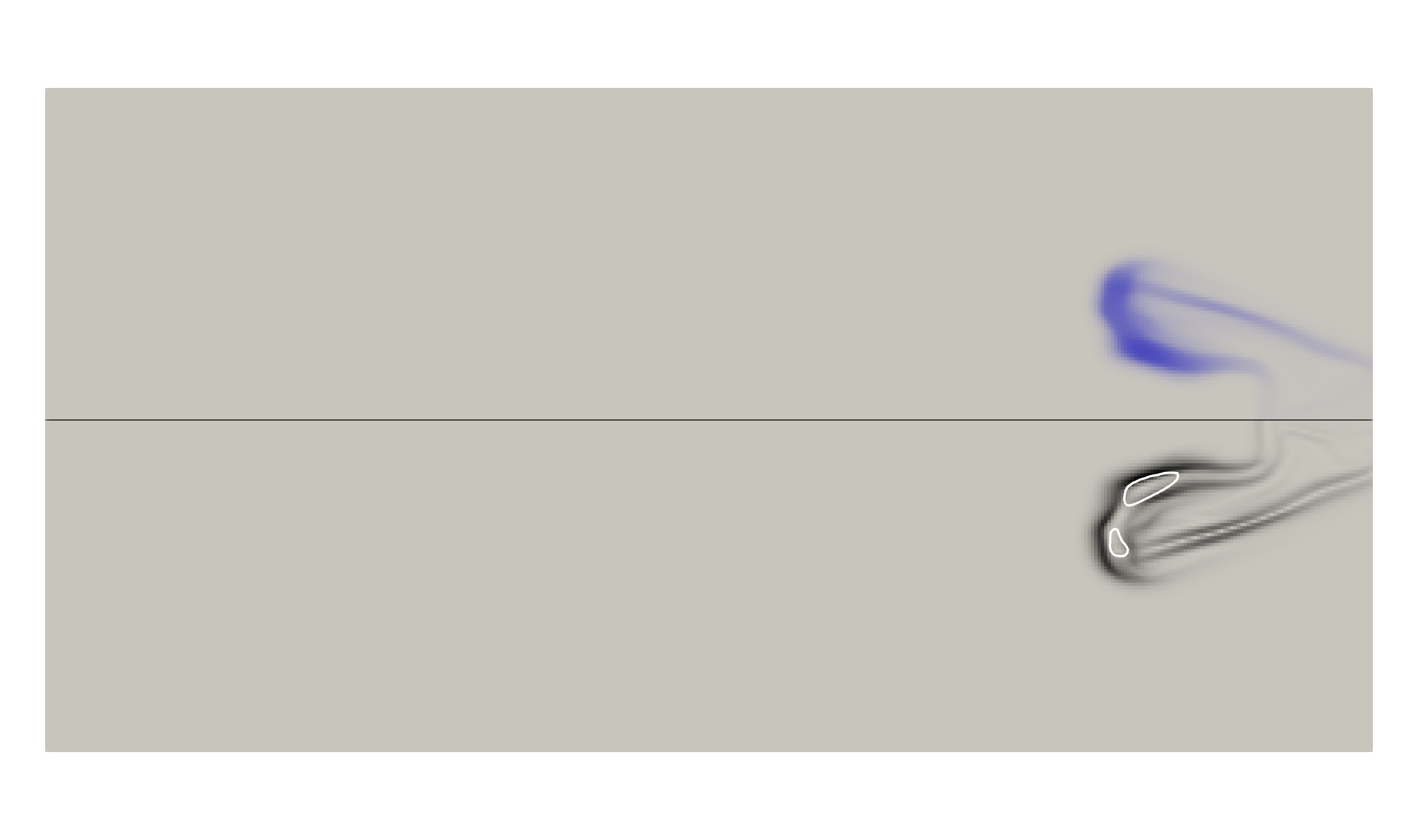}\\
    \includegraphics[width=.4\linewidth]{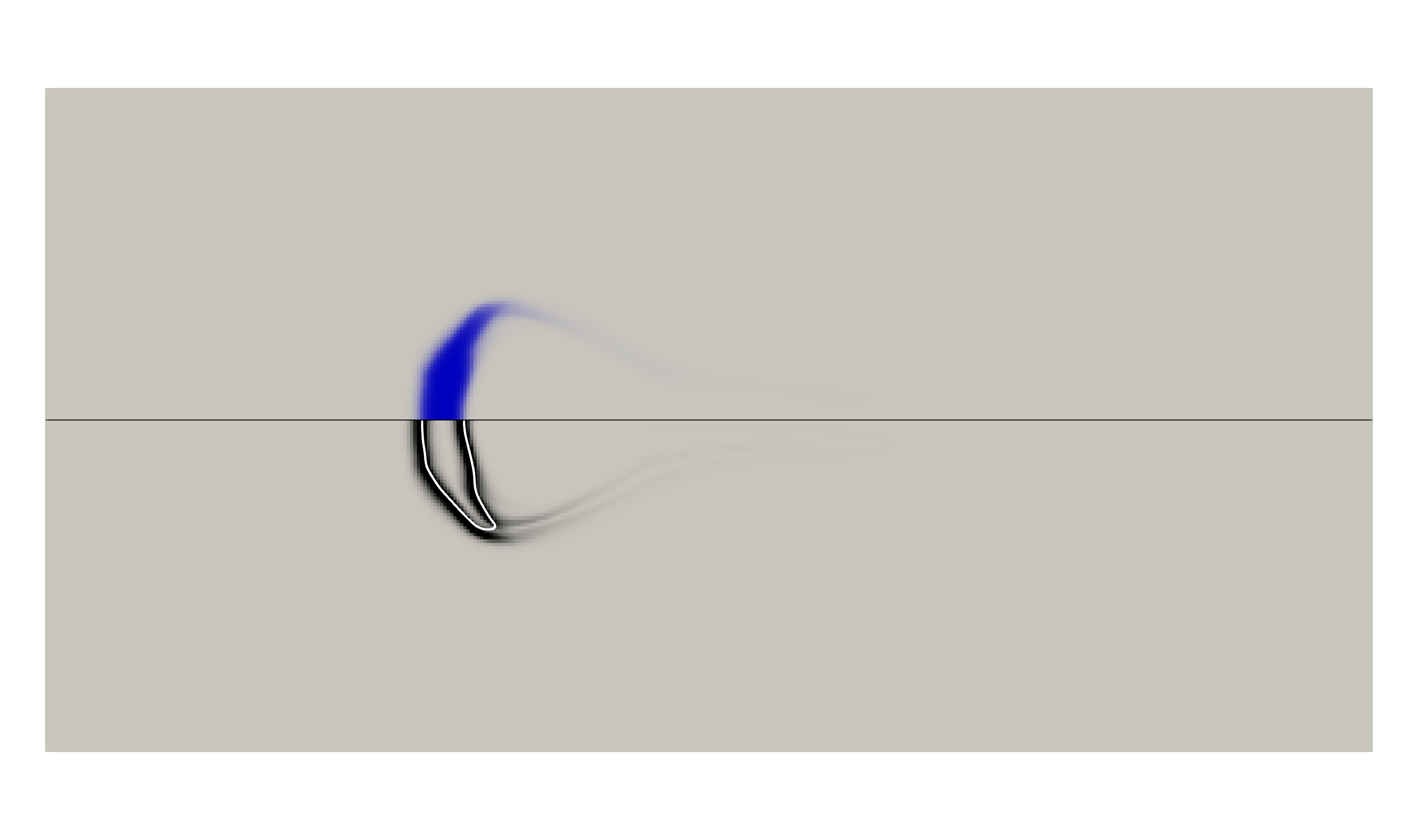}
    \includegraphics[width=.4\linewidth]{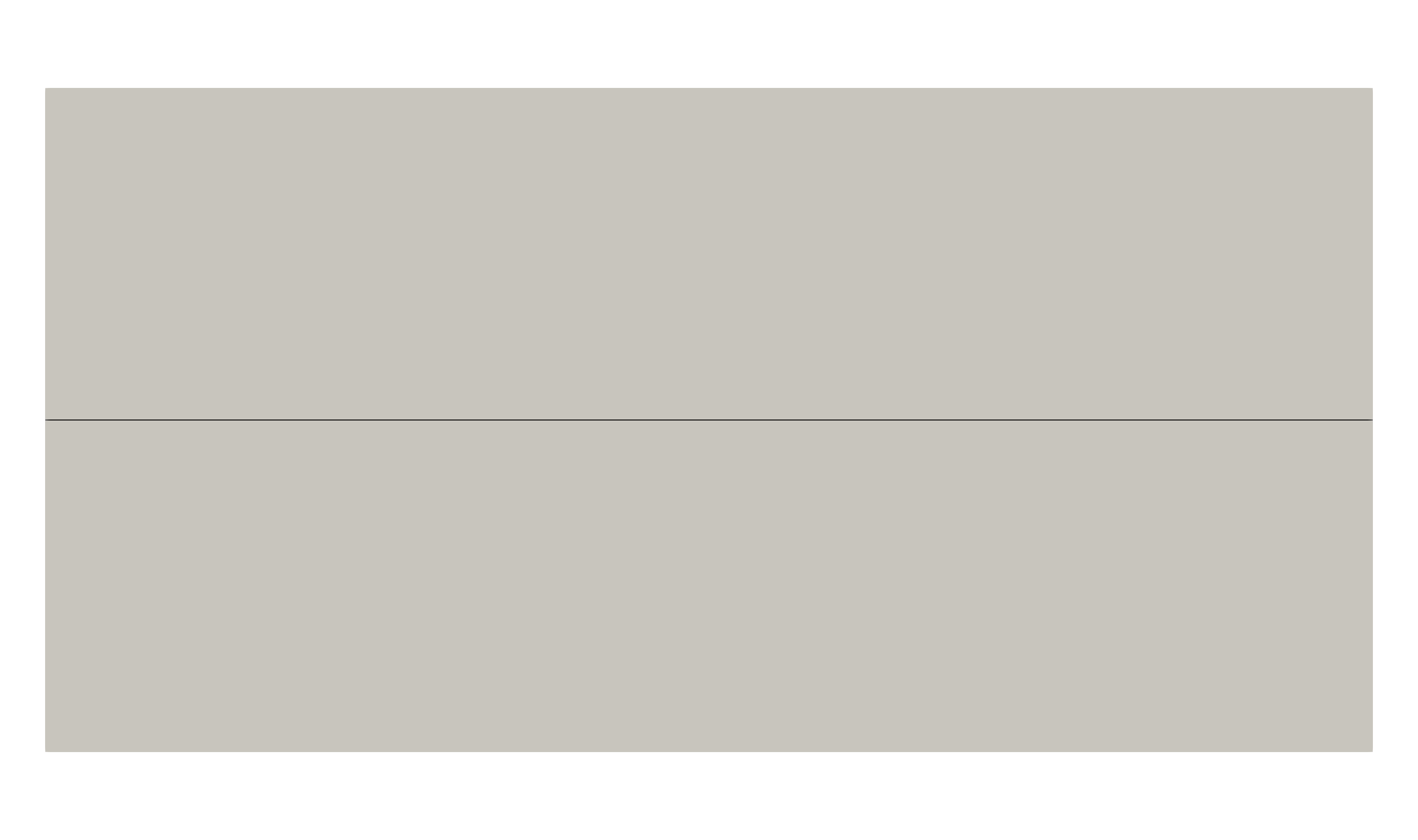}
    \caption{Liquid column deformation without inter-scale transfer. Liquid volume fraction $\overline{\alpha}_1\in(0,1)$
    {\protect\tikz \protect\node [rectangle, left color=paraviewwhite, right color=blue, anchor=north, minimum width=1cm, minimum height=0.1cm] (box) at (0,0){};}
    (top) and estimator of the IAD $\Vert~\nabla~\overline{\alpha}_1~\Vert~\in~(0,16)$~{\protect\tikz \protect\node [rectangle, left color=paraviewwhite, right color=black, anchor=north, minimum width=1cm, minimum height=0.1cm] (box) at (0,0){};} (bottom) with the iso-line $\overline{\alpha}_1=0.5$ (white, bottom). Snapshots are taken each $0.25$ s from $t=0$ s to $t=2.5$ s from top to bottom and left to right.}
    \label{fig:dynamics-no-transfer}
\end{figure}

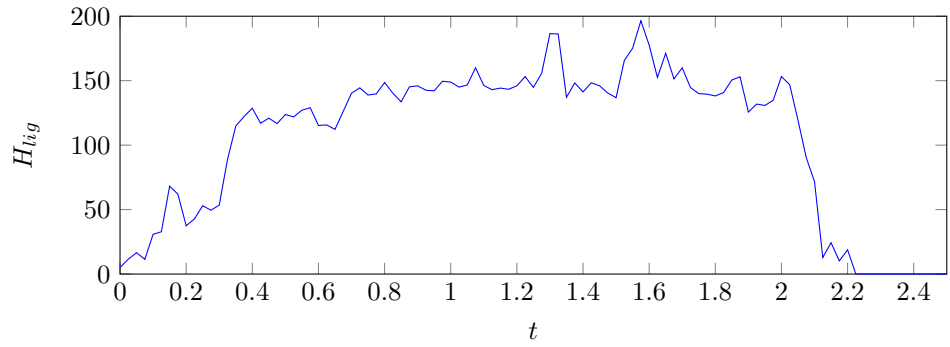
\begin{figure}
    \centering
    \begin{tikzpicture}
        \centering
        \begin{axis}[
                width=.7\textwidth,
                height=5cm,
                xmin=0,
                xmax=2.5,
                ymin=0,
                ymax=200,
                xlabel={$t$},
                ylabel={$H_{lig}$},
            ]
            \addplot[blue] table [x={time}, y={Hmax}] {no-transfer.txt};
        \end{axis}
    \end{tikzpicture}
    \caption{Evolution in time of the mean curvature $H_{lig}$ when inter-scale transfer is deactivated.}
    \label{fig:H-evo-no-transfer}
\end{figure}

\begin{figure}
    \centering
    \adjincludegraphics[trim={{.2\width} {.25\height} 0 {.25\height}},clip,width=.4\linewidth]{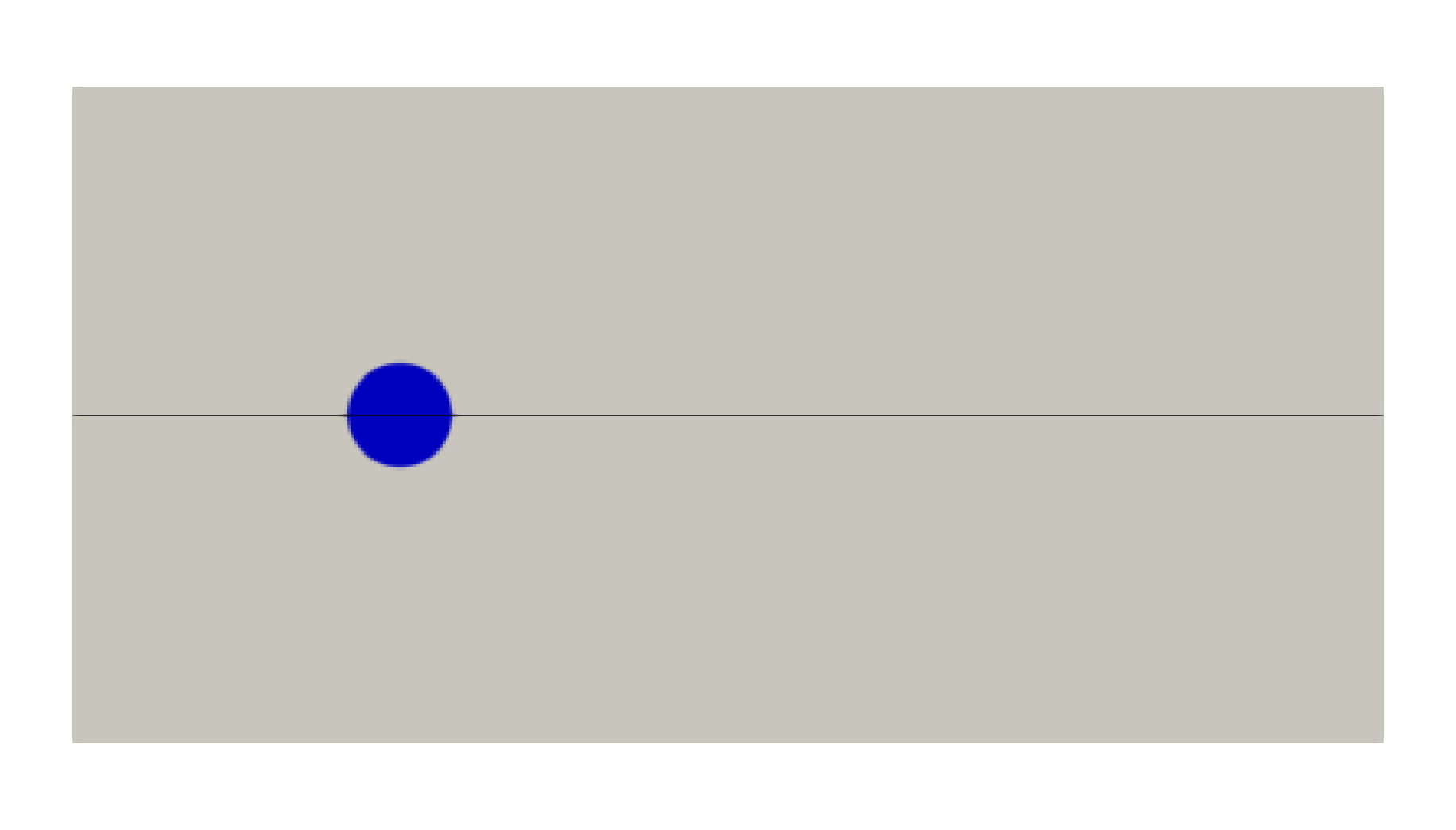}
    \adjincludegraphics[trim={{.2\width} {.25\height} 0 {.25\height}},clip,width=.4\linewidth]{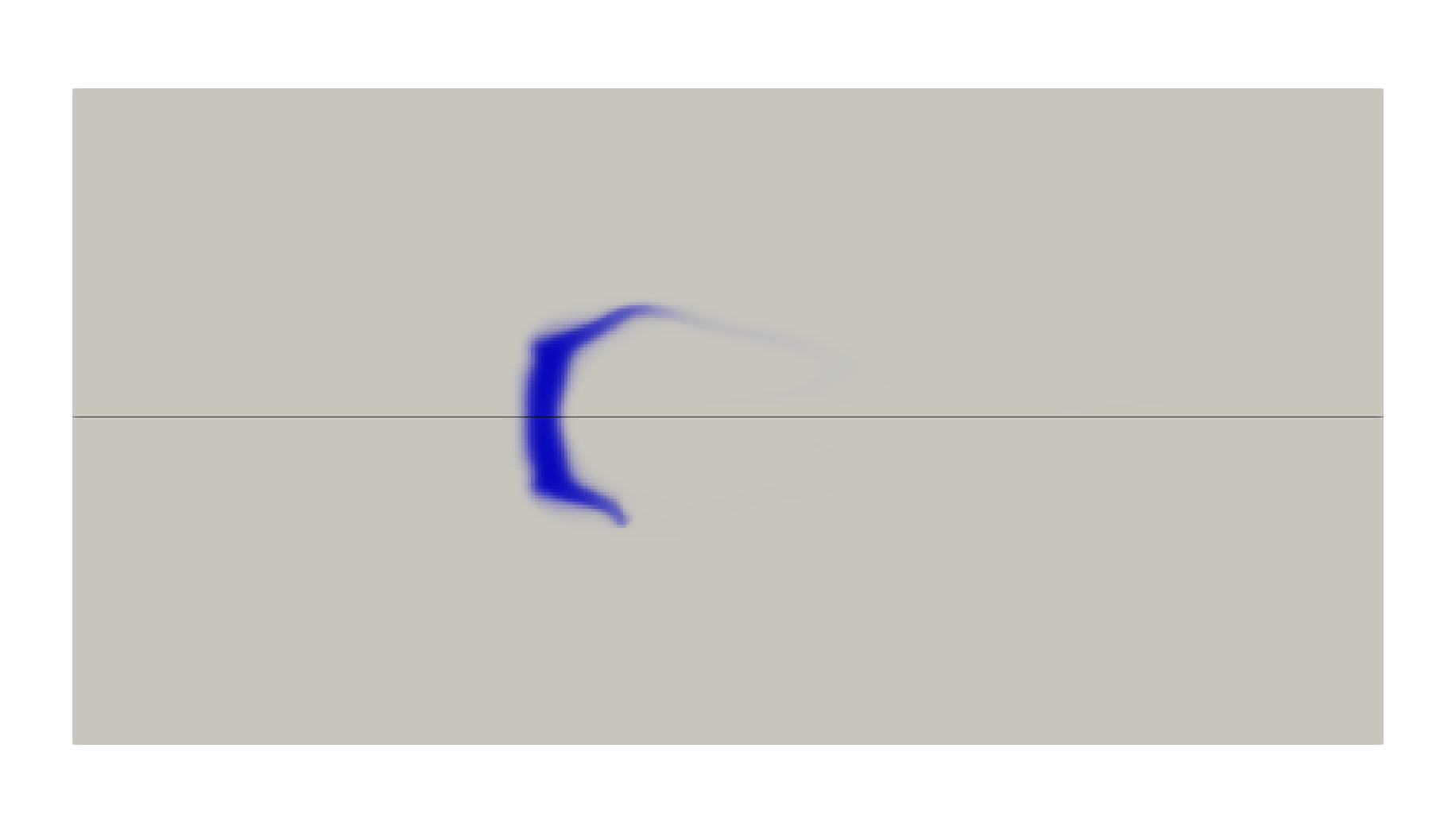}\\[5pt]
    \adjincludegraphics[trim={{.2\width} {.25\height} 0 {.25\height}},clip,width=.4\linewidth]{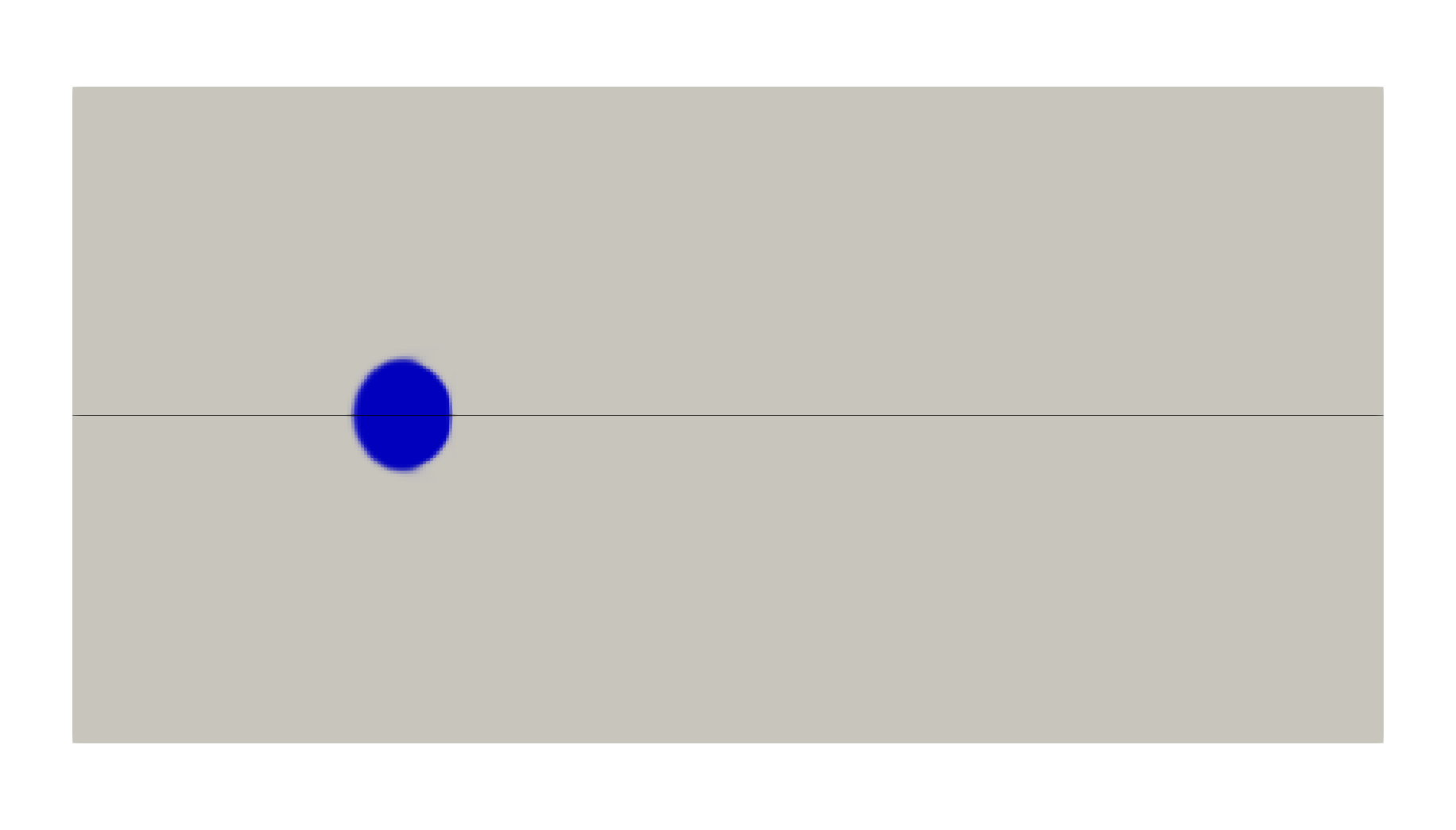}
    \adjincludegraphics[trim={{.2\width} {.25\height} 0 {.25\height}},clip,width=.4\linewidth]{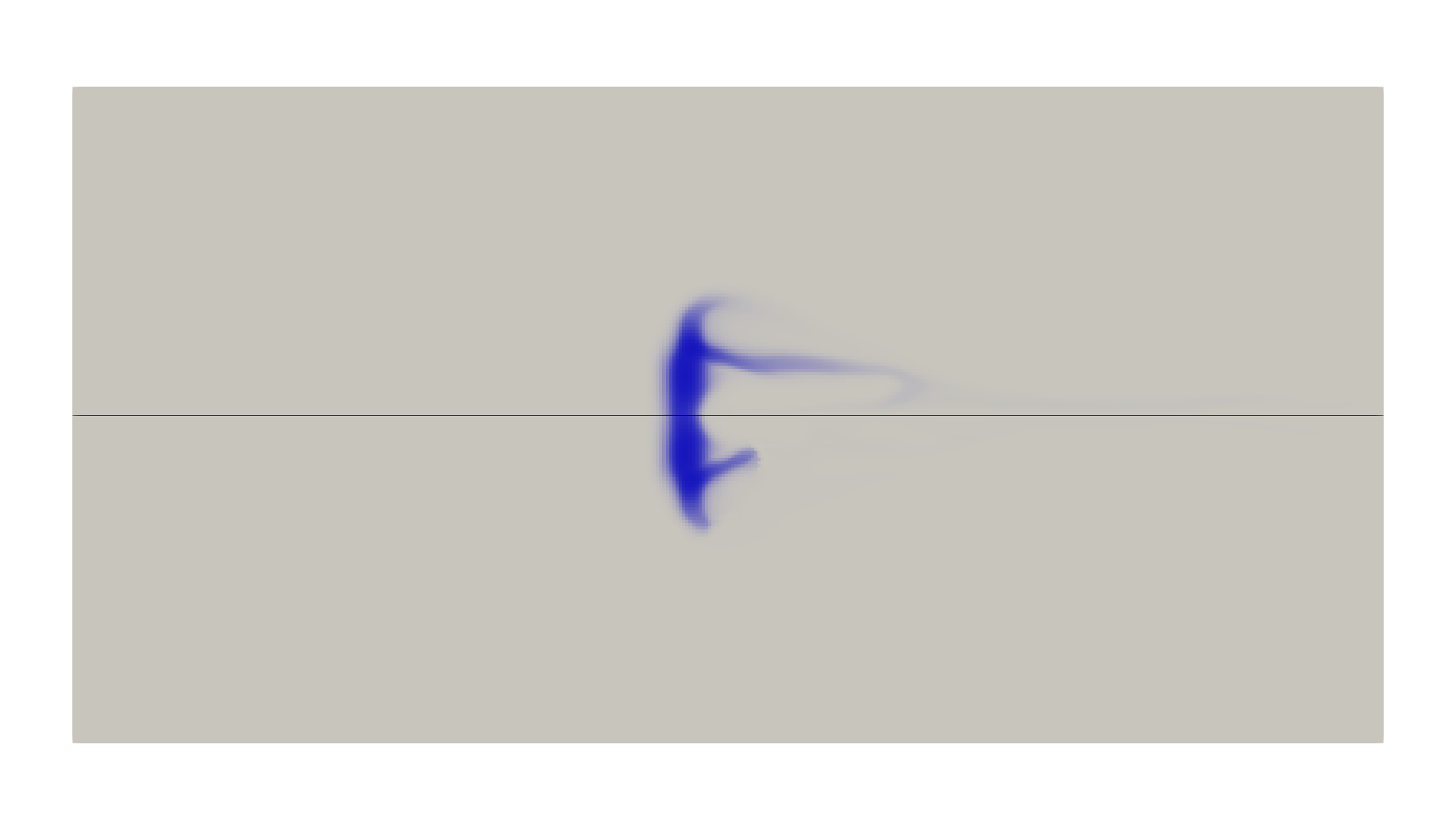}\\[5pt]
    \adjincludegraphics[trim={{.2\width} {.25\height} 0 {.25\height}},clip,width=.4\linewidth]{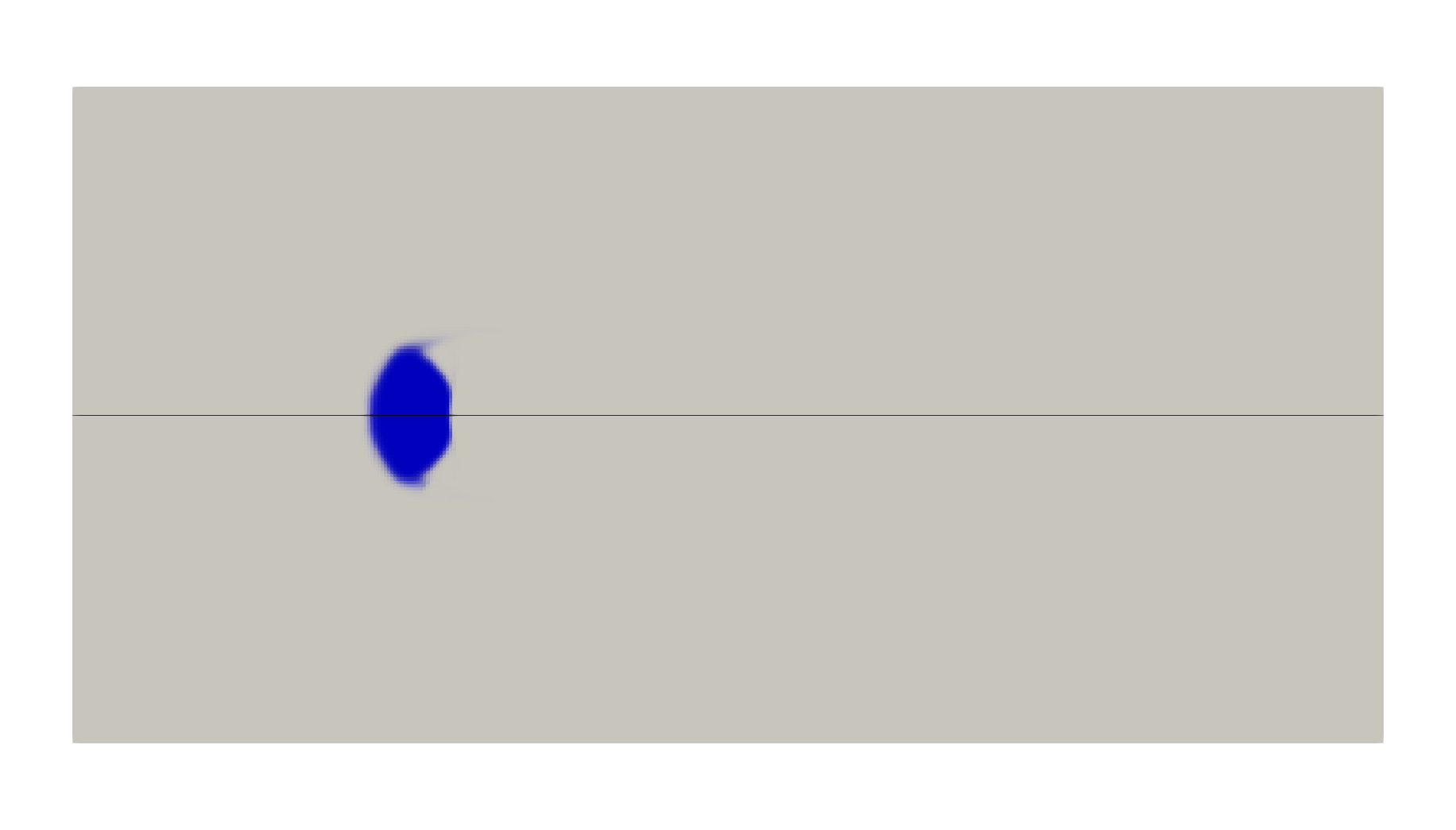}
    \adjincludegraphics[trim={{.2\width} {.25\height} 0 {.25\height}},clip,width=.4\linewidth]{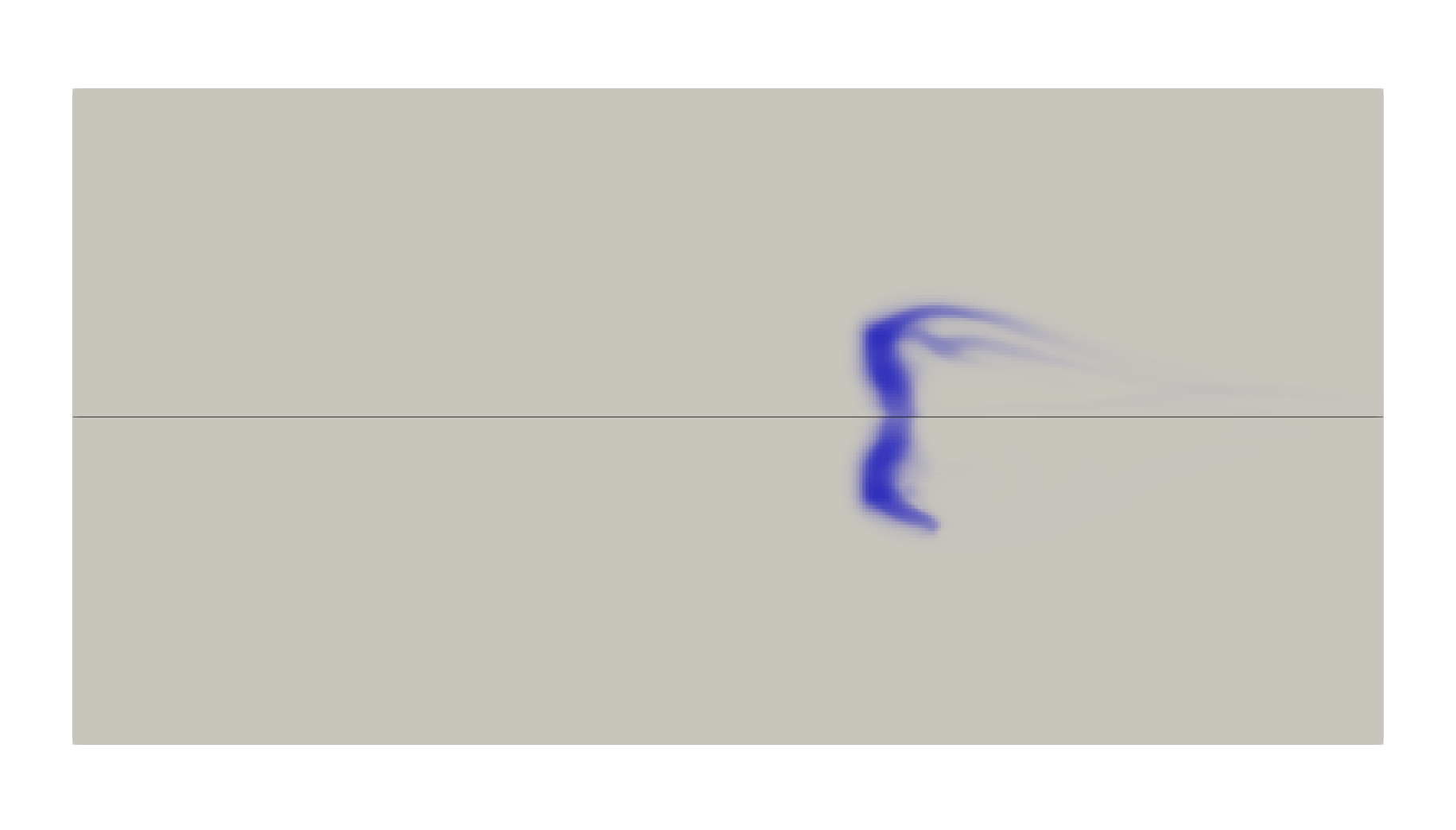}\\[5pt]
    \adjincludegraphics[trim={{.2\width} {.25\height} 0 {.25\height}},clip,width=.4\linewidth]{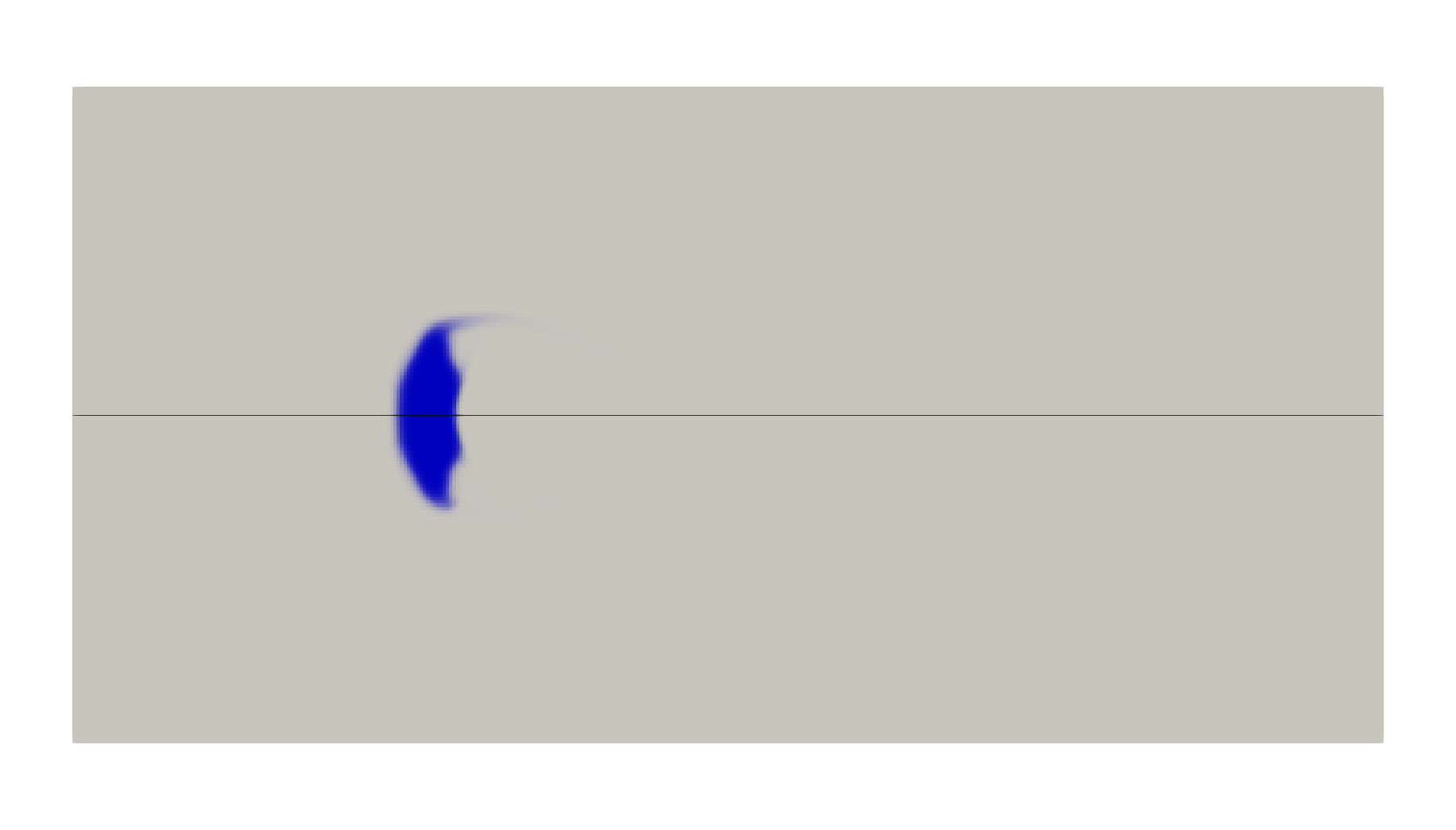}
    \adjincludegraphics[trim={{.2\width} {.25\height} 0 {.25\height}},clip,width=.4\linewidth]{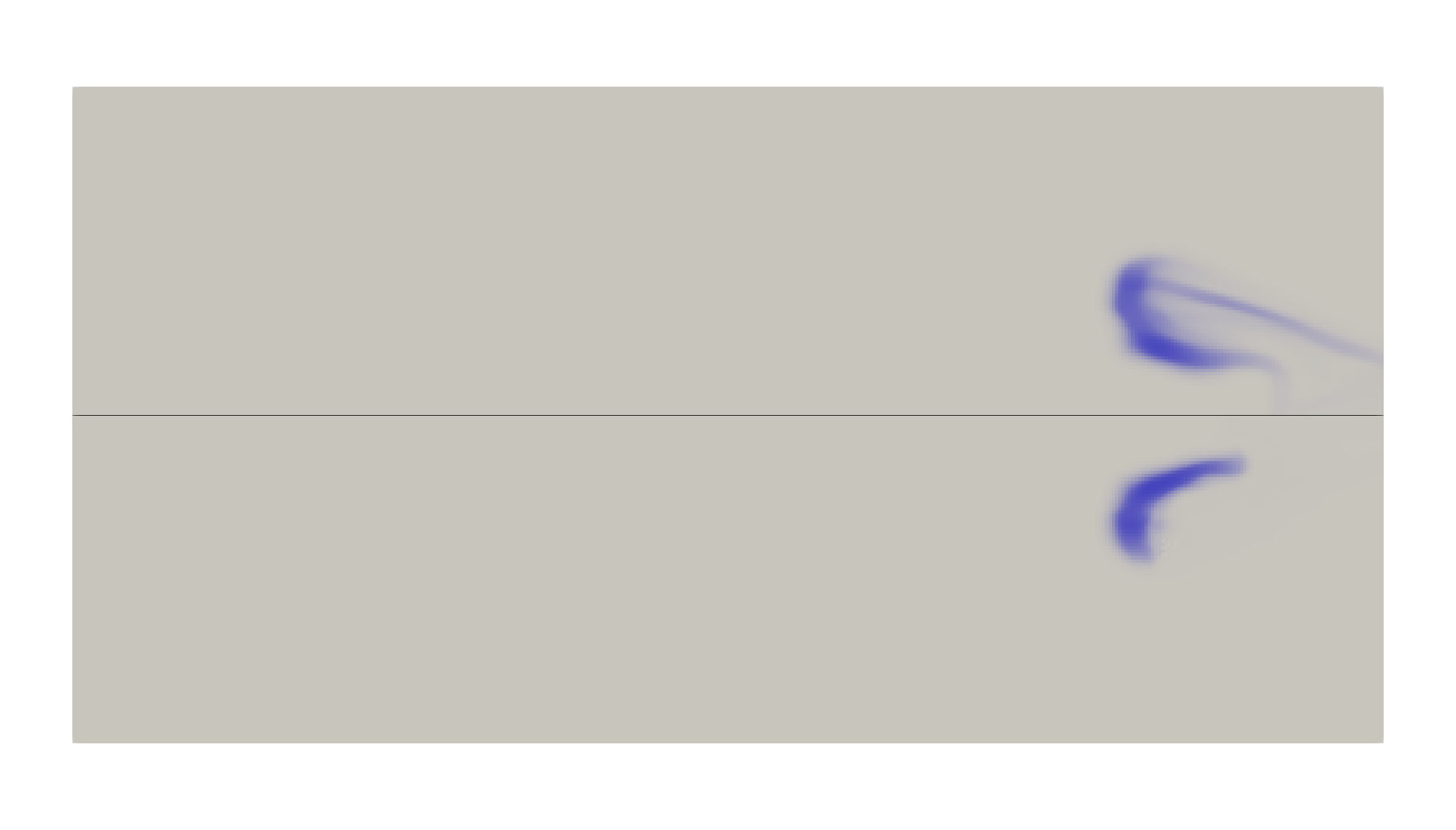}\\[5pt]
    \adjincludegraphics[trim={{.2\width} {.25\height} 0 {.25\height}},clip,width=.4\linewidth]{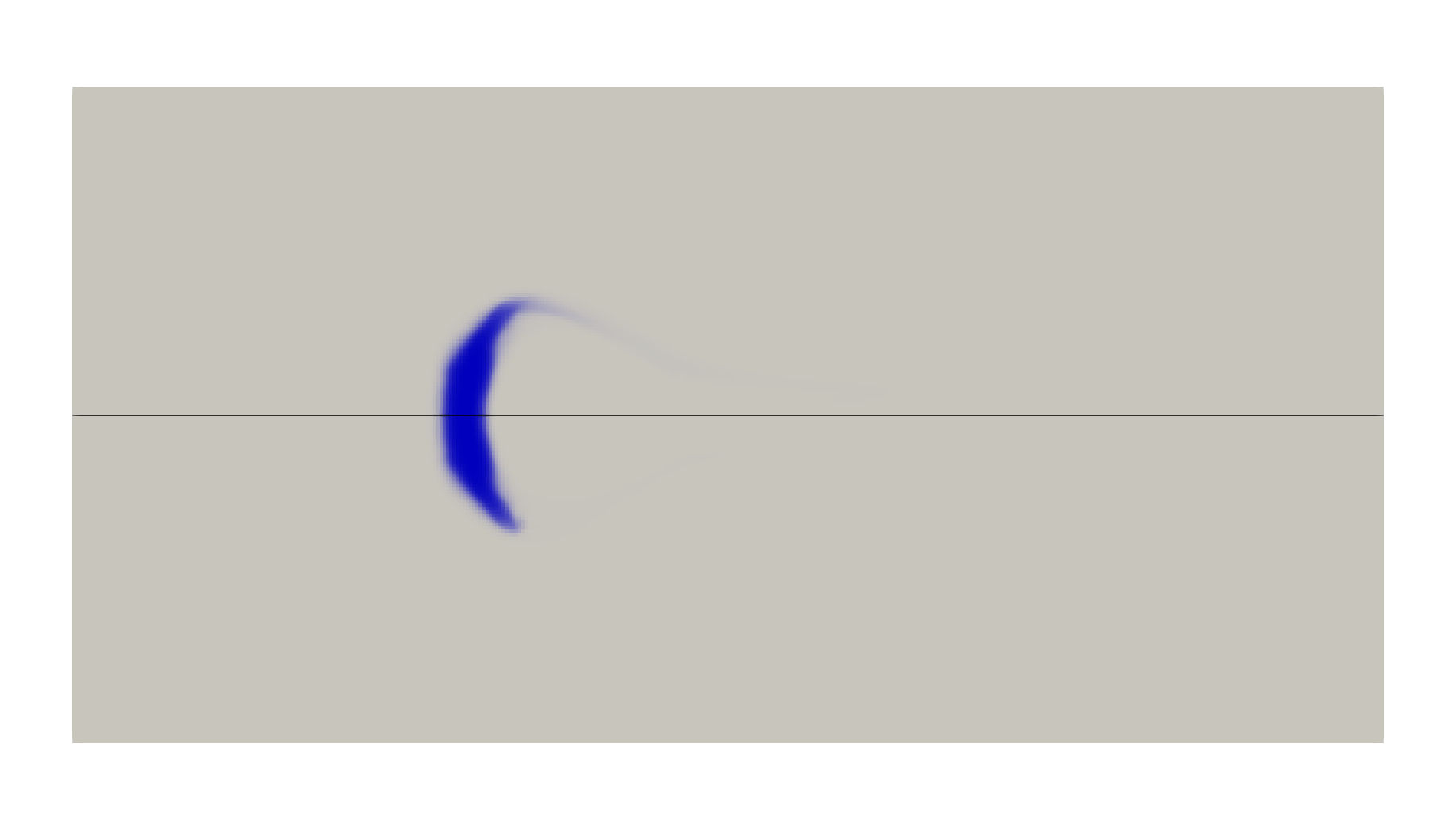}
    \adjincludegraphics[trim={{.2\width} {.25\height} 0 {.25\height}},clip,width=.4\linewidth]{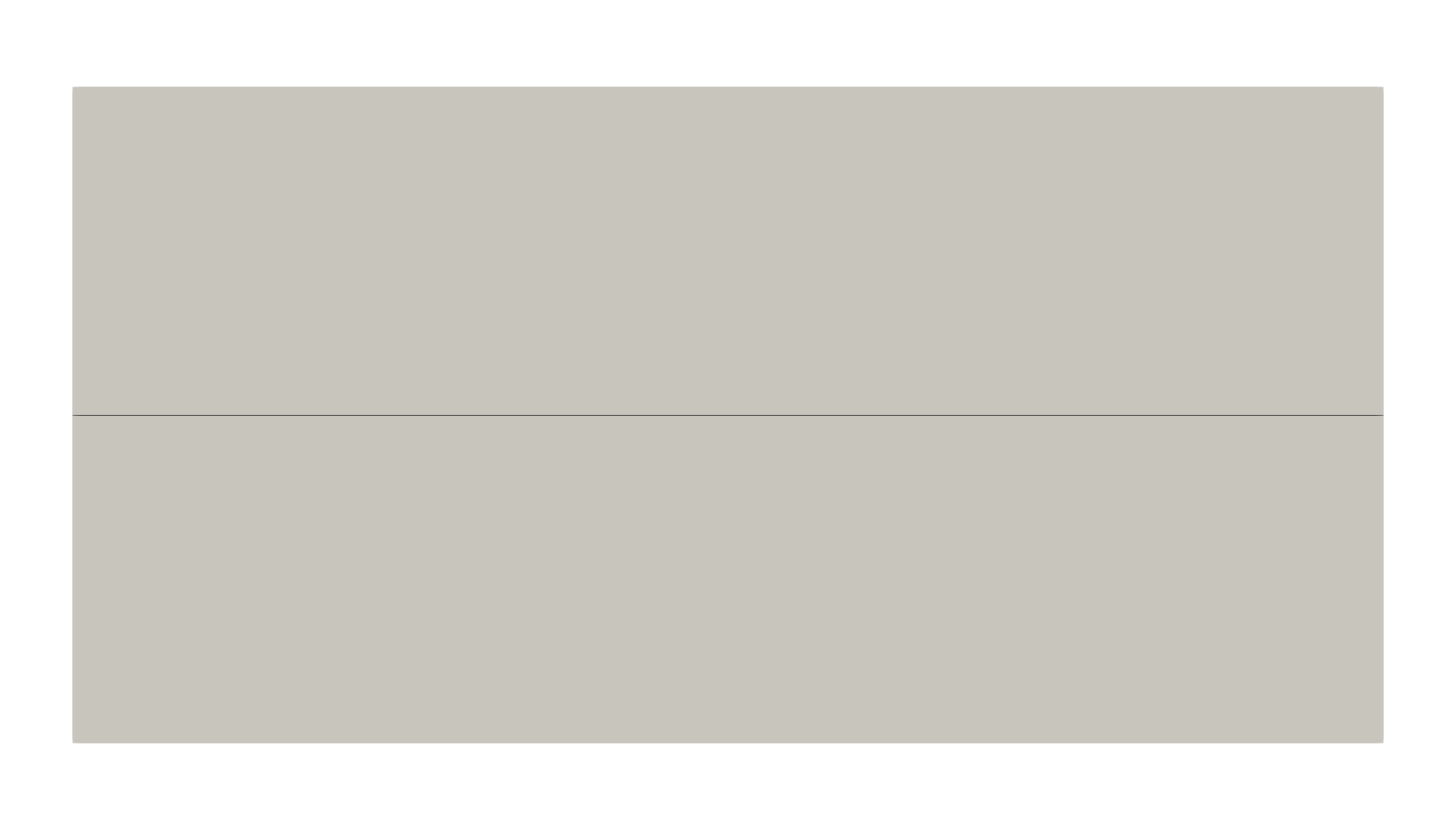}
    \caption{Large-scale liquid volume fraction $\overline{\alpha}_1\in(0,1)$
    {\protect\tikz \protect\node [rectangle, left color=paraviewwhite, right color=blue, anchor=north, minimum width=1cm, minimum height=0.1cm] (box) at (0,0){};}
    without mass transfer (top) and with mass transfer (bottom). Snapshots are taken each $0.25$ s from $t=0$ s to $t=2.5$ s from top to bottom and left to right.}
    \label{fig:compare-alpha}
\end{figure}

\begin{figure}
    \centering
    \adjincludegraphics[trim={{.2\width} {.25\height} 0 {.25\height}},clip,width=.4\linewidth]{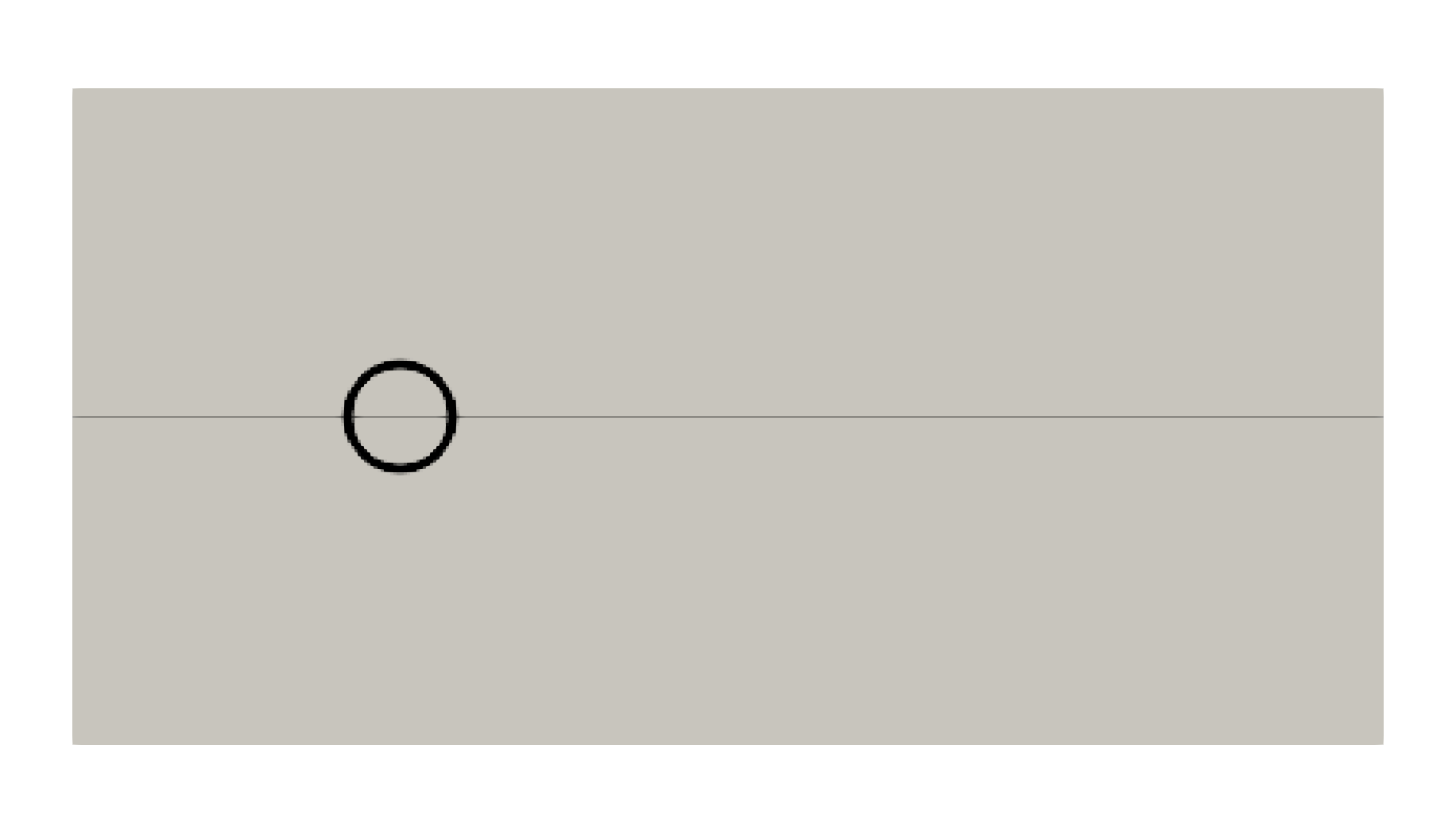}
    \adjincludegraphics[trim={{.2\width} {.25\height} 0 {.25\height}},clip,width=.4\linewidth]{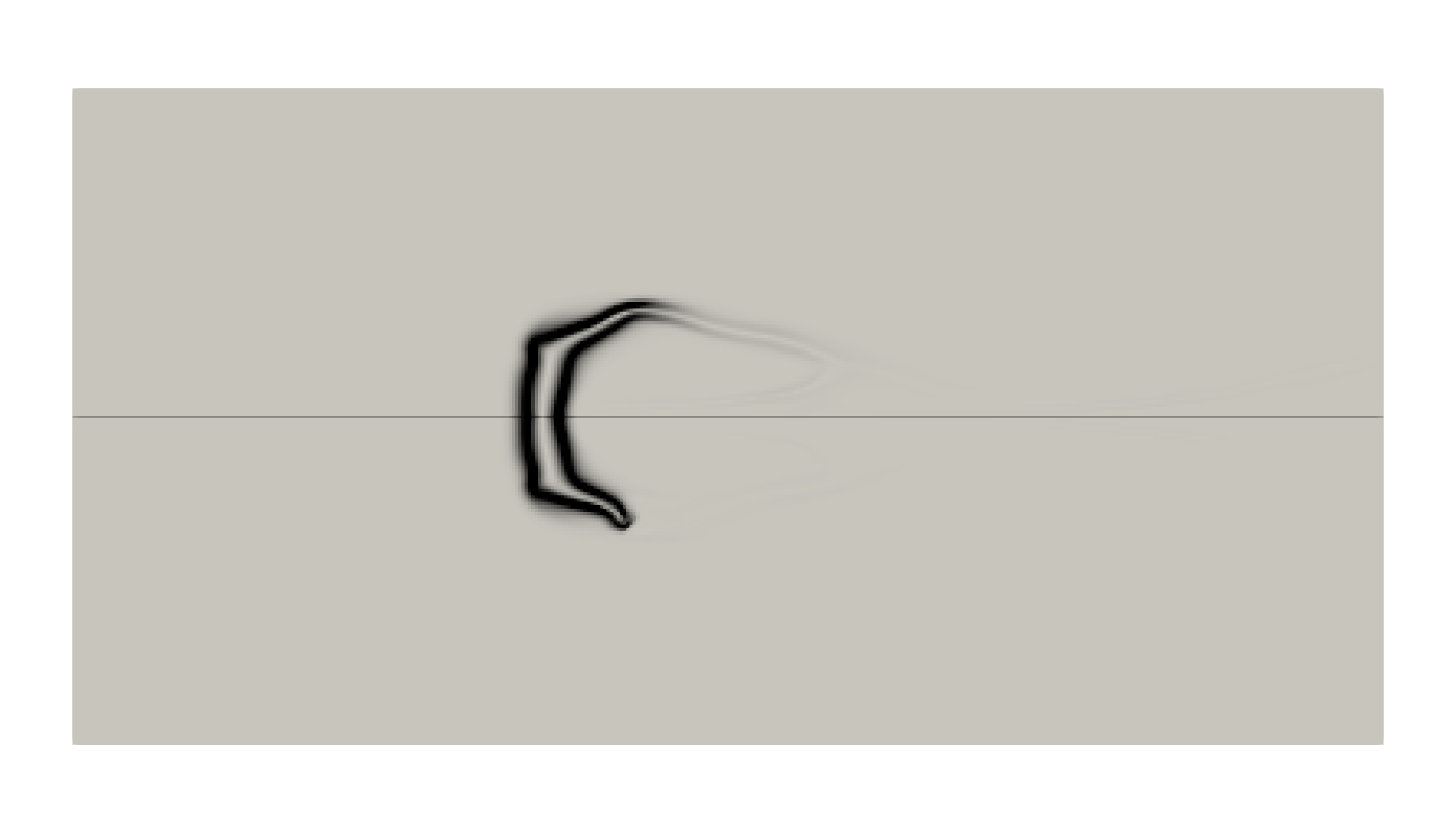}\\[5pt]
    \adjincludegraphics[trim={{.2\width} {.25\height} 0 {.25\height}},clip,width=.4\linewidth]{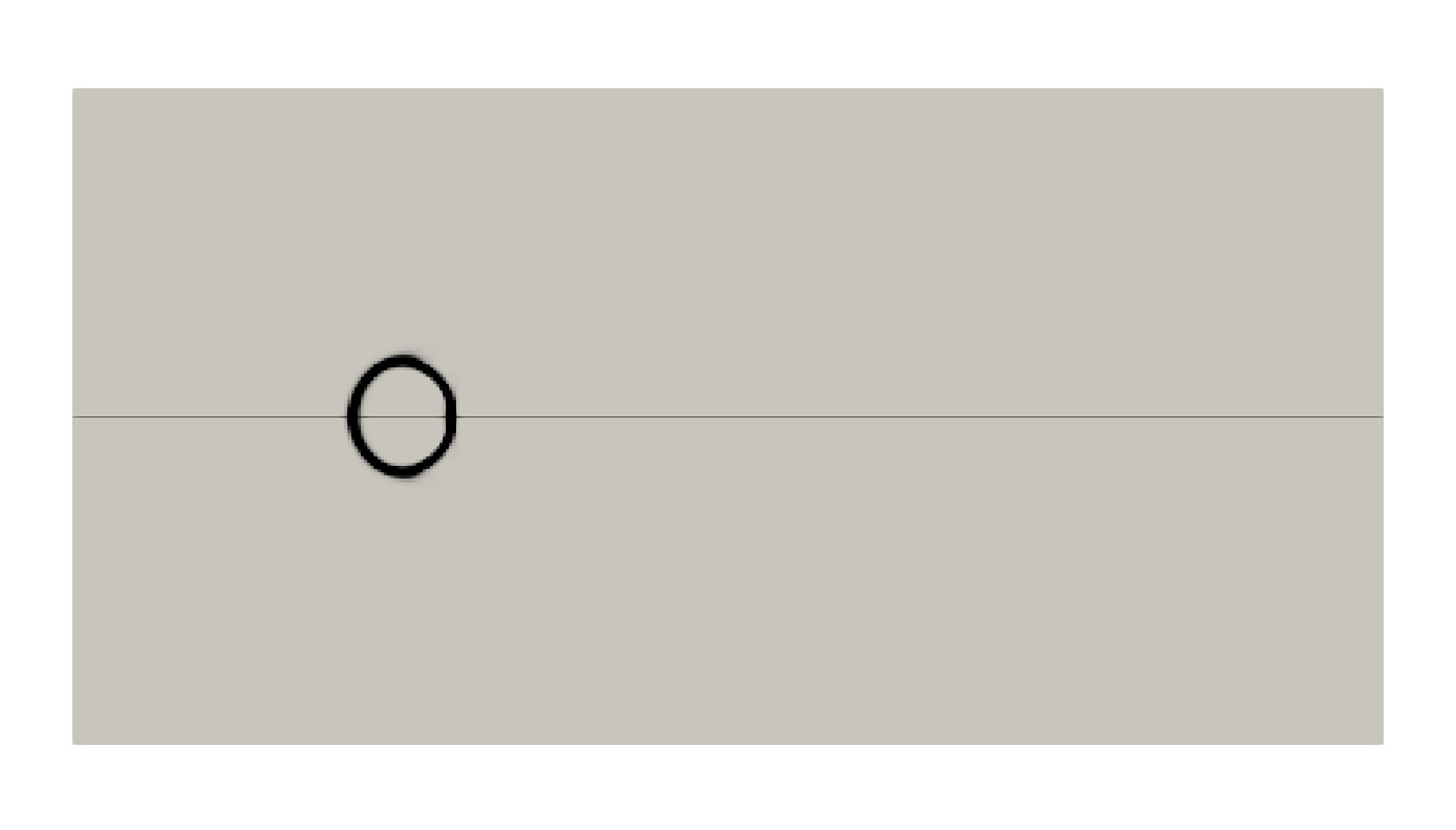}
    \adjincludegraphics[trim={{.2\width} {.25\height} 0 {.25\height}},clip,width=.4\linewidth]{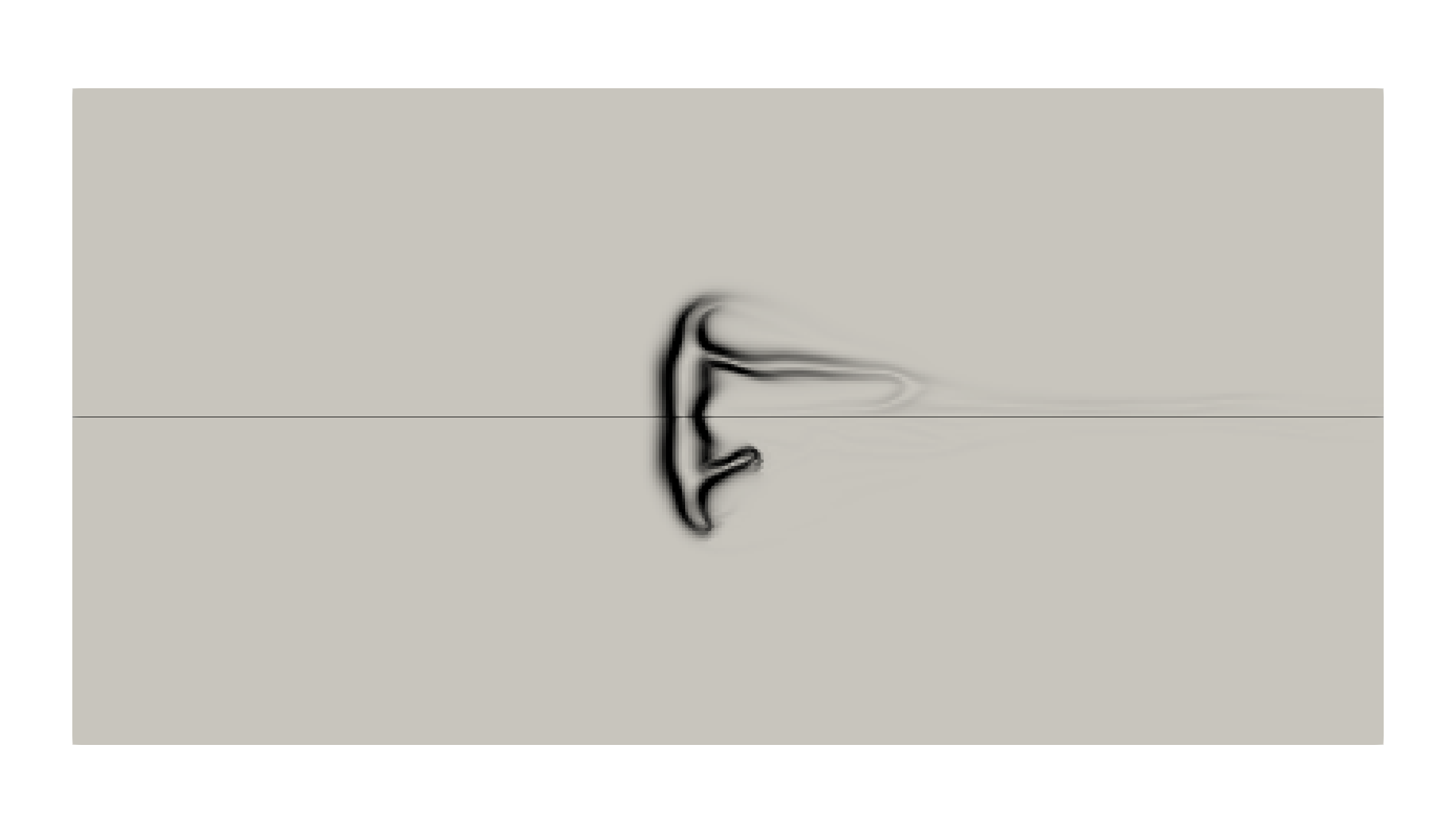}\\[5pt]
    \adjincludegraphics[trim={{.2\width} {.25\height} 0 {.25\height}},clip,width=.4\linewidth]{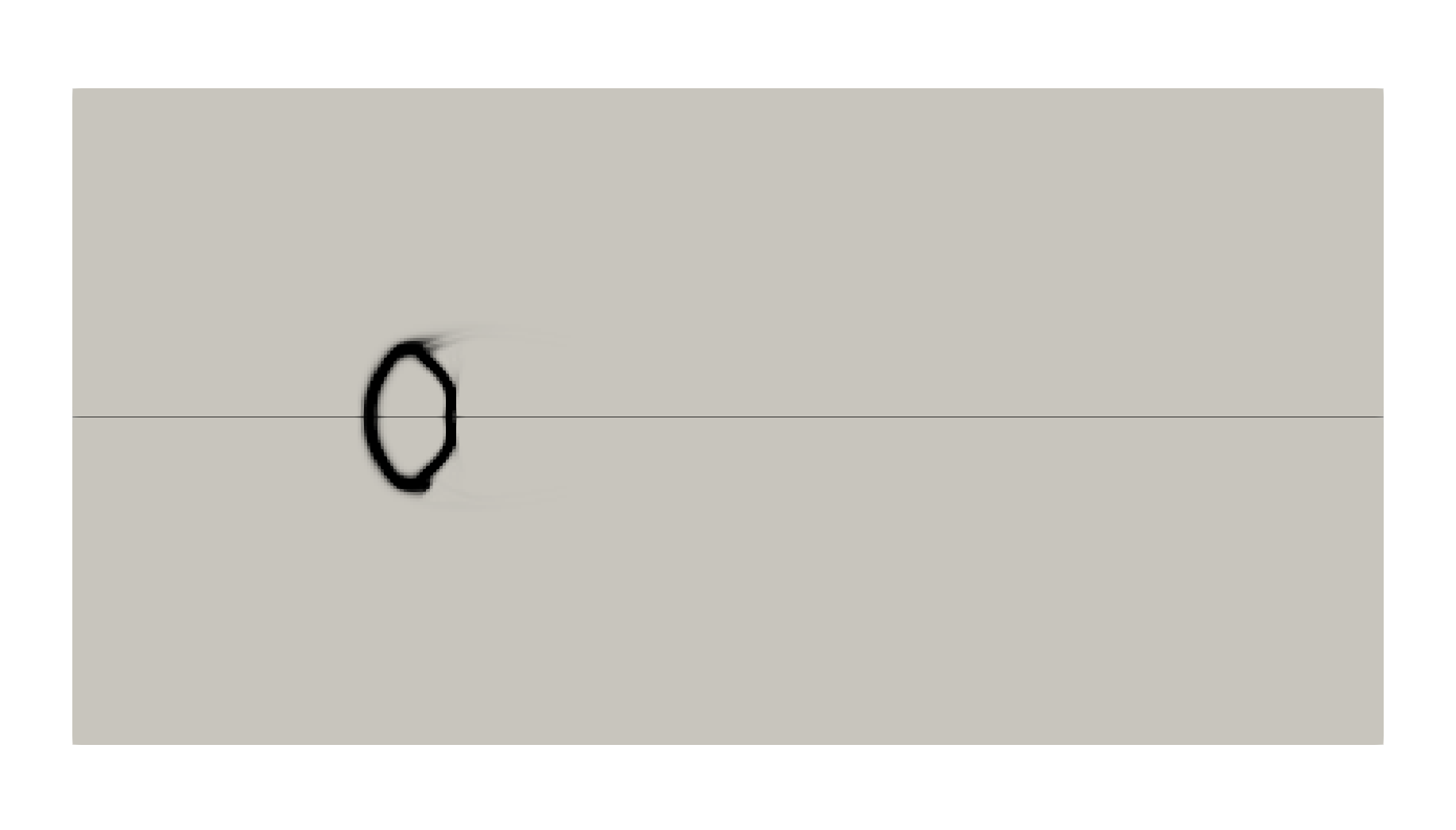}
    \adjincludegraphics[trim={{.2\width} {.25\height} 0 {.25\height}},clip,width=.4\linewidth]{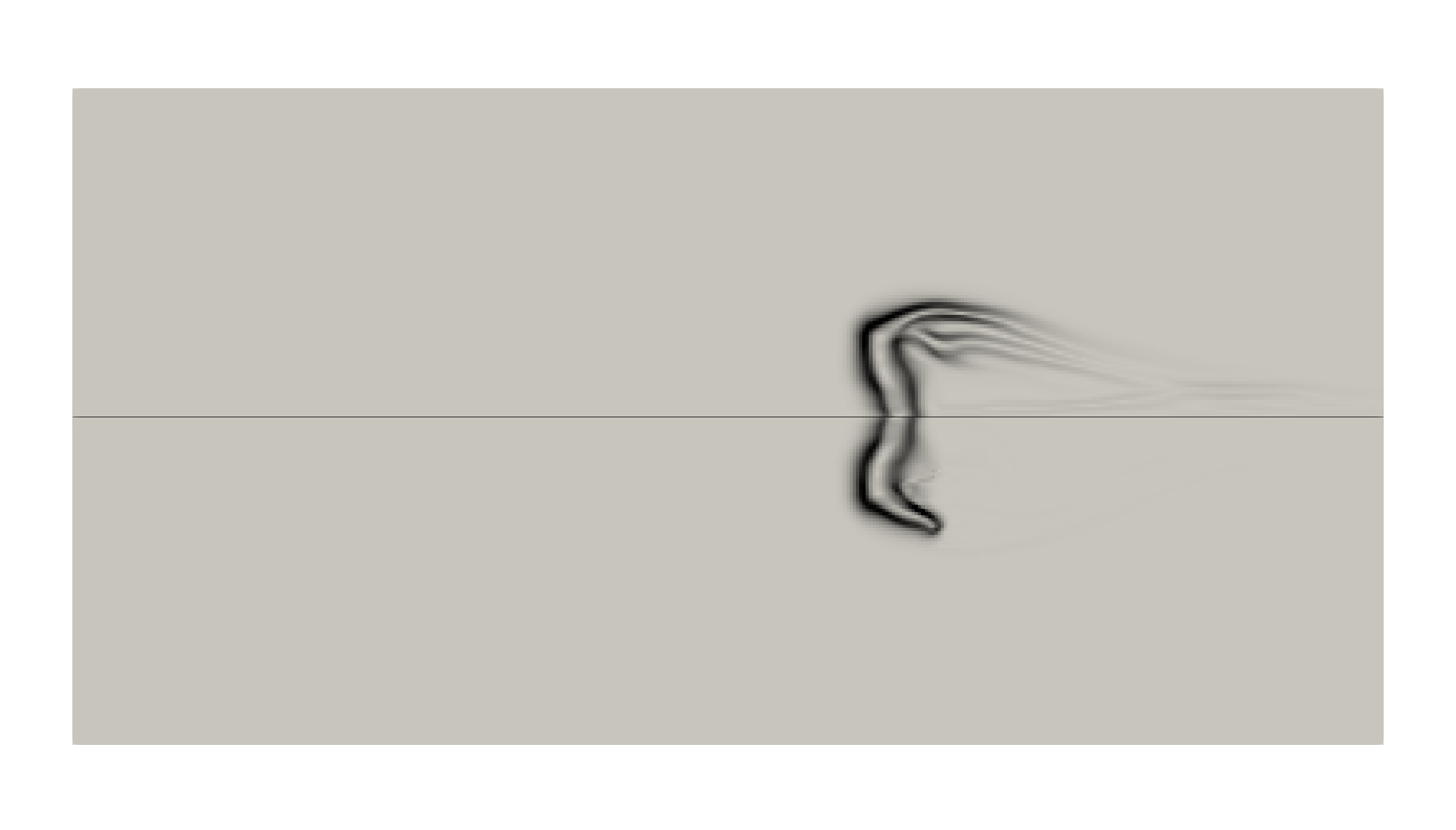}\\[5pt]
    \adjincludegraphics[trim={{.2\width} {.25\height} 0 {.25\height}},clip,width=.4\linewidth]{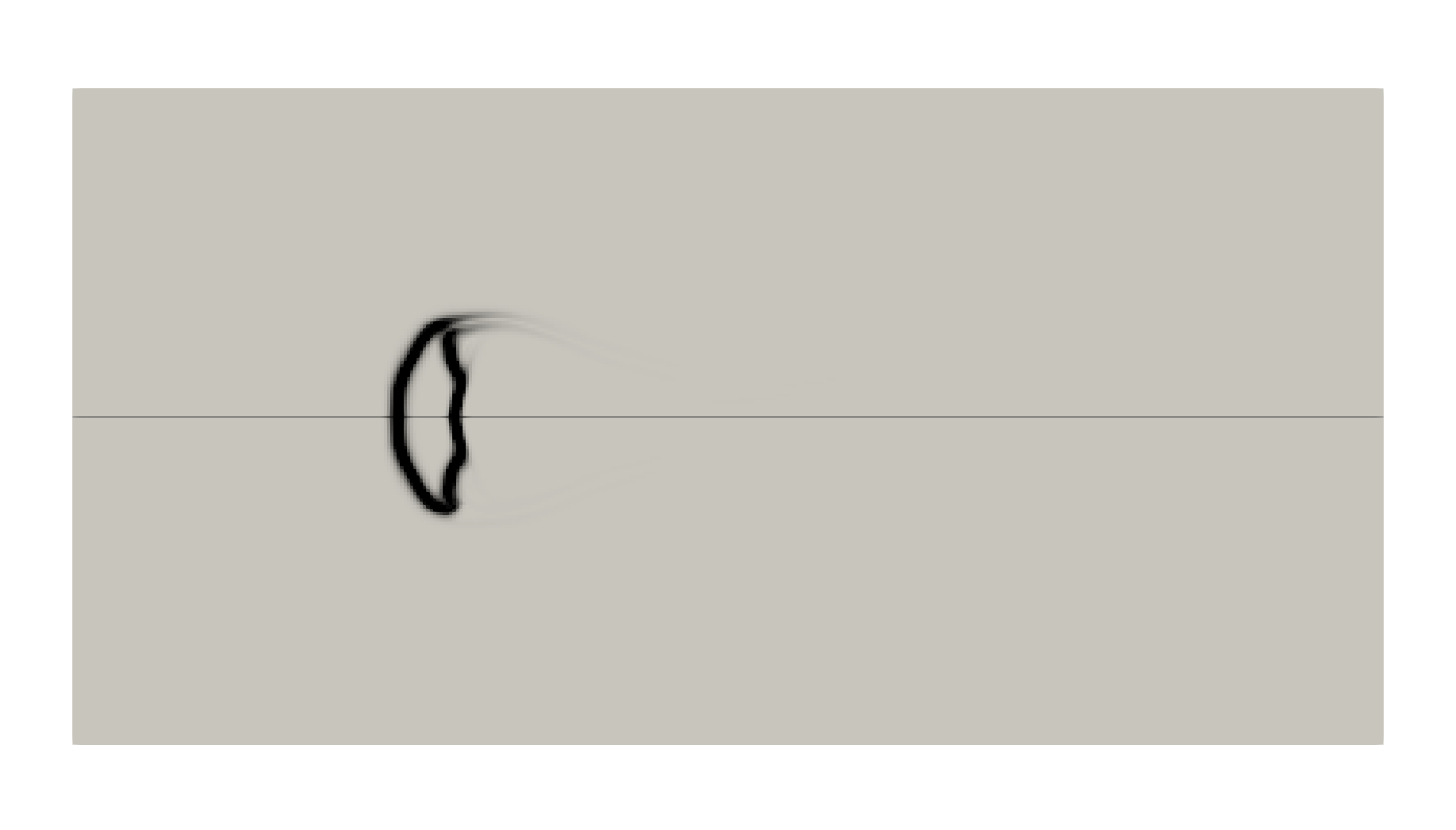}
    \adjincludegraphics[trim={{.2\width} {.25\height} 0 {.25\height}},clip,width=.4\linewidth]{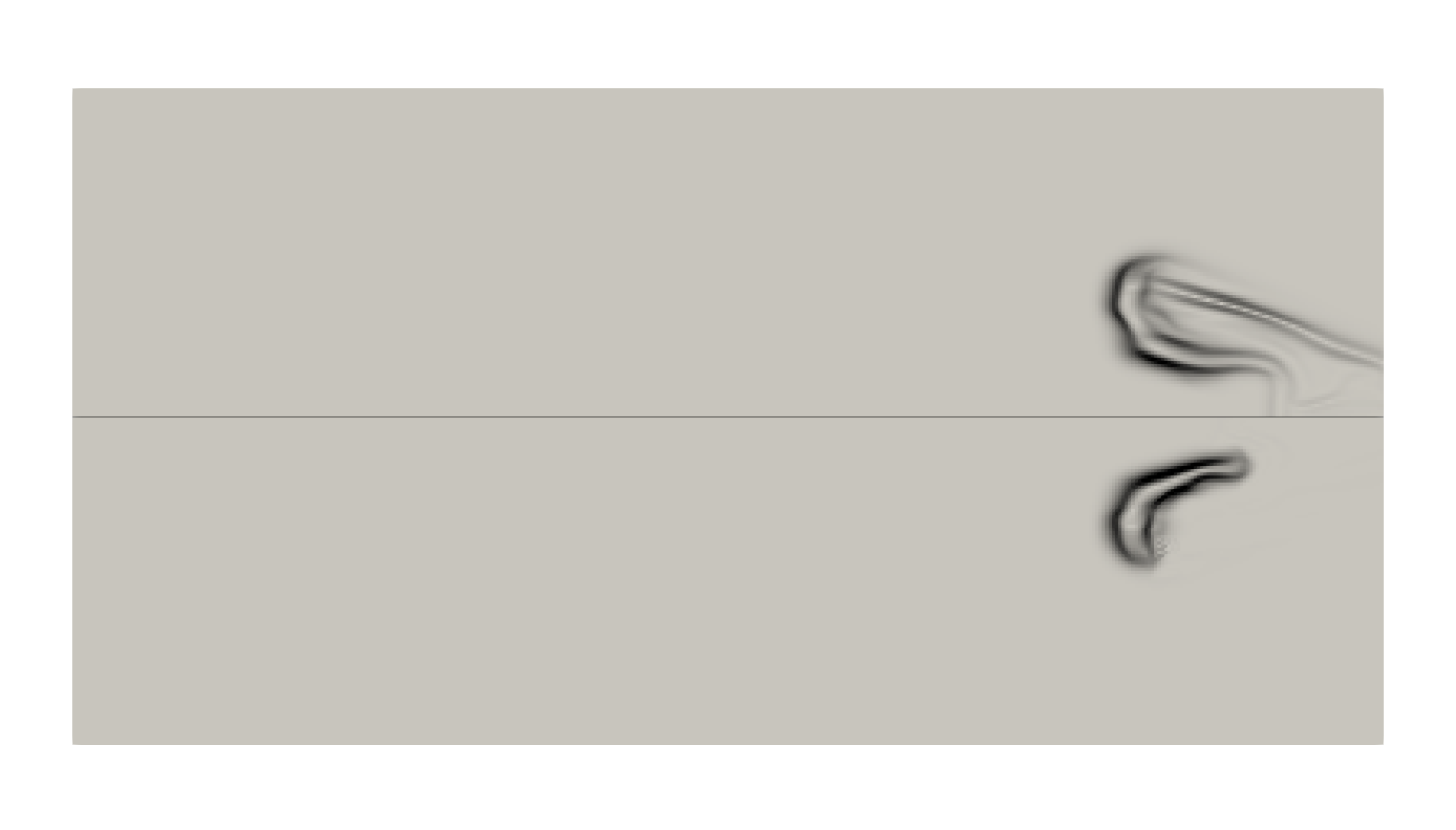}\\[5pt]
    \adjincludegraphics[trim={{.2\width} {.25\height} 0 {.25\height}},clip,width=.4\linewidth]{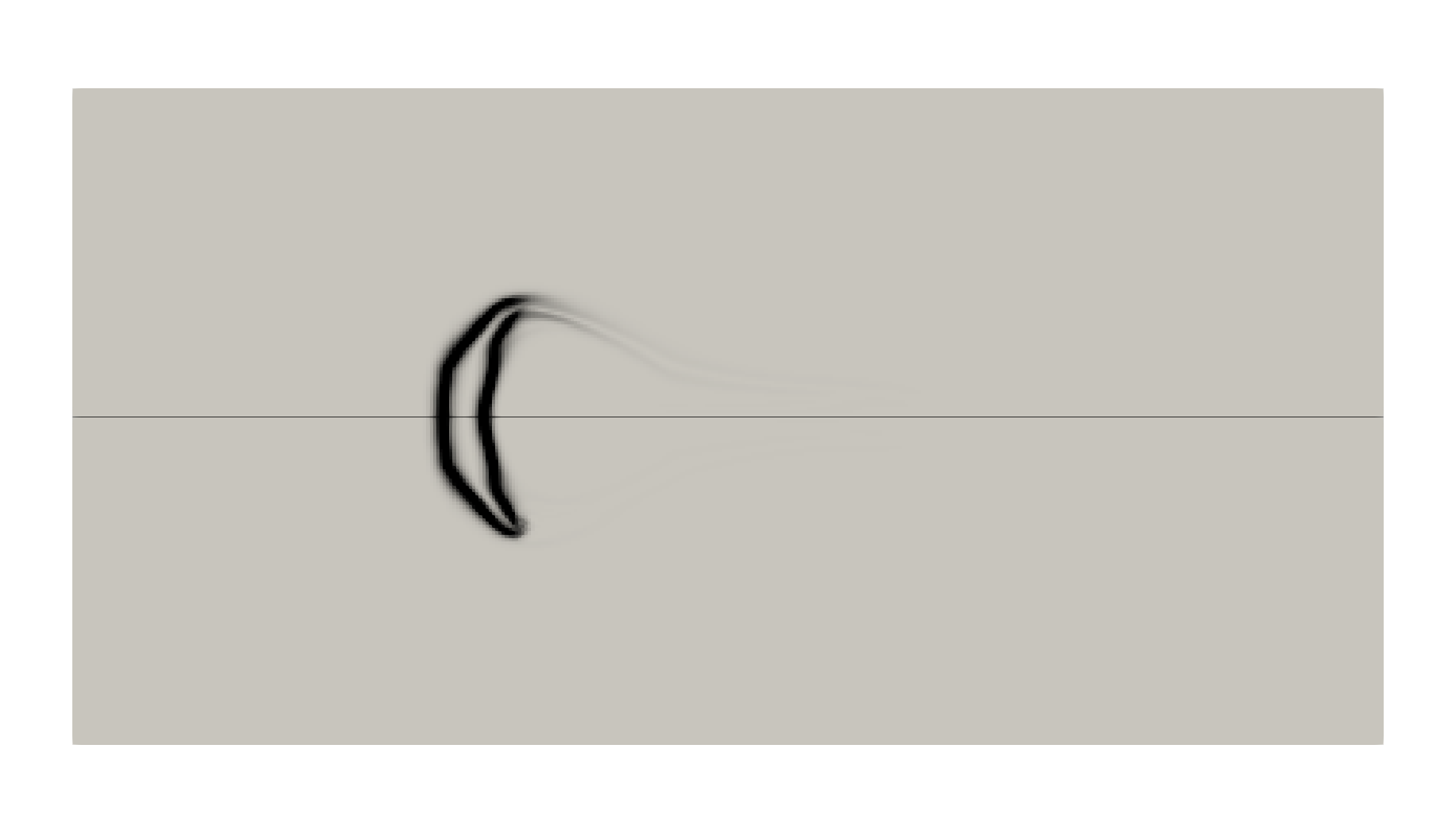}
    \adjincludegraphics[trim={{.2\width} {.25\height} 0 {.25\height}},clip,width=.4\linewidth]{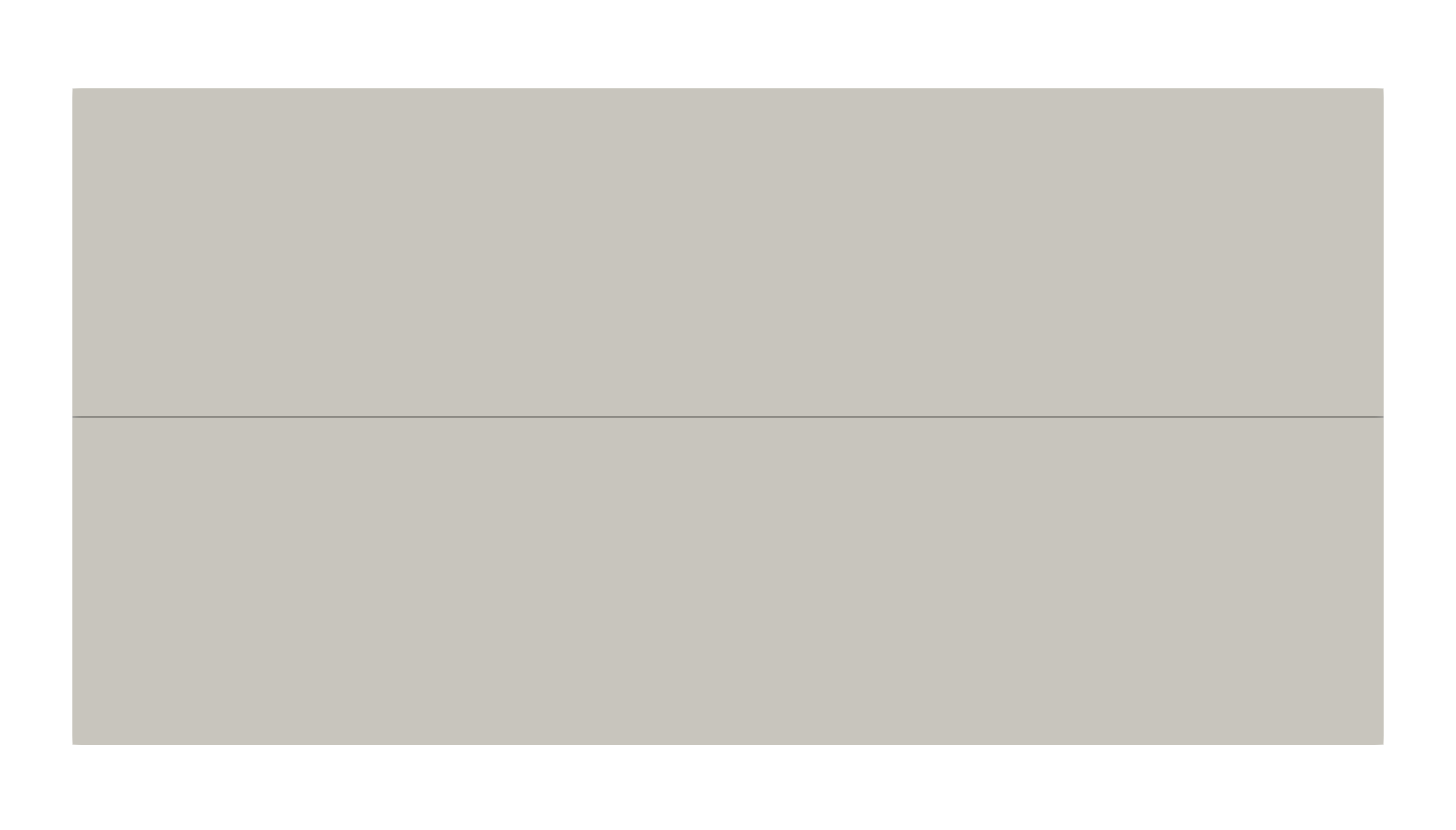}
    \caption{Large-scale IAD $\Vert\bnabla\overline{\alpha}_1\Vert\in(0,16)$
    {\protect\tikz \protect\node [rectangle, left color=paraviewwhite, right color=black, anchor=north, minimum width=1cm, minimum height=0.1cm] (box) at (0,0){};}
    without mass transfer (top) and 
    with mass transfer (bottom). Snapshots are taken each $0.25$ s from $t=0$ s to $t=2.5$ s from top to bottom and left to right.}
    \label{fig:compare-IAD}
\end{figure}

\begin{figure}[!ht]
    \centering
    \begin{tikzpicture}
        \centering
        \begin{axis}[
                width=.7\textwidth,
                height=5cm,
                xmin=0,
                xmax=2.5,
                ymin=0,
                ymax=200,
                xlabel={$t$},
                ylabel={$H_{lig}$},
            ]
            \addplot[blue] table [x={time}, y={Hmax}] {no-transfer.txt};
            \addplot[red] table [x={time}, y={Hmax}] {40-transfer.txt};
            \addplot[mark=none, black, samples=2, dashed] {40};
        \end{axis}
    \end{tikzpicture}
    \caption{Evolution in time of the mean curvature $H_{lig}$ when inter-scale transfer is either activated (red) or deactivated (blue). Threshold $H_{max}=40$ m${}^{-1}$ is represented with the black dashed line.}
    \label{fig:H-evo-compare}
\end{figure}
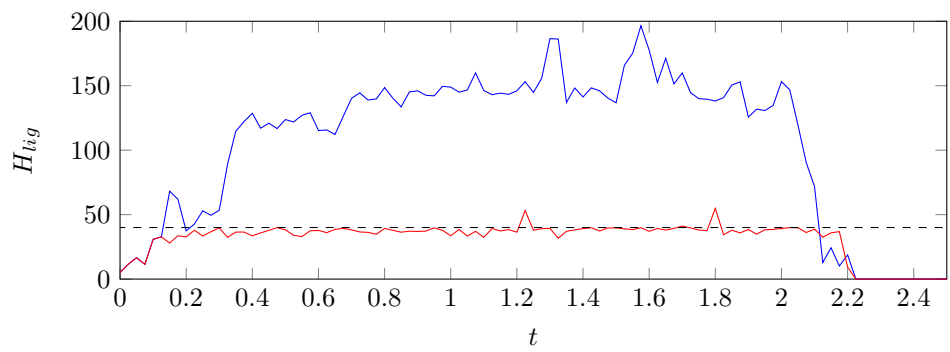

\begin{figure}
    \centering
    \includegraphics[width=.4\linewidth]{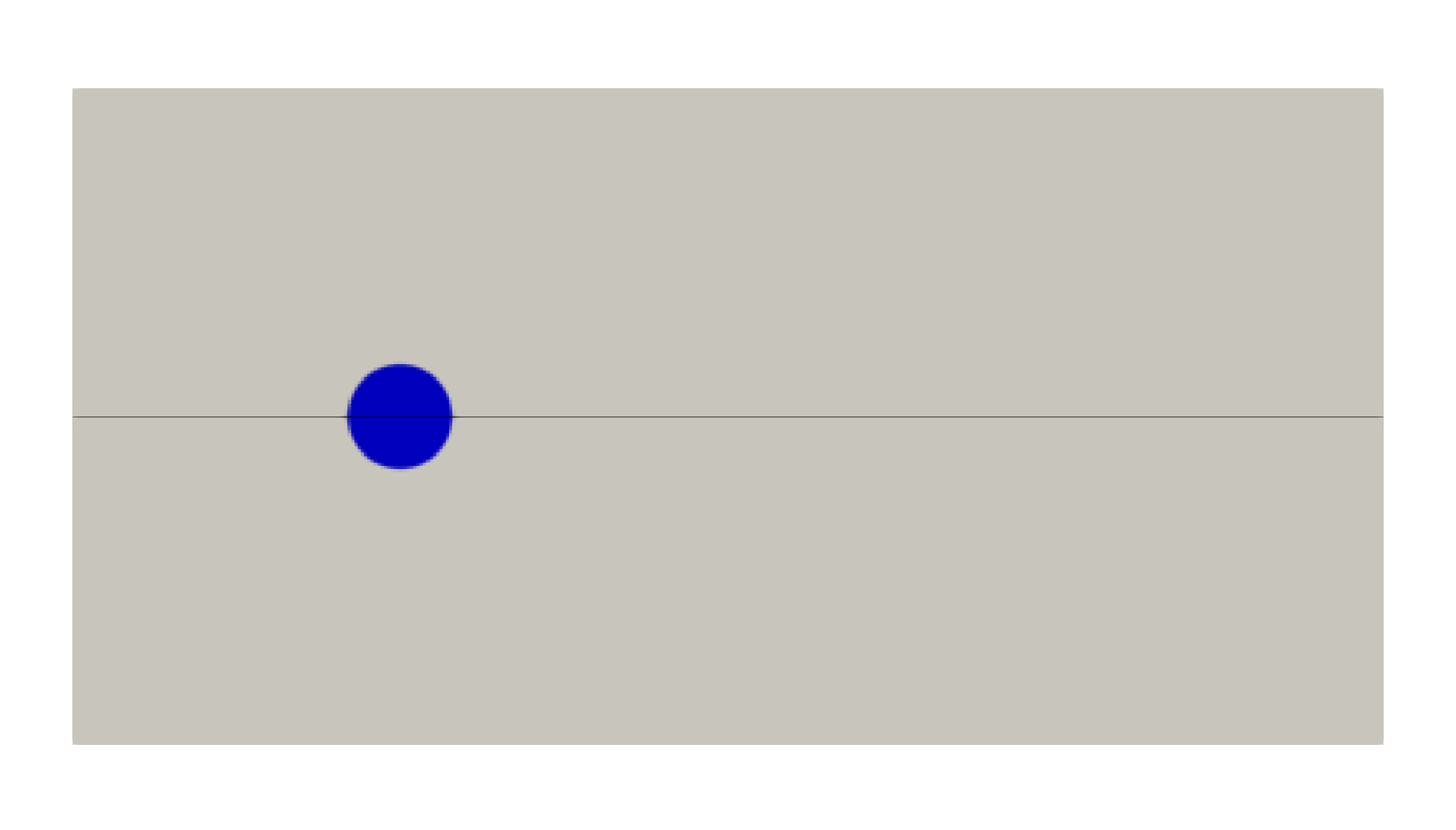}
    \includegraphics[width=.4\linewidth]{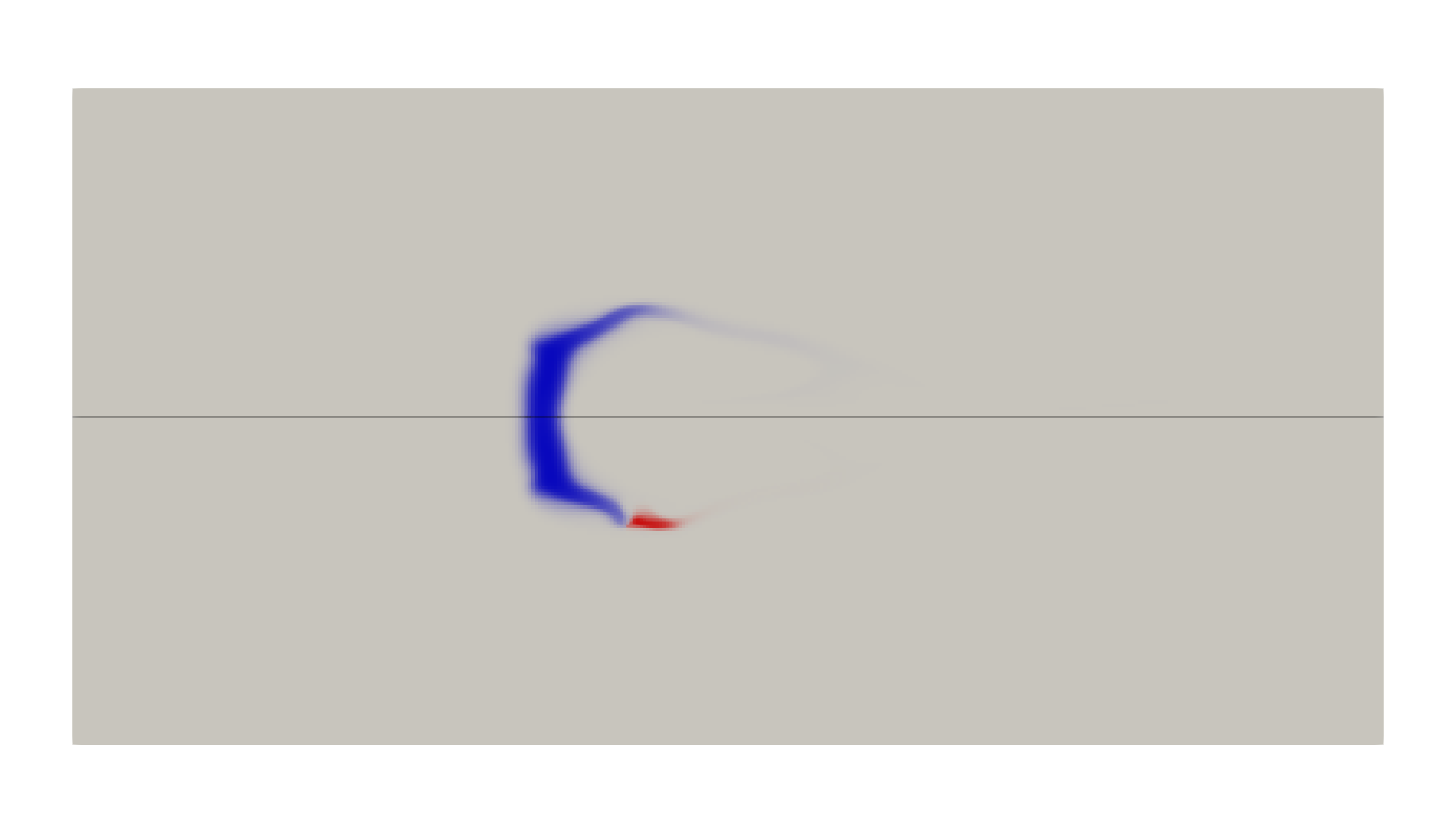}\\
    \includegraphics[width=.4\linewidth]{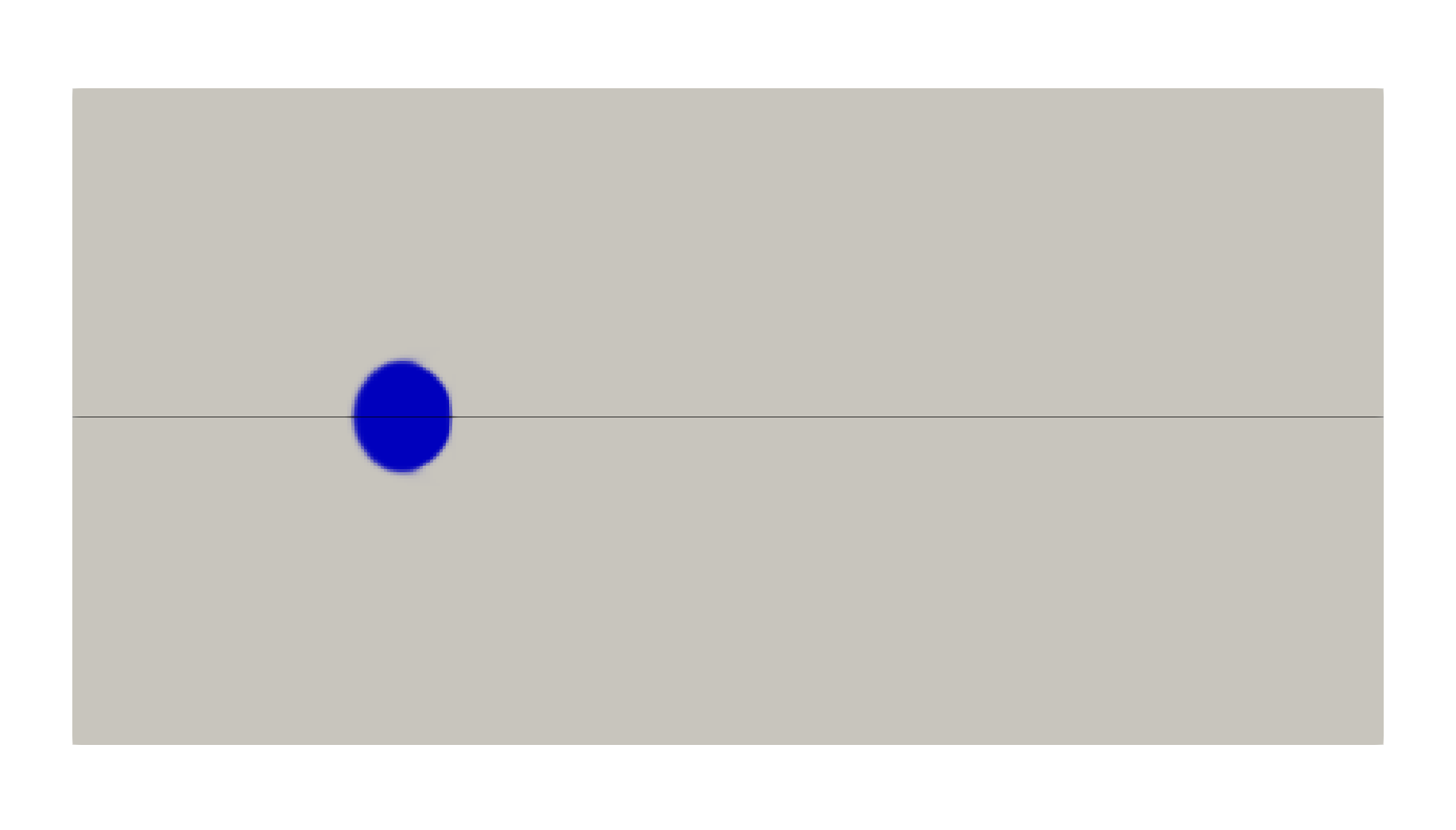}
    \includegraphics[width=.4\linewidth]{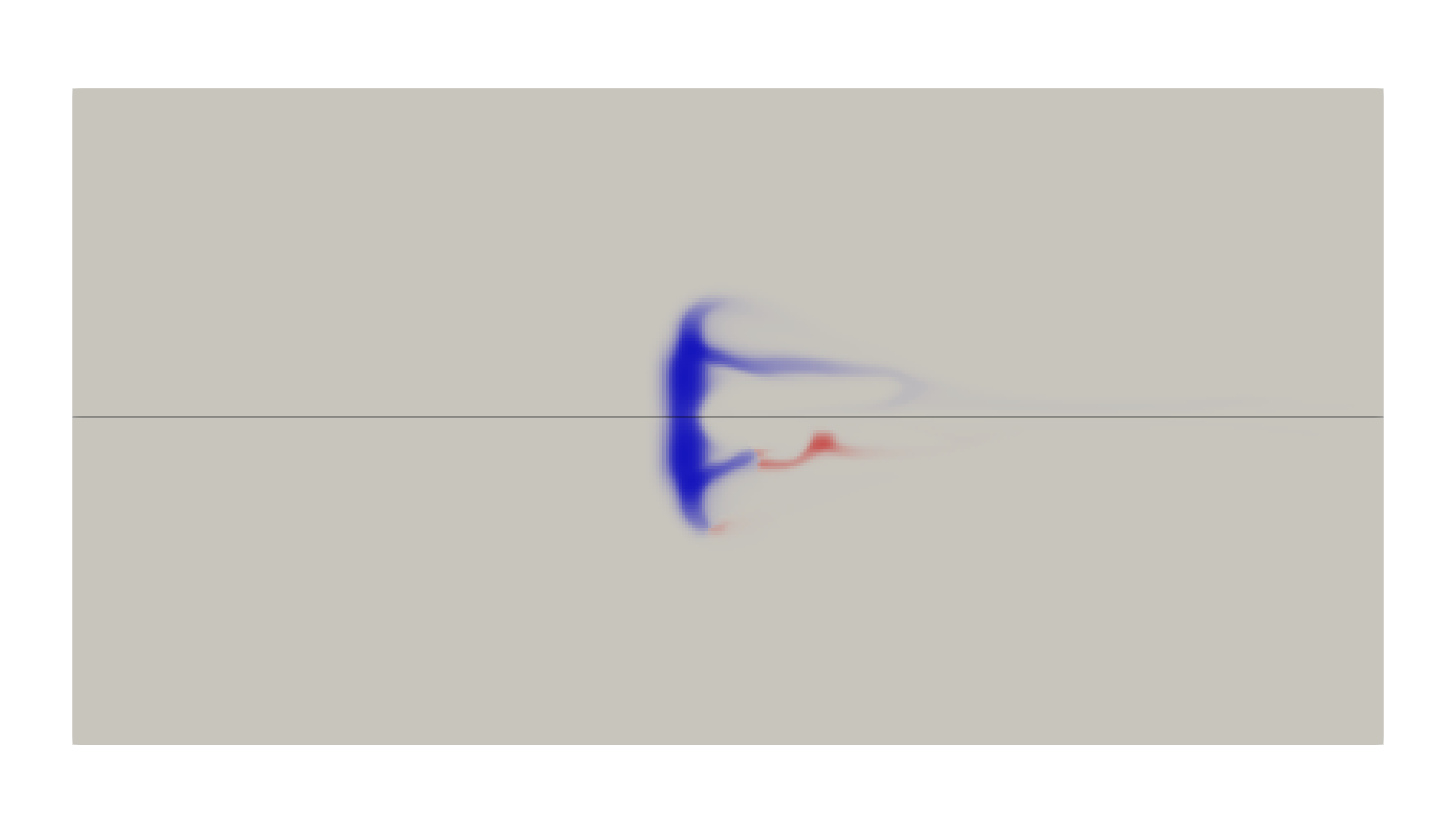}\\
    \includegraphics[width=.4\linewidth]{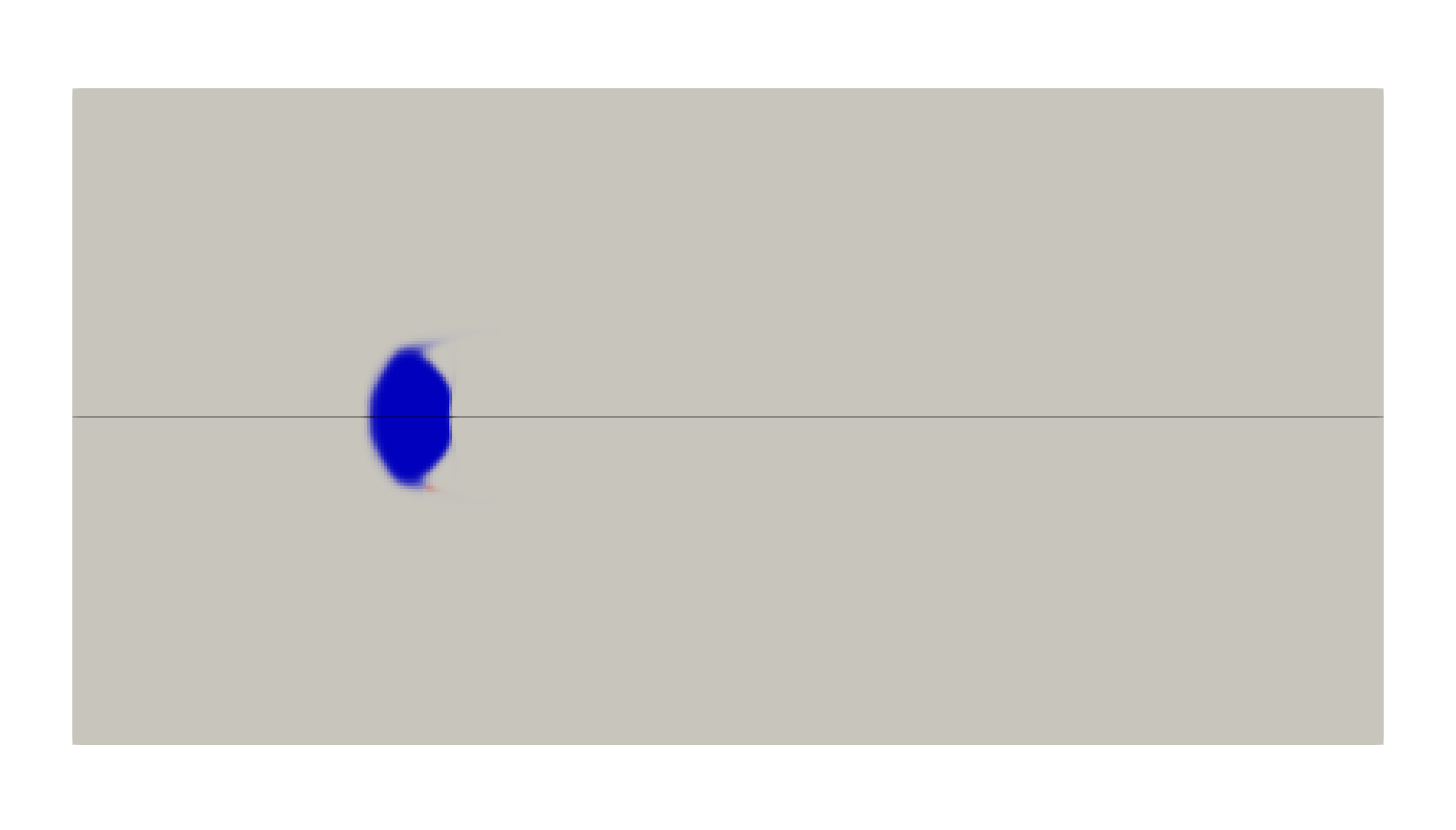}
    \includegraphics[width=.4\linewidth]{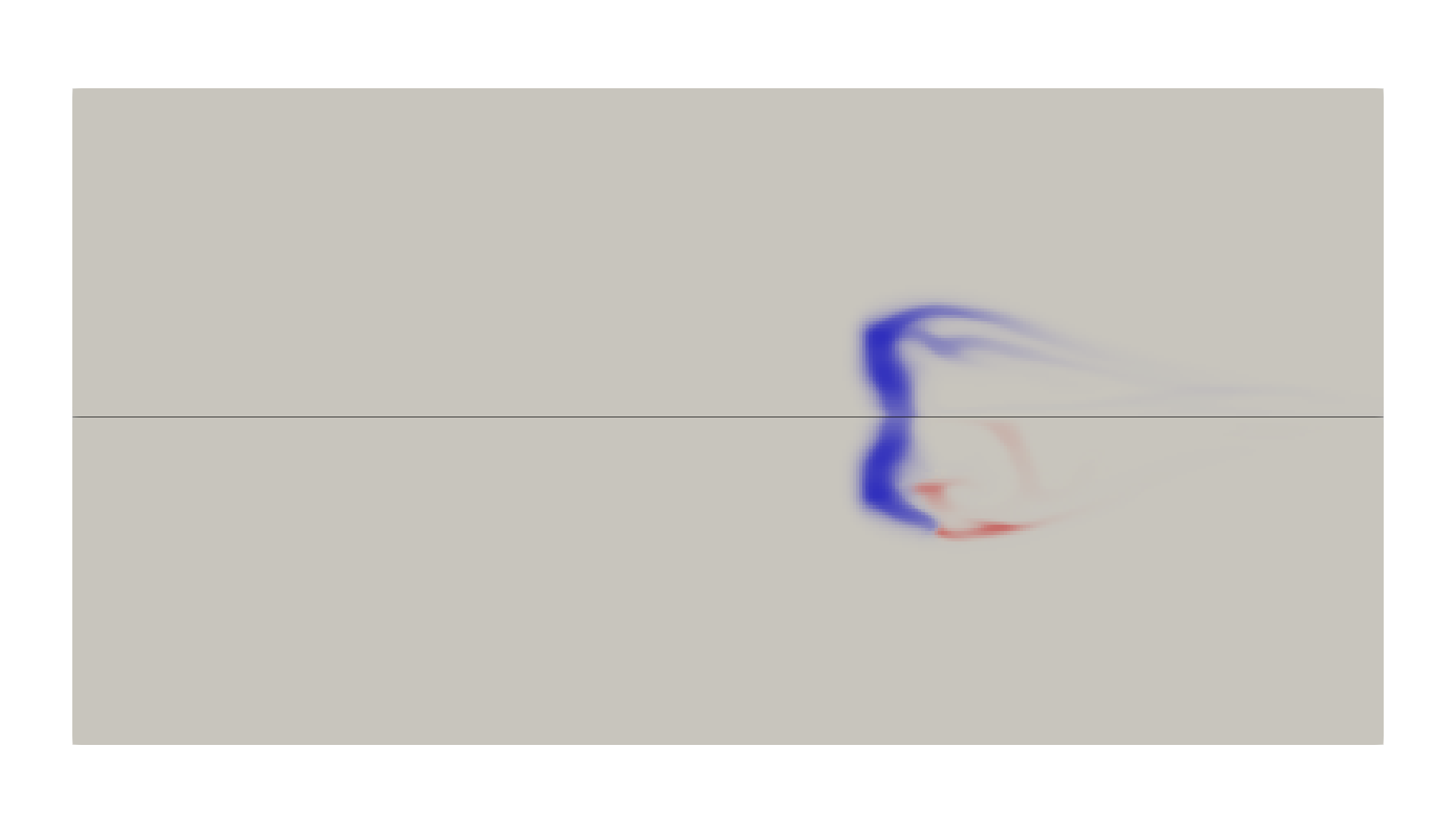}\\
    \includegraphics[width=.4\linewidth]{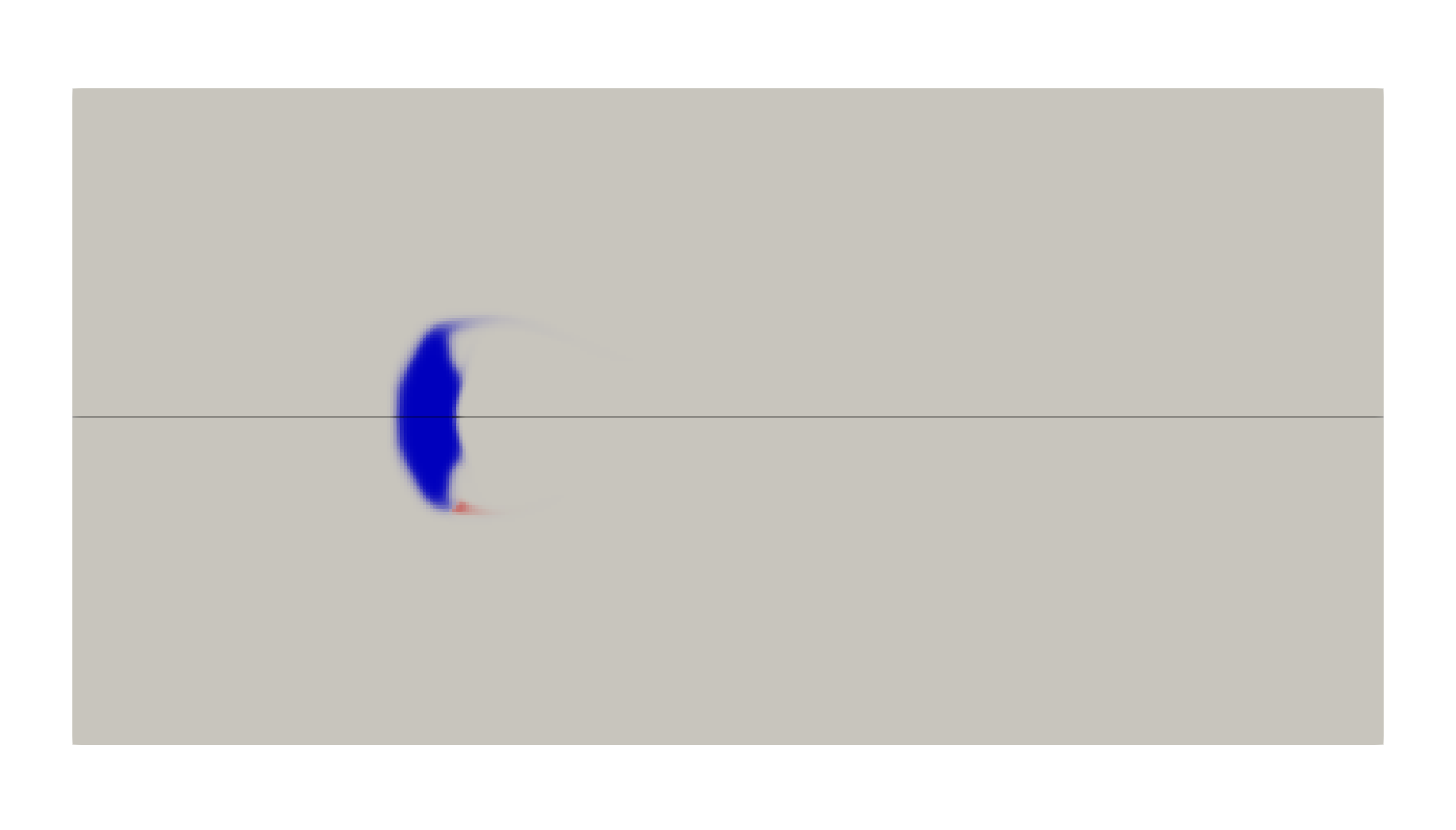}
    \includegraphics[width=.4\linewidth]{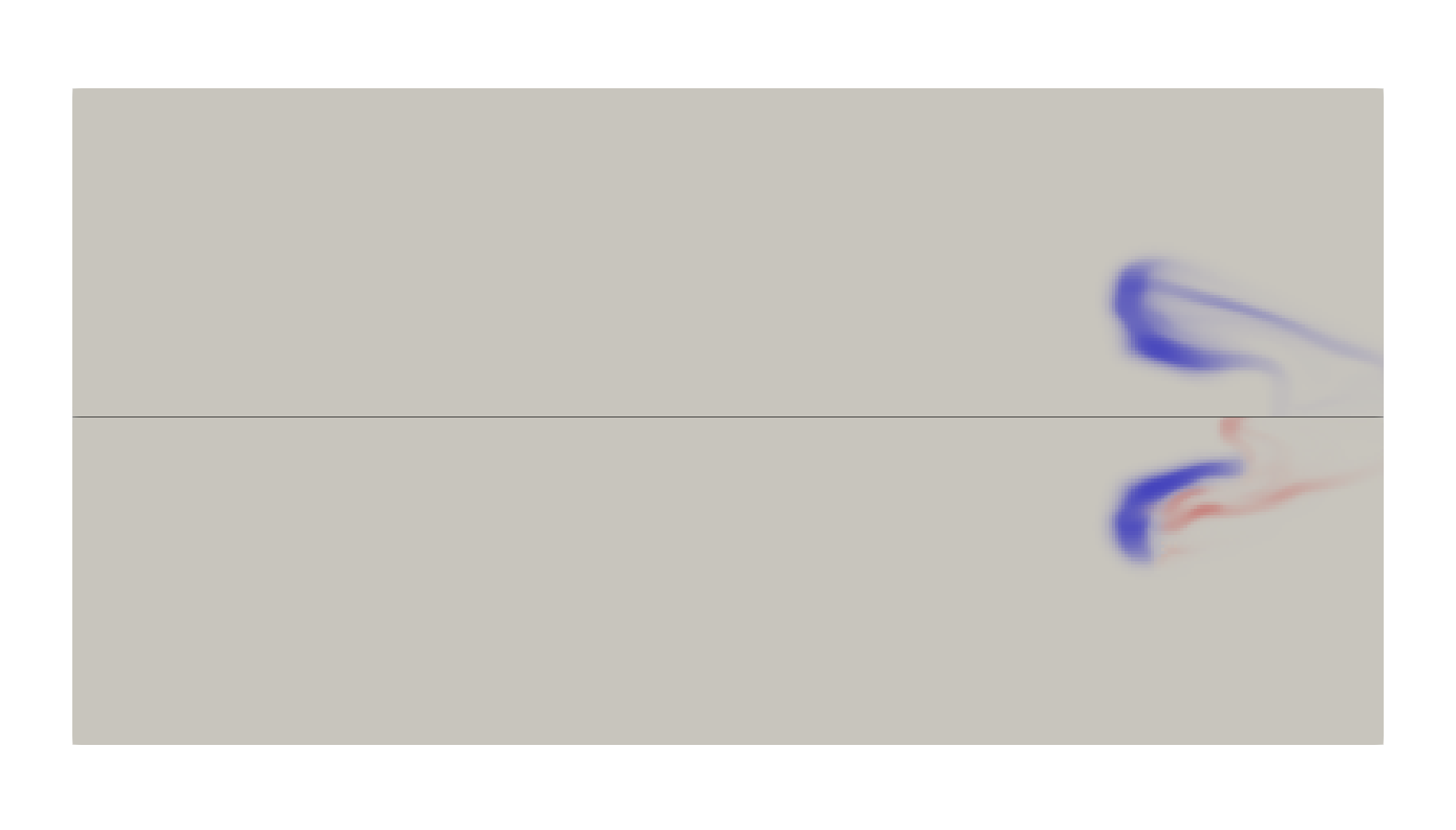}\\
    \includegraphics[width=.4\linewidth]{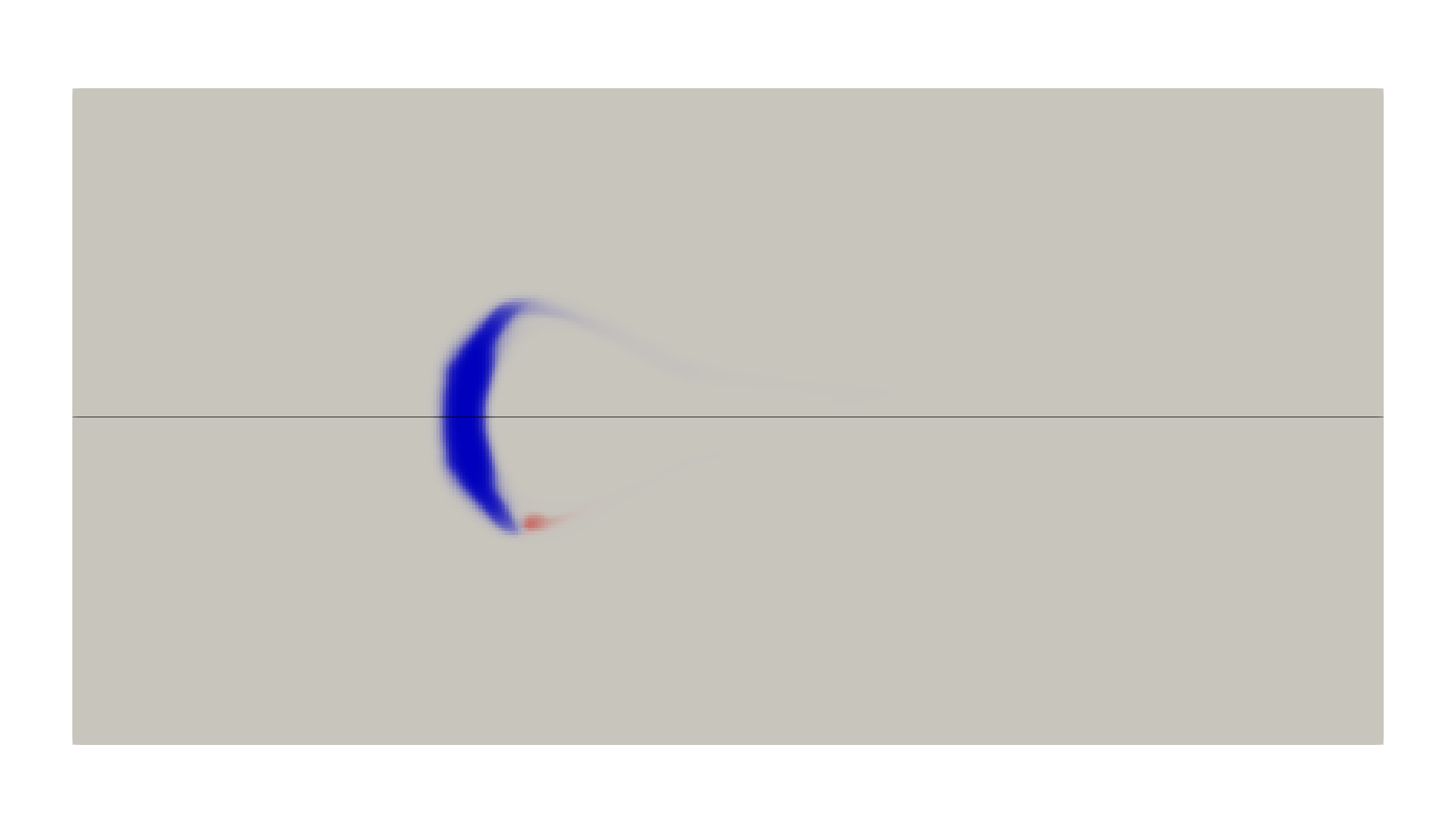}
    \includegraphics[width=.4\linewidth]{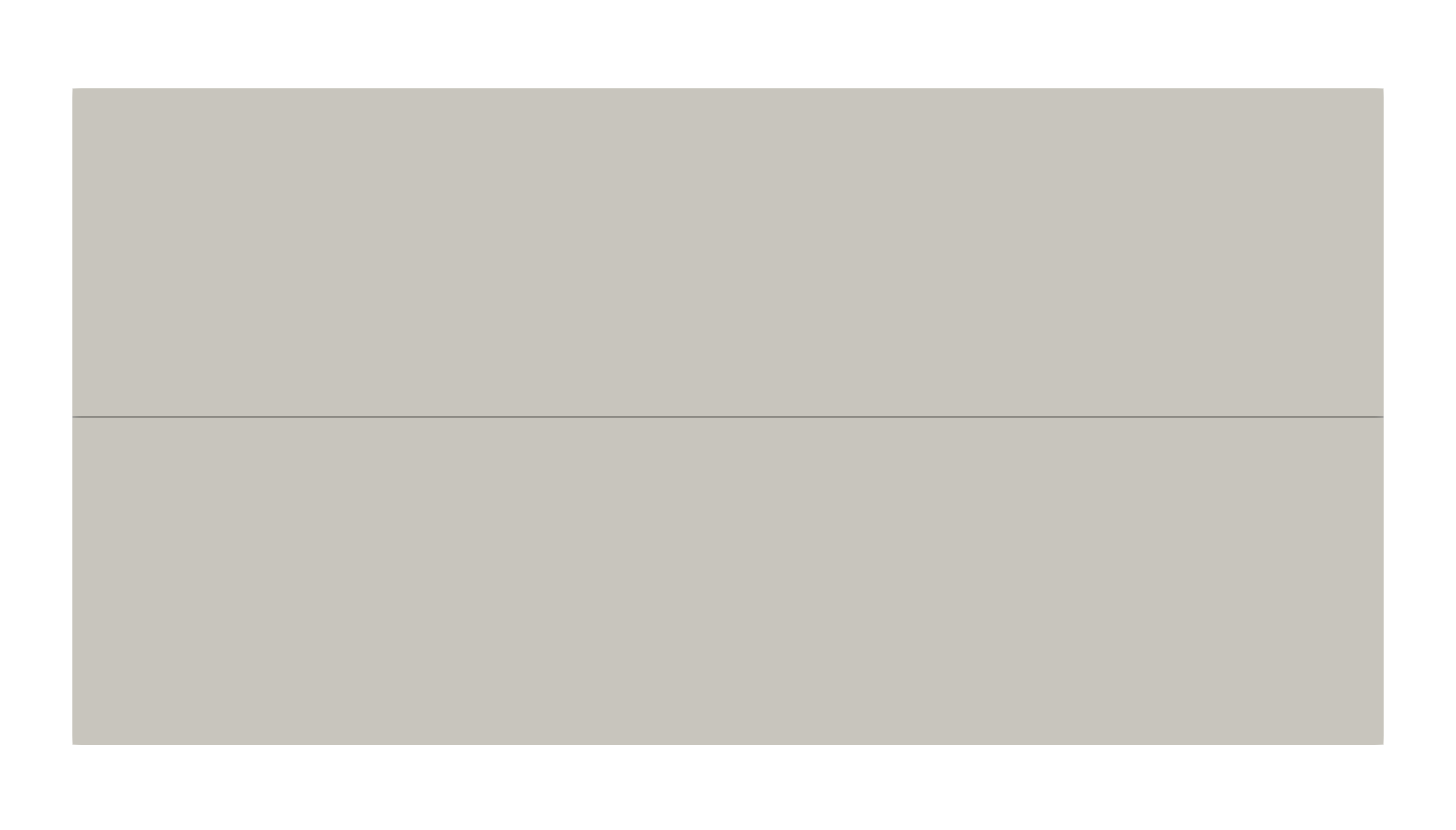}
    \caption{Large-scale liquid effective density $\alpha_1\rho_1\in(0,10^3)$ without inter-scale transfer
    {\protect\tikz \protect\node [rectangle, left color=paraviewwhite, right color=black, anchor=north, minimum width=1cm, minimum height=0.1cm] (box) at (0,0){};} (top),
    with inter-scale transfer
    {\protect\tikz \protect\node [rectangle, left color=paraviewwhite, right color=blue, anchor=north, minimum width=1cm, minimum height=0.1cm] (box) at (0,0){};} (bottom),
    and small-scale liquid effective density $\alpha_1^d\rho_1^d\in(0,3.8\times 10^2)$
    with inter-scale transfer
    {\protect\tikz \protect\node [rectangle, left color=paraviewwhite, right color=red, anchor=north, minimum width=1cm, minimum height=0.1cm] (box) at (0,0){};}
    (bottom).
    Snapshots are taken each $0.25$ s from $t=0$ s to $t=2.5$ s from top to bottom and left to right.}
    \label{fig:compare-mass}
\end{figure}

\begin{figure}[!ht]
    \centering
    \includegraphics[width=.4\linewidth]{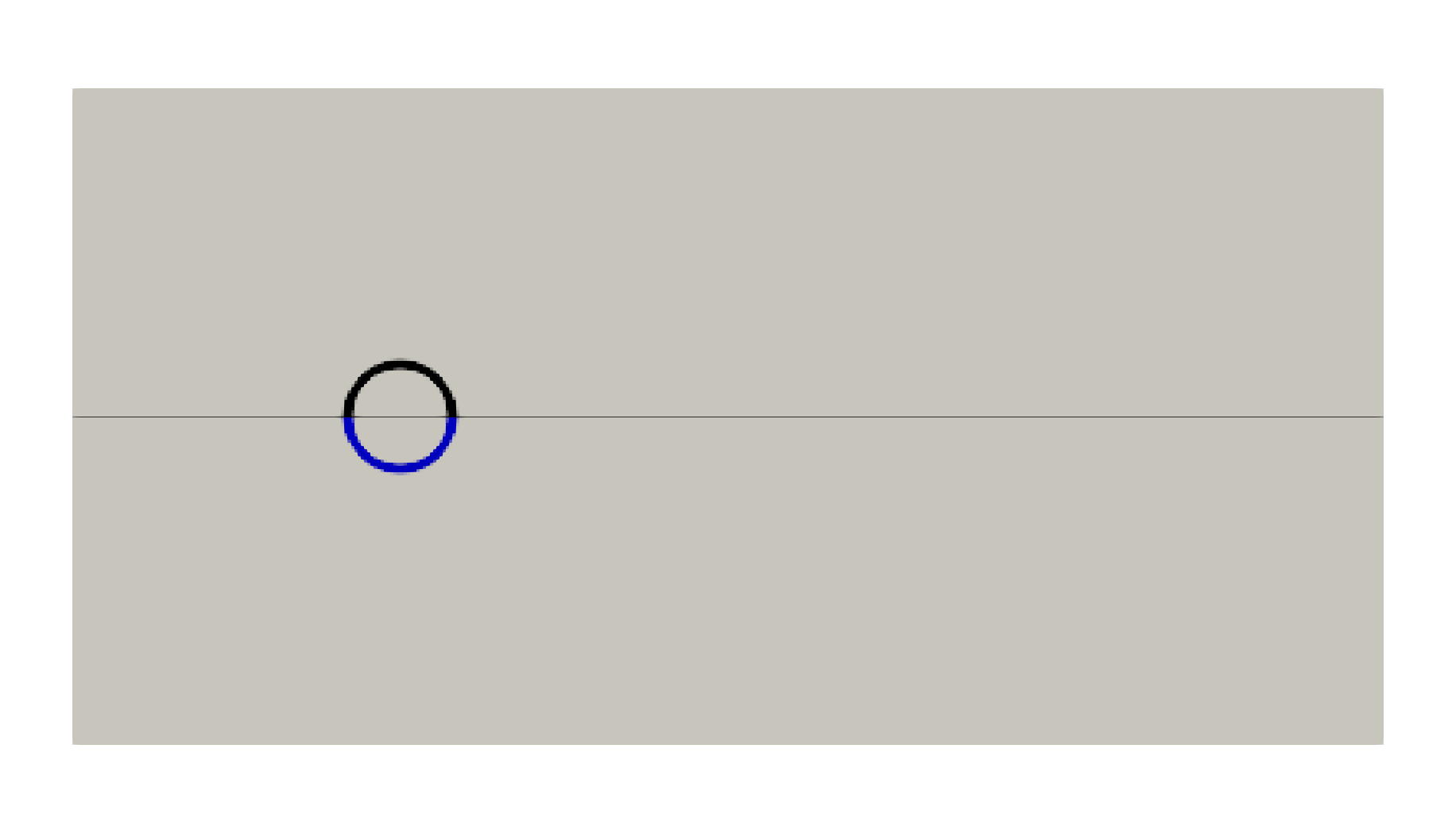}
    \includegraphics[width=.4\linewidth]{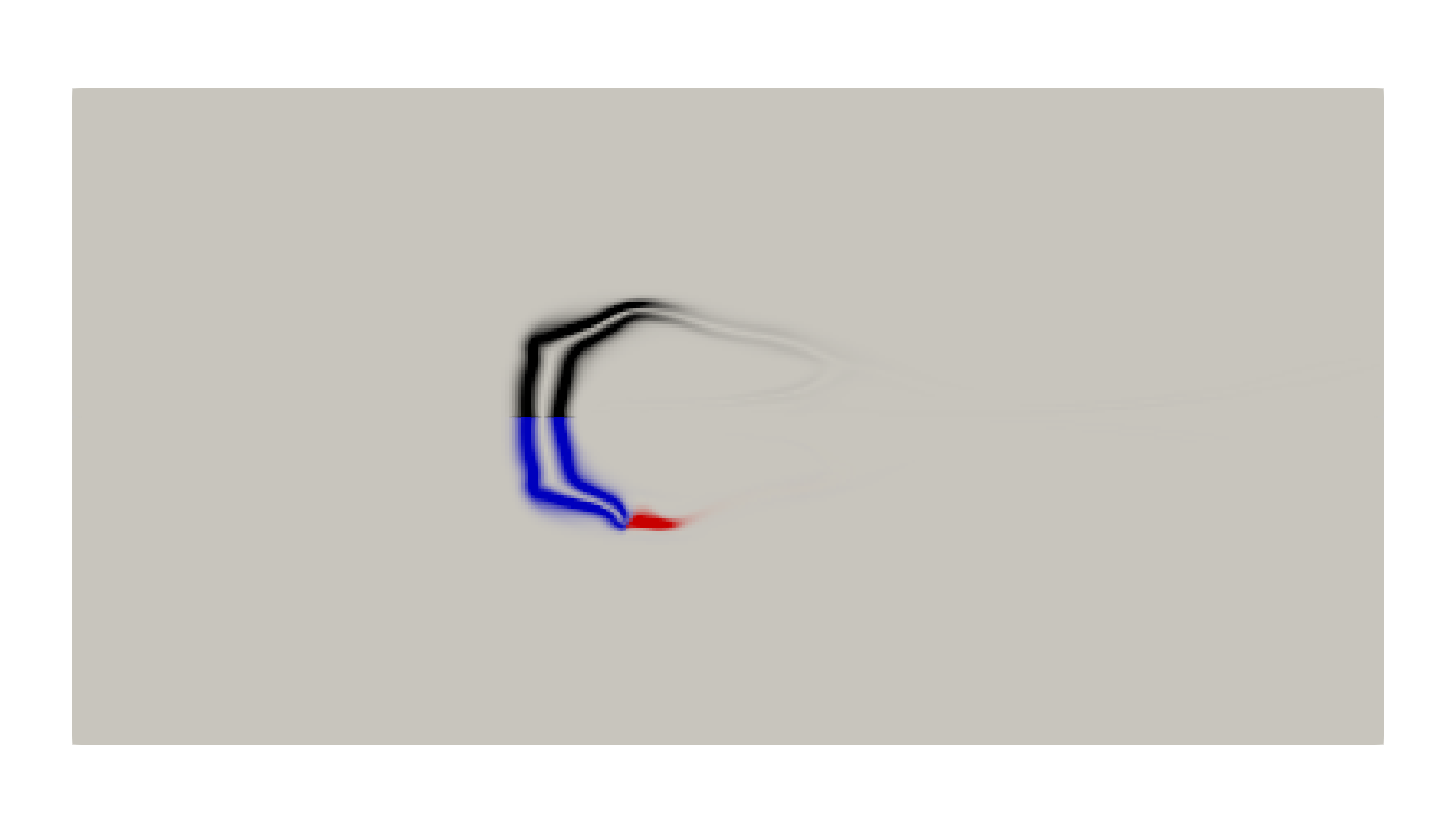}\\
    \includegraphics[width=.4\linewidth]{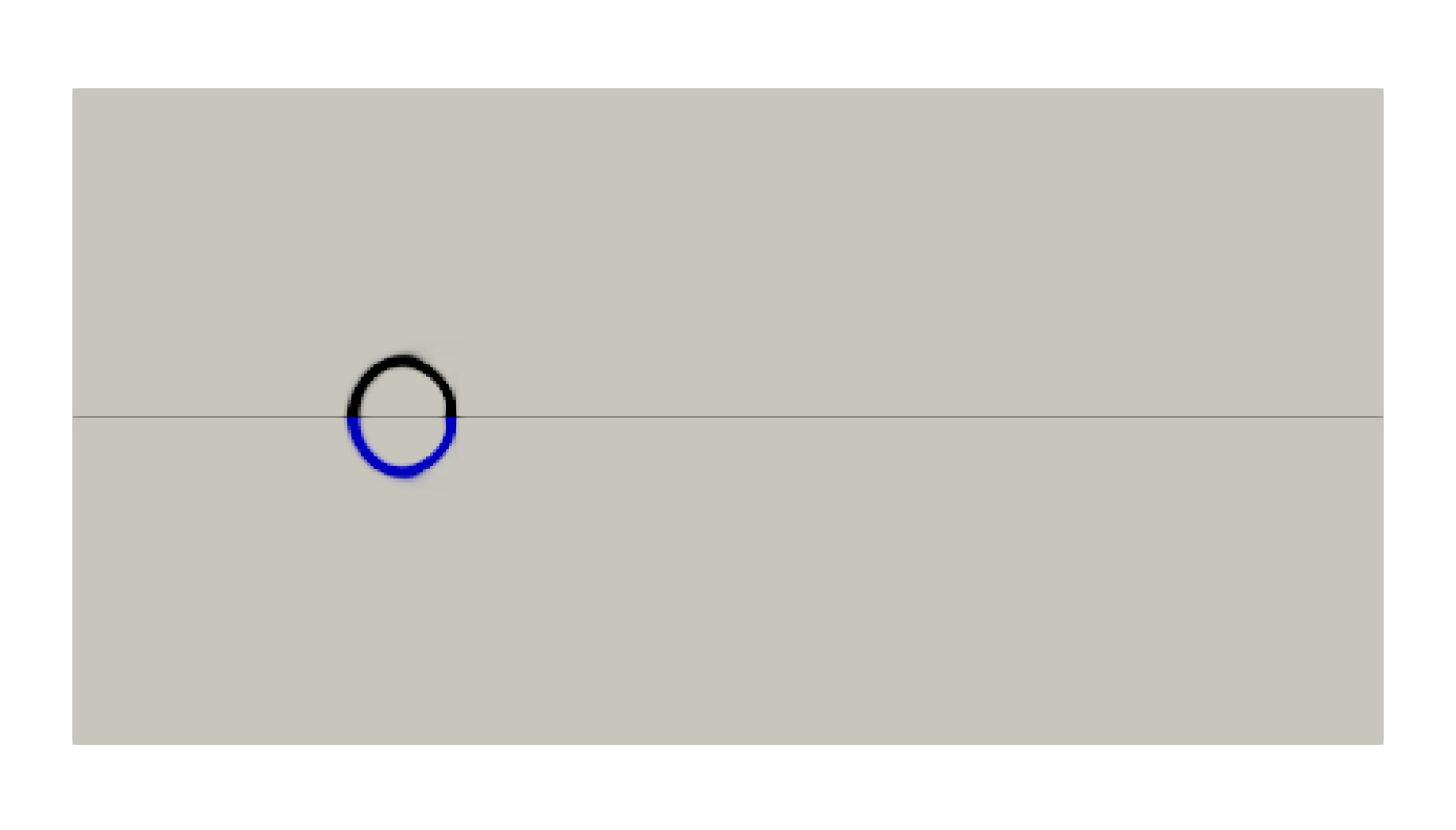}
    \includegraphics[width=.4\linewidth]{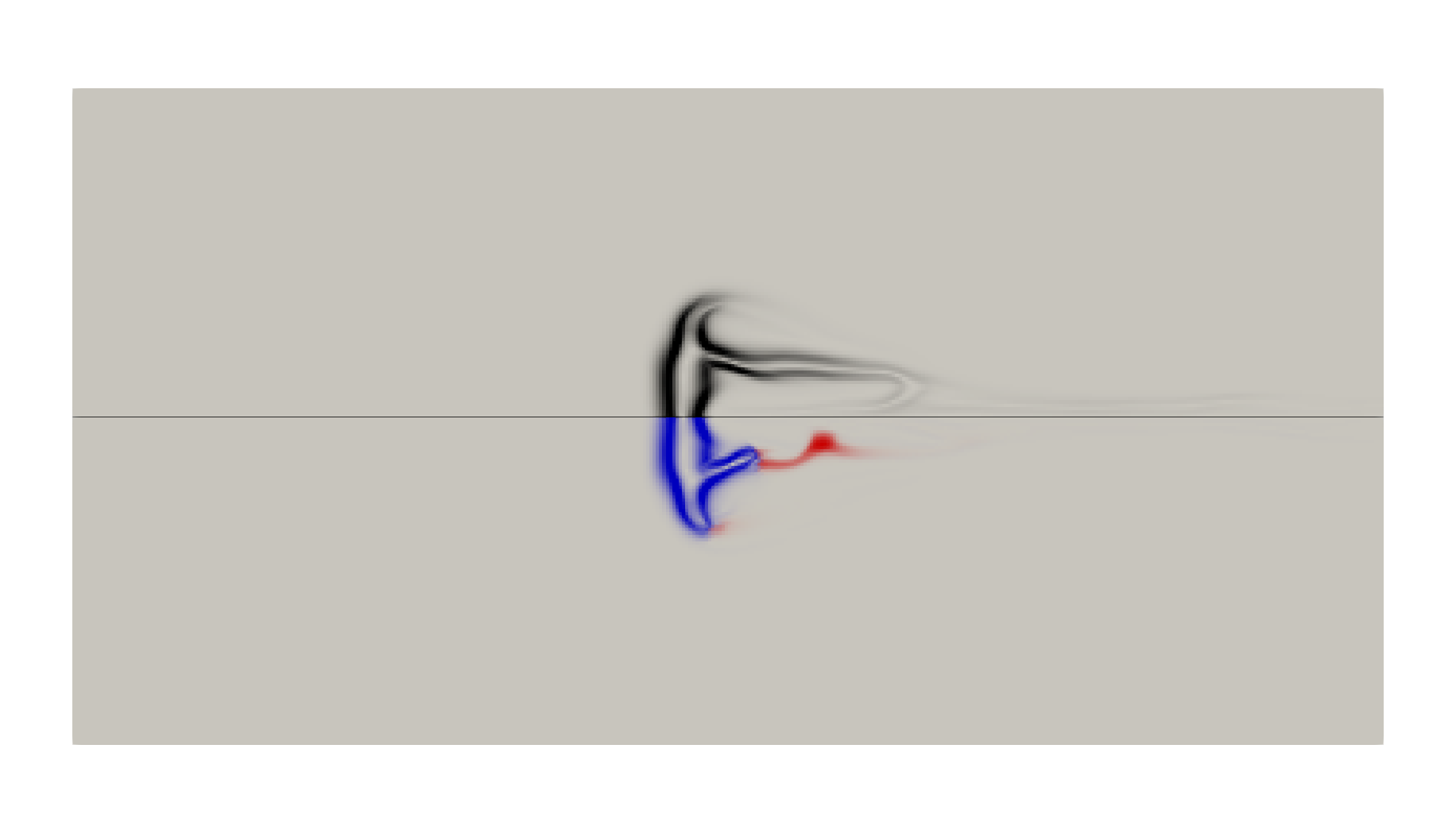}\\
    \includegraphics[width=.4\linewidth]{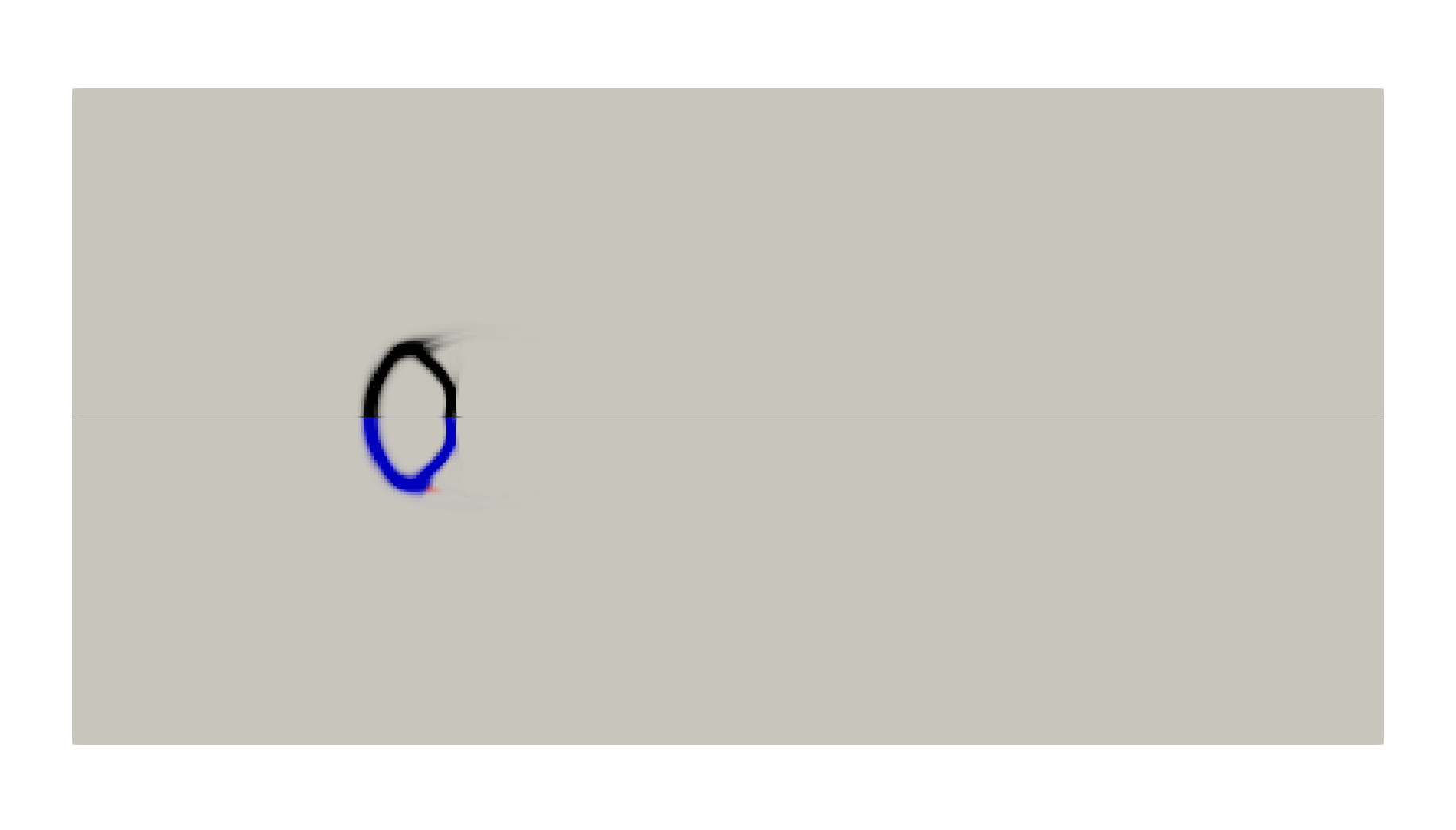}
    \includegraphics[width=.4\linewidth]{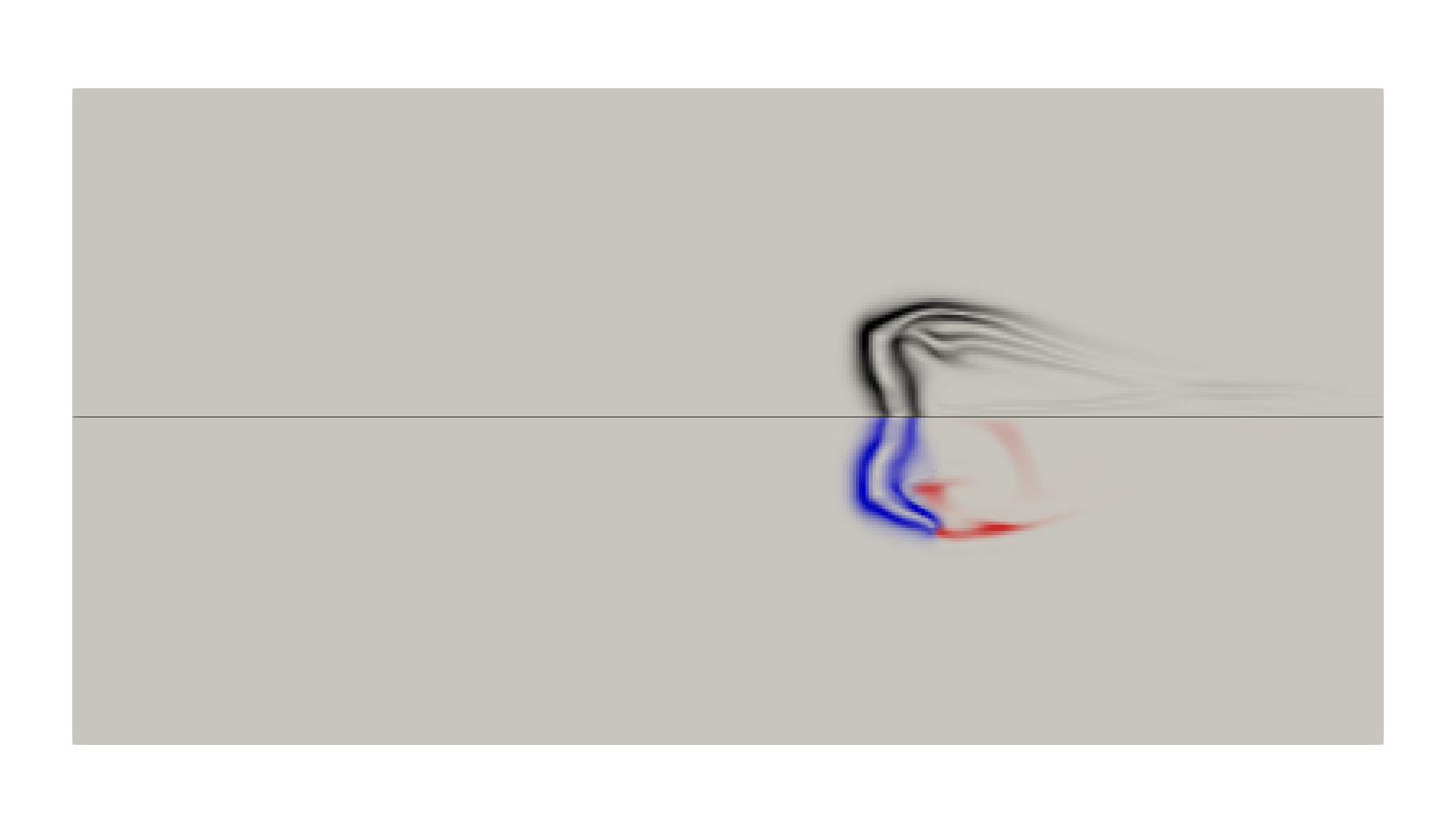}\\
    \includegraphics[width=.4\linewidth]{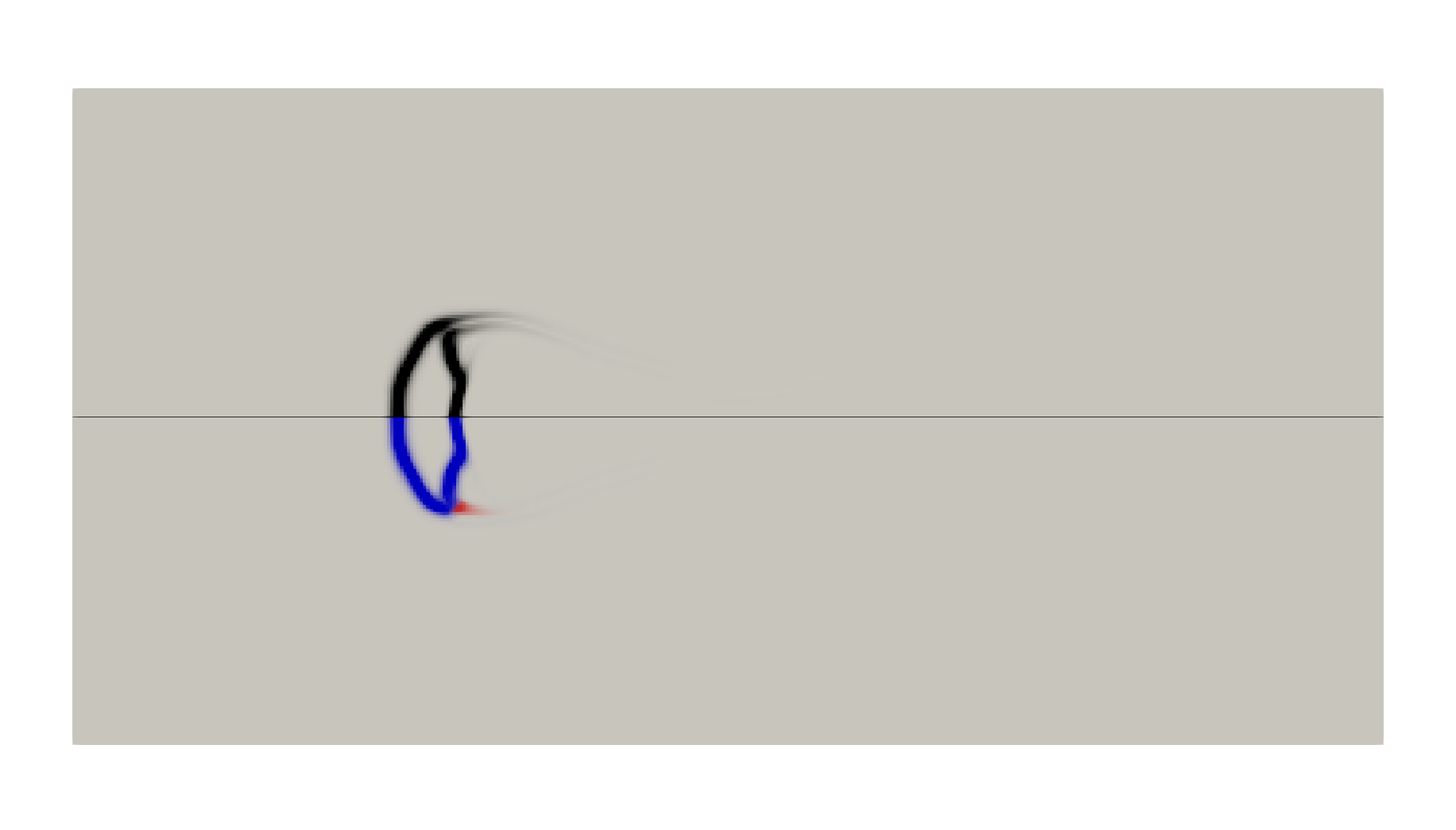}
    \includegraphics[width=.4\linewidth]{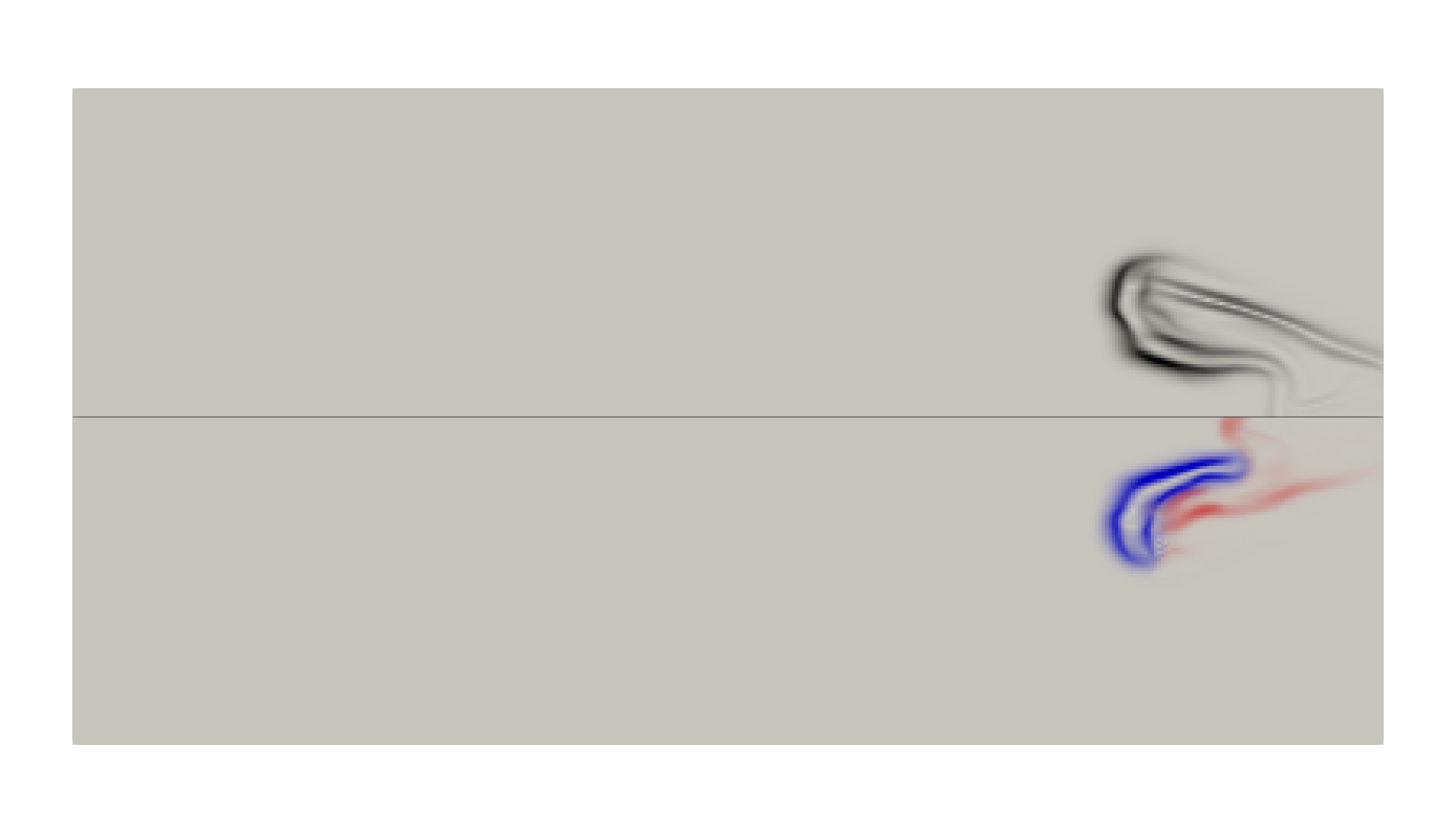}\\
    \includegraphics[width=.4\linewidth]{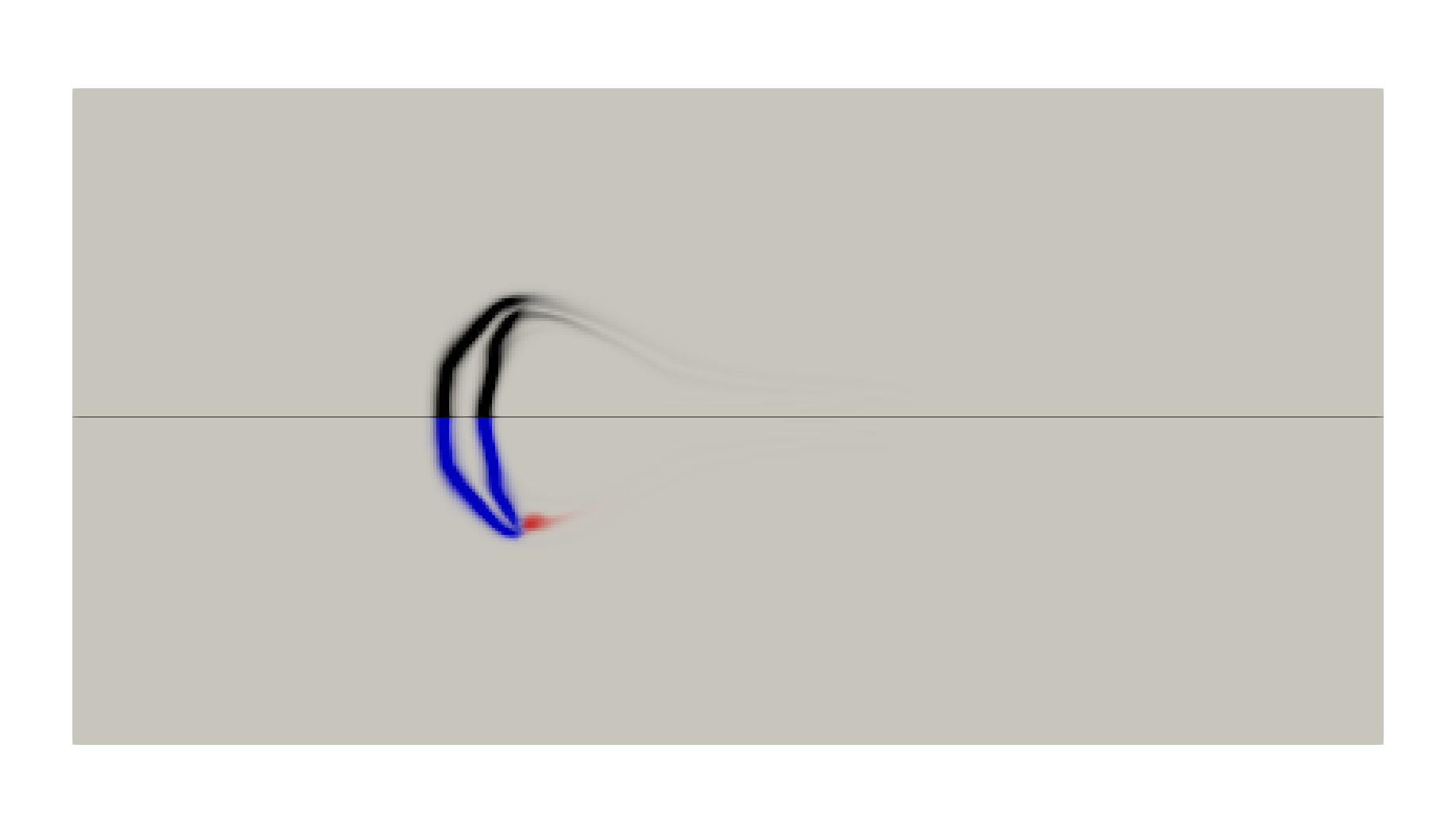}
    \includegraphics[width=.4\linewidth]{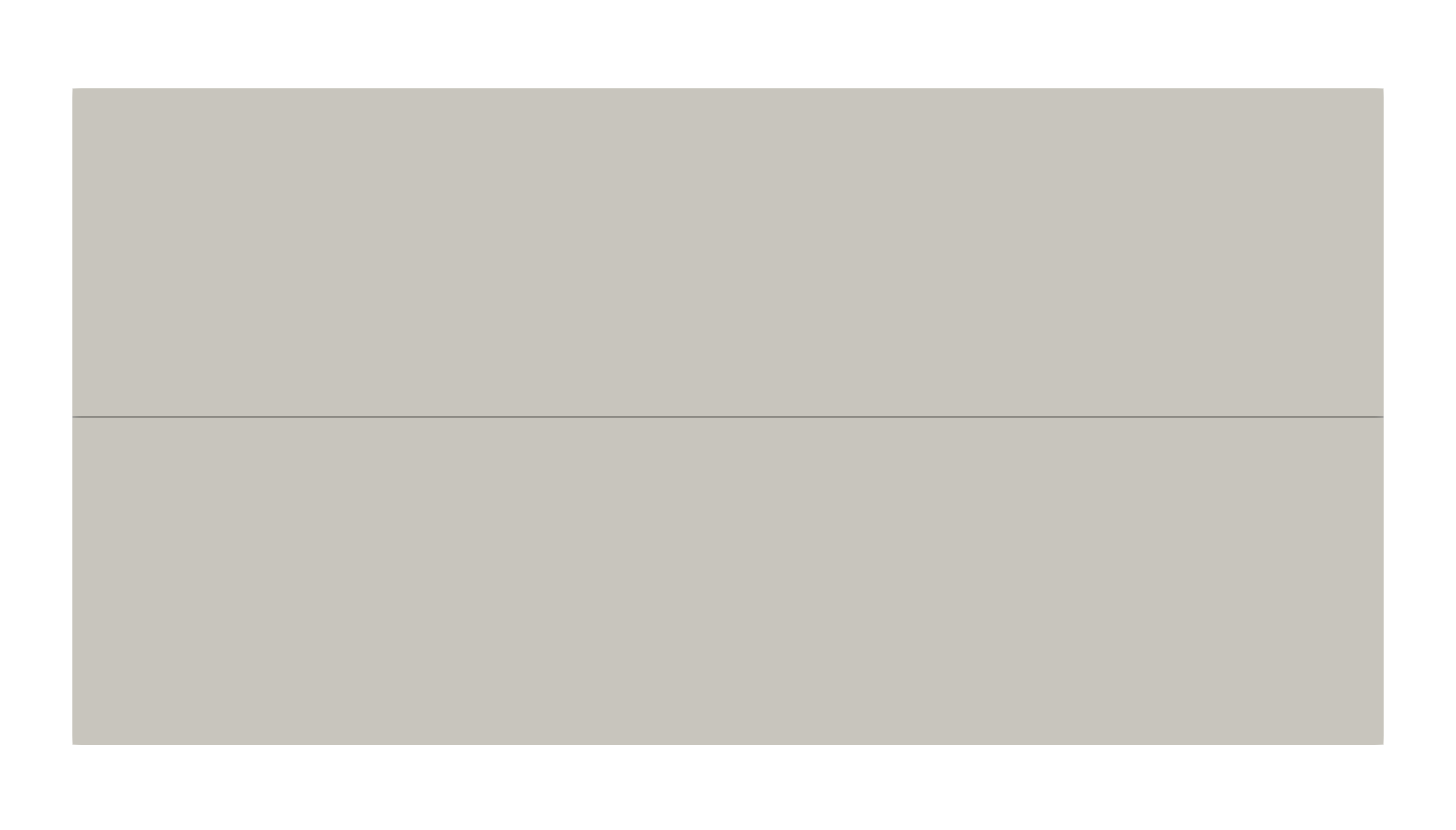}
    \caption{Large-scale IAD $\Vert\bnabla\overline{\alpha}_1\Vert\in(0,1.6)$
    without inter-scale transfer
    {\protect\tikz \protect\node [rectangle, left color=paraviewwhite, right color=black, anchor=north, minimum width=1cm, minimum height=0.1cm] (box) at (0,0){};}
    (top),
    with inter-scale transfer
    {\protect\tikz \protect\node [rectangle, left color=paraviewwhite, right color=blue, anchor=north, minimum width=1cm, minimum height=0.1cm] (box) at (0,0){};}
    (bottom),
    and small-scale IAD $\Sigma\in(0,3.4)$ with inter-scale transfer
    {\protect\tikz \protect\node [rectangle, left color=paraviewwhite, right color=red, anchor=north, minimum width=1cm, minimum height=0.1cm] (box) at (0,0){};}
    (bottom). Snapshots are taken each $0.25$ s from $t=0$ s to $t=2.5$ s from top to bottom and left to right.}
    \label{fig:compare-IAD_2_scale}
\end{figure}

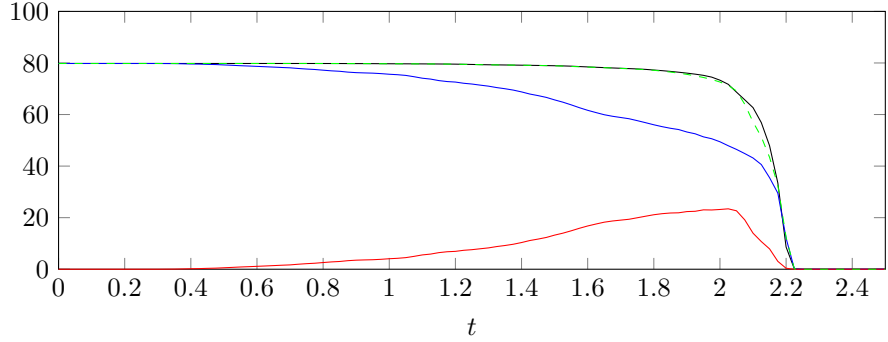
\begin{figure}[!ht]
    \centering
    \begin{tikzpicture}
        \centering
        \begin{axis}[
                width=.7\textwidth,
                height=5cm,
                xmin=0,
                xmax=2.5,
                ymin=0,
                ymax=100,
                xlabel={$t$},
            ]
            \addplot[black] table [x={time}, y={arho1}] {integrated.txt};
            \addplot[blue] table [x={time}, y={arho1LS}] {integrated.txt};
            \addplot[red] table [x={time}, y={arho1d}] {integrated.txt};
            \addplot [green, dashed]
            table [x=time ,y expr=\thisrow{arho1LS}+\thisrow{arho1d}] {integrated.txt};
        \end{axis}
    \end{tikzpicture}
    \caption{Evolution in time of liquid effective density for large-scale $\alpha_1\rho_1$ and no inter-scale transfer (black),
    for large-scale $\alpha_1\rho_1$ and inter-scale transfer (blue),
    for small-scale $\alpha_1^d\rho_1^d$ and inter-scale transfer (red),
    for both scales $\alpha_1\rho_1+\alpha_1^d\rho_1^d$ and inter-scale transfer (dashed green),
    }
    \label{fig:compare-mass-evo}
\end{figure}

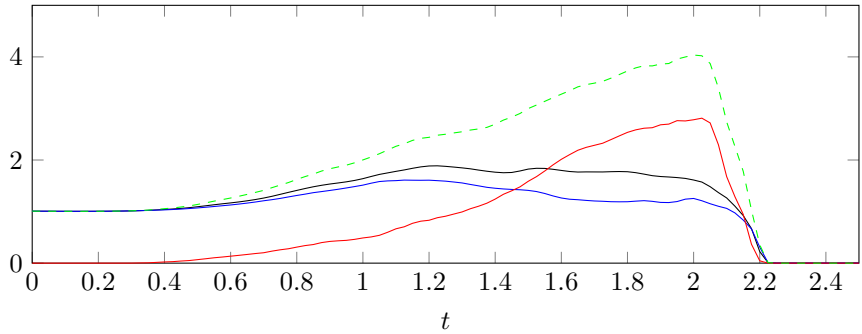
\begin{figure}
    \centering
    \begin{tikzpicture}
        \centering
        \begin{axis}[
                width=.7\textwidth,
                height=5cm,
                xmin=0,
                xmax=2.5,
                ymin=0,
                ymax=5,
                xlabel={$t$},
            ]
            \addplot[black] table [x={time}, y={IAD}] {integrated.txt};
            \addplot[blue] table [x={time}, y={IADLS}] {integrated.txt};
            \addplot[red] table [x={time}, y={capSigma}] {integrated.txt};
            \addplot [green, dashed]
            table [x=time ,y expr=\thisrow{IADLS}+\thisrow{capSigma}] {integrated.txt};
        \end{axis}
    \end{tikzpicture}
    \caption{Evolution in time of large-scale IAD $\Vert\bnabla\overline{\alpha}_1\Vert$ and no inter-scale transfer (black),
    for large-scale $\Vert\bnabla\overline{\alpha}_1\Vert$ and inter-scale transfer (blue),
    for small-scale $\Sigma$ and inter-scale transfer (red),
    for both scales $\Vert\bnabla\overline{\alpha}_1\Vert+\Sigma$ and inter-scale transfer (dashed green),
    }
    \label{fig:compare-IAD-evo}
\end{figure}



\end{document}